\documentclass[]{nature_mod}
\usepackage{amsmath,amsfonts}

\usepackage{txfonts}
\usepackage{color}
\usepackage{xr}
\usepackage{graphicx}
\usepackage{tocloft}
\usepackage{rotating}
\usepackage{booktabs}
\usepackage[table]{xcolor}
\usepackage{verbatim}

\usepackage{hyperref}
\hypersetup{pdfpagemode=None,%
            plainpages=false,%
            pdfstartview=FitH,%
            pdffitwindow=true,%
            colorlinks,%
            linkcolor=[rgb]{0,0,0},%
            citecolor=[rgb]{0,0,0},%
            urlcolor=black,%
            pdfauthor={},%
            pdftitle={},%
            pdfsubject={},%
            pdfkeywords={},%
      }

\definecolor{MyRed2}{rgb}{0.7,0.2,0.2}
\definecolor{MyBlue}{rgb}{0.2,0.2,0.7}
\definecolor{MyGreen}{rgb}{0.2,0.7,0.2}

\def\letter#1{\textbf{#1,}}
\def\letterD#1{\textbf{#1}}

\definecolor{tableShade}{HTML}{F1F5FA} 
\definecolor{MyGray}{rgb}{0.95,0.95,0.95}

\newcommand{\ra}[1]{
\renewcommand{\arraystretch}{#1}}
\newcommand{\rr}{\raggedright}

\title{Link communities reveal multi-scale complexity in networks}

\author{Yong-Yeol Ahn$^{1,2}$, James P.~Bagrow$^{1,2}$ \& Sune Lehmann$^{3,4}$}

\begin{document}

\maketitle

\begin{affiliations}
 \item Center for Complex Network Research, Department of Physics, Northeastern University, Boston, MA 02115.
 \item Center for Cancer Systems Biology, Dana-Farber Cancer Institute, Harvard University, Boston, MA 02215.
 \item Institute for Quantitative Social Science, Harvard University, Cambridge MA, 02138.
 \item College of Computer and Information Science, Northeastern University, Boston MA, 02115
\end{affiliations}

\begin{abstract}
Networks have become a key approach to understanding systems of interacting objects, unifying the study of diverse phenomena including biological organisms and human society\cite{newman_structure_2006, caldarelli_scale-free_2007, dorogovtsev_critical_2008}. One crucial step when studying the structure and dynamics of networks is to identify com\-muni\-ties\cite{newmanGirvanCommsPNAS,Fortunato201075}: groups of related nodes that correspond to functional subunits such as protein complexes\cite{krogan_APMS_2006,gavin_APMS_2006} or social spheres\cite{wassermanFaustBookSocNetAnalysis,palla_cpm_2005,pallaQuantifying_2008}. Communities in networks often overlap\cite{palla_cpm_2005,pallaQuantifying_2008} such that nodes simultaneously belong to several groups. Meanwhile, many networks are known to possess hierarchical organisation, where communities are recursively grouped into a hierarchical structure\cite{ravasz_science_2002,salespardo_extracting_2007,clauset_nature_2008}. However, the fact that many real networks have communities with pervasive overlap, where each and every node belongs to more than one group, has the consequence that a global hierarchy of nodes cannot capture the relationships between overlapping groups. Here we reinvent communities as groups of links rather than nodes and show that this unorthodox approach successfully reconciles the antagonistic organising principles of  overlapping communities and hierarchy. In contrast to the existing literature, which has entirely focused on grouping nodes, link communities naturally incorporate overlap while revealing hierarchical organization. We find relevant link communities in many networks, including major biological networks such as protein-protein interaction\cite{krogan_APMS_2006,gavin_APMS_2006,yu_ppi_2008} and metabolic networks\cite{ravasz_science_2002,guimera_functional_2005,feist_ecoli_2007}, and show that a large social network\cite{onnela_structure_2007,pallaQuantifying_2008,gonzales_uncovering_2008} contains hierarchically organized community structures spanning inner-city to regional scales while maintaining pervasive overlap. Our results imply that link communities are fundamental building blocks which reveal overlap and hierarchical organization in networks to be two aspects of the same phenomenon.
\end{abstract} 

\begin{figure}\centering
\includegraphics[width=.6\textwidth]{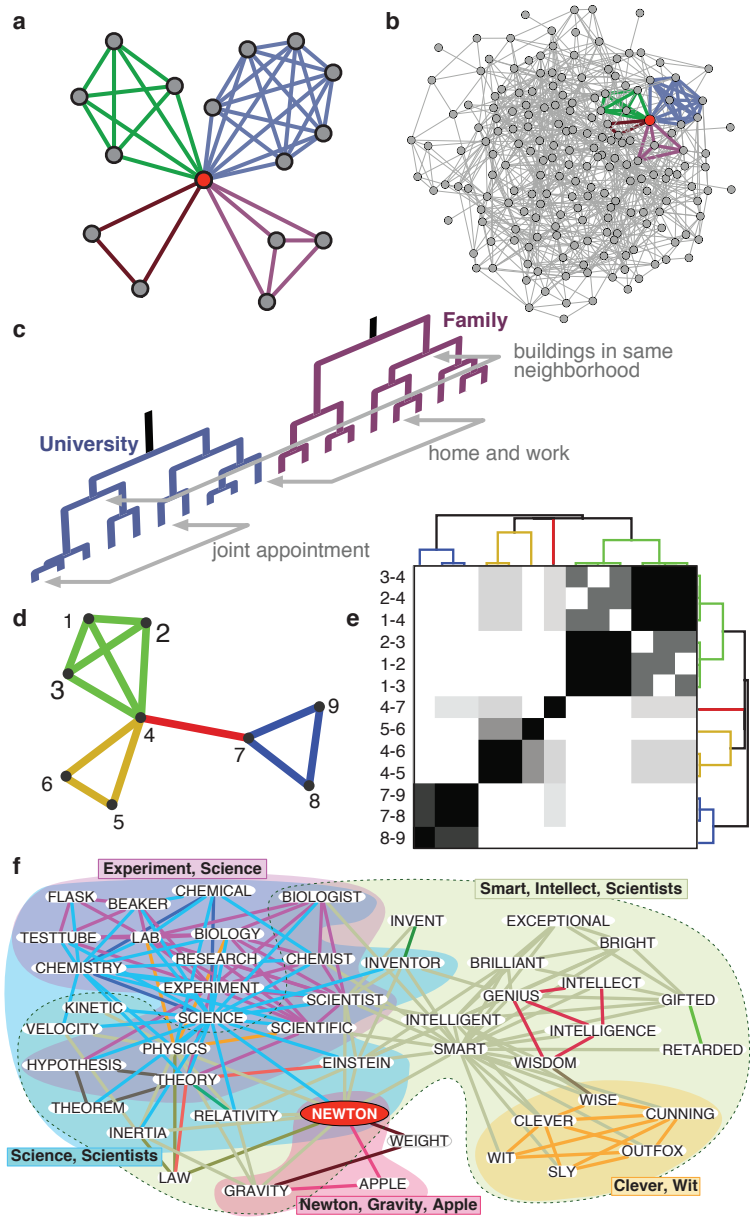}
\caption[]{\textbf{Overlapping communities lead to dense networks and prevent the discovery of a single node hierarchy.}  \letter{a} Local structure in many networks is simple: an individual node sees the communities it belongs to.  \letter{b} Complex global structure emerges when every node is in the situation displayed in \letterD{a}.  \letter{c} Pervasive overlap hinders the discovery of hierarchical organization because nodes cannot occupy multiple leaves of a node dendrogram, preventing a single tree from encoding the full hierarchy. \letter{d} An example showing link communities (colours in \letterD{d}), the link similarity matrix (\letterD{e}; darker entries show more similar pairs of links) and the link dendrogram (\letterD{e}). \letter{f} Link communities from the full word association network around the word `Newton'. Link colours represent communities and filled regions provide a guide for the eye. Link communities capture concepts related to science and allow substantial overlap. Note that the words were produced by experiment participants during free word associations. \label{fig:overlap}}
\end{figure}

Although no common definition has been agreed upon, it is widely accepted that a community should have more internal than external connections\cite{radicchi-definition-2004,newman_finding_2004,rosvall_infomap_2008,PhysRevLett.93.218701,Li:arXiv0807.0521,lancichinetti_detecting_2009}. Counterintuitively, highly overlapping communities can have many more external than internal connections (Fig.~\ref{fig:overlap}a, b). Because pervasive overlap breaks even this fundamental assumption, a new approach is needed.

The discovery of hierarchy and community organization has always been considered a problem of determining the correct membership (or memberships) of each node. Notice that, whereas nodes belong to multiple groups (individuals have families, co-workers and friends; Fig.~1c), links often exist for one dominant reason (two people are in the same family, work together or have common interests). Instead of assuming that a community is a set of nodes with many links between them, we consider a community to be a set of closely interrelated links.

Placing each link in a single context allows us to reveal hierarchical and overlapping relationships simultaneously. We use hierarchical clustering with a similarity between links to build a dendrogram where each leaf is a link from the original network and branches represent link communities (Fig.~1d,e and Methods). In this dendrogram, links occupy unique positions whereas nodes naturally occupy multiple positions, owing to their links. We extract link communities at multiple levels by cutting this dendrogram at various thresholds. Each node inherits all memberships of its links and can thus belong to multiple, overlapping communities. Even though we assign only a single membership per link, link communities can also capture multiple relationships between nodes, because multiple nodes can simultaneously belong to several communities together.

The link dendrogram provides a rich hierarchy of structure, but to obtain the most relevant communities it is necessary to determine the best level at which to cut the tree. For this purpose, we introduce a natural objective function, the partition density, $D$, based on link density inside communities; unlike modularity\cite{newman_finding_2004}, $D$ does not suffer from a resolution limit\cite{FortunatoBarthelemy07_ModularityResolution} (Methods). Computing $D$ at each level of the link dendrogram allows us to pick the best level to cut (although meaningful structure exists above and below that threshold). It is also possible to optimize $D$ directly. We can now formulate overlapping community discovery as a well-posed optimization problem, accounting for overlap at every node without penalizing that nodes participate in multiple communities. 

As an illustrative example, Fig.~1f shows link communities around the word `Newton' in a network of commonly associated English words. (See Supplementary Information, section S6, for details on networks used throughout the text.) The `clever, wit' community is correctly identified inside the `smart/intellect' community. The words `Newton' and `Gravity' both belong to the `smart/intellect', `weight' and `apple' communities, illustrating that link communities capture multiple relationships between nodes. See Supplementary Information, section S3.6, for further visualizations.

Having unified hierarchy and overlap, we provide quantitative, real-world evidence that a link-based approach is superior to existing, node-based approaches. Using data-driven performance measures, we analyse link communities found at the maximum partition density in real-world networks, compared with node communities found by three widely used and successful methods: clique percolation\cite{palla_cpm_2005}, greedy modularity optimization\cite{clauset_2004_finding} and Infomap\cite{rosvall_infomap_2008}. Clique percolation is the most prominent overlapping community algorithm, greedy modularity optimization is the most popular modularity-based\cite{newman_finding_2004} technique and Infomap is often considered the most accurate method available\cite{lancichinetti-comparison-2009}.

We compiled a test group of 11 networks covering many domains of active research and representing the wide body of available data (Supplementary Table 2). These networks vary from small to large, from sparse to dense, and from those with modular structure to those with highly overlapping structure. We highlight a few data sets of particular scientific importance: The mobile phone network is the most comprehensive proxy of a large-scale social network currently in existence\cite{onnela_structure_2007,gonzales_uncovering_2008}; the metabolic network iAF1260, from \emph{Escherichia coli} K-12 MG1655 strain, is one of the most elaborate reconstructions currently available\cite{feist_ecoli_2007}; and the three protein-protein interaction networks of \emph{Saccharomyces cerevisiae} are the most recent and complete protein-protein interaction data yet published\cite{yu_ppi_2008}.

These networks possess rich metadata that allow us to describe the structural and functional roles of each node. For example, the biological roles of each protein in the protein-protein interaction network can be described by a controlled vocabulary (Gene Ontology terms\cite{go}). By calculating metadata-based similarity measures between nodes (Methods and Supplementary Information, section S5), we can determine the quality of communities by the similarity of the nodes they contain (`community quality'). Likewise, we can use metadata to estimate the expected amount of overlap around a node, testing the quality of the discovered overlap according to the metadata (`overlap quality'). For example, metabolites that participate in more metabolic pathways are expected to belong to more communities than metabolites that participate in fewer pathways. Some methods may find high-quality communities but only for a small fraction of the network; coverage measures describe how much of the network was classified by each algorithm (`community coverage') and how much overlap was discovered (`overlap coverage'). Each community algorithm is tested by comparing its output with the metadata, to determine how well the discovered community structure reflects the metadata, according to the four measures. Each measure is normalized such that the best method attains a value of one. `Composite performance' is the sum of these four normalized measures, such that the maximum achievable score is four. Full details are in Methods and Supplementary Information, sections S5 and S6.

Figure 2 displays the results of this quantitative comparison, showing that link communities reveal more about every network's metadata than other tested methods. Not only is our approach the overall leader in every network, it is also the winner in most individual aspects of the composite performance for all networks, particularly the quality measures. The performance of link communities stands out for dense networks, such as the metabolic and word association networks, which are expected to have pervasively overlapping structure.

It is instructive to examine further the statistics of link communities in the metabolic and mobile phone networks (Fig.~3). The community size distribution at the optimum value of $D$ is heavy tailed for both networks, whereas the number of communities per node distinguishes them (Fig.~3, insets): Mobile phone users are limited to a smaller range of community memberships, most likely as a result of social and time constraints. Meanwhile, the membership distribution of the metabolic network displays the universality of currency metabolites (water, ATP and so on) through the large number of communities they participate in. Notable previous work\cite{ravasz_science_2002,guimera_functional_2005} removed currency metabolites before identifying meaningful community structure. The statistics presented here match current knowledge about the two systems, further confirming the communities' relevance.

Having established that link communities at the maximal partition density are meaningful and relevant, we now show that the link dendrogram reveals meaningful communities at different scales. Figure 4a-c shows that mobile phone users in a community are spatially co-located. Figure 4a maps the most likely geographic locations of all users in the network; several cities are present. In Fig.~4b, we show (insets) several communities at different cuts above the optimum threshold, revealing small, intra-city communities. Below the optimum threshold, larger, yet still spatially correlated, communities exist (Fig.~4c). Because we expect a tight-knit community to have only small geographical dispersion, the clustered structures on the map indicate that the communities are meaningful. The geographical correlation of each community does not suddenly break down, but is sustained over a wide range of thresholds. In Fig.~4d, we look more closely at the social network of the largest community in Fig.~4c, extracting the structure of its largest subcommunity along with its remaining hierarchy and revealing the small-scale structures encoded in the link dendrogram. This example provides evidence for the presence of spatial, hierarchical organization at a societal scale. To validate the hierarchical organization of communities quantitatively throughout the dendrogram, we use a randomized control dendrogram that quantifies how community quality would evolve if there were no hierarchical organization beyond a certain point. Figure 4e shows that the quality of the actual communities decays much more slowly than the control, indicating that real link dendrograms possess a large range of high quality community structures. The quantitative results of Fig.~4 are typical for the full test group, implying that rich, meaningful community structure is contained within the link dendrogram. Additional results supporting these conclusions are presented in Supplementary Information, section S7.

Many cutting-edge networks are far from complete. For example, an ambitious project to map all protein-protein interactions in yeast is currently estimated to detect approximately 20\% of connections\cite{yu_ppi_2008}. As the rate of data collection continues to increase, networks become denser and denser, overlap becomes increasingly pervasive and approaches specifically designed to untangle complex, highly overlapping structure become essential. More generally, the shift in perspective from nodes to links represents a fundamentally new way to study complex systems. Here we have taken steps towards understanding the consequences of a link-based approach, but its full potential remains unexplored. Our work has primarily focused on the highly overlapping community structure of complex networks, but, as we have shown, the hierarchy that organizes these overlapping communities holds great promise for further study.

While finalizing this manuscript, we have been made aware of a similar approach developed independently by T. S. Evans and R. Lambiotte\cite{evans_line_2009,evans_line2_2009}.


\begin{methods}
\subsection*{Link communities}
For an undirected, unweighted network, we denote the set of node $i$ and its neighbours as $n_+(i)$. Limiting ourselves to link pairs that share a node, expected to be more similar than disconnected pairs, we find the similarity, $S$ between links $e_{ik}$ and $e_{jk}$ to be
\begin{equation}
	S(e_{ik}, e_{jk}) =   \frac{\left| n_+(i) \cap n_+(j) \right|}{\left| n_+(i) \cup n_+(j) \right|}
	\label{eqn:jaccardSimFM}
\end{equation}
Shared node $k$ does not appear in $S$ because it provides no additional information and introduces bias. Single-linkage hierarchical clustering builds a link dendrogram from equation (2) (ties in $S$ are agglomerated simultaneously). Cutting this dendrogram at some clustering threshold---for example the threshold with maximum partition density (see below)---yields link communities. See Supplementary Information for details, generalizations to multipartite and weighted graphs, and the usage of other algorithms.

\subsection*{Partition density}
For a network with $M$ links and $N$ nodes, $P=\{P_1,\ldots,P_C\}$ is a partition of the links into $C$ subsets.  The number of links in subset $P_c$ is $m_c = \left|P_ c\right|$.  The number of induced nodes, all nodes that those links touch, is  $n_c = \left|\cup_{e_{ij} \in P_c} \{i,j\}\right|$.  Note that $\sum_c m_c = M$ and $\sum_c n_c \geq N$ (assuming no unconnected nodes). The link density $D_c$ of community $c$ is
\begin{equation}
D_c=\frac{m_c - (n_c - 1)}{n_c(n_c-1)/2 - (n_c -1 )}.
\label{eqn:ApartDensityFM}
\end{equation}
This is the number of links in $P_c$, normalized by the minimum and maximum numbers of links possible between those nodes, assuming they remain connected. (We assume $D_c = 0$ if $n_c=2$.) The partition density, $D$, is the average of $D_c$, weighted by the fraction of present links:
\begin{equation}\label{eq:fullpdFM}
D  =\frac{2}{M}\sum_c m_c \frac{m_c -(n_c-1)}{(n_c-2)(n_c-1)}.
\end{equation}
Equation~(3) does not possess a resolution limit\cite{FortunatoBarthelemy07_ModularityResolution} since each term is local in $c$.

\subsection*{Community validation}
Nontrivial communities possess 3+ nodes.  We use metadata `enrichment' to assess community quality, comparing how similar nodes are within nontrivial communities relative to all nodes (global baseline). Overlap quality is the mutual information between the number of nontrivial memberships and the overlap metadata (Supplementary Table 2). Community coverage is the fraction of nodes belonging to 1$+$ nontrivial communities. Overlap coverage, because methods with equal community coverage can extract different amounts of overlap, is the average number of nontrivial memberships per node (equivalent to community coverage for non-overlapping methods). See Supplementary Information for details.

\subsection*{Validation of hierarchical structure}
To test whether the hierarchical structure is valid beyond some threshold, $t_*$, we introduce the following control. First we compute the similarities $S(e_{ik},e_{jk})$ for all connected edge pairs $(e_{ik},e_{jk})$, as normal.  We then perform our standard single-linkage hierarchical clustering, merging all edge pairs in descending order of $S$ for $S \geq t_*$, fixing the community structure at $t = t_*$.  This randomization only alters the merging order, and ensures that the rate of edge pair merging is preserved, because the same similarities are clustered. This strictly controls not only the merging rate but also the similarity distributions and the high-quality community structure found at $t_*$. This procedure ensures that the dendrogram is properly randomized while other salient features are conserved. Full details are in Supplementary Information, section~S7.4.
\end{methods}

\spacing{1}

\begin{addendum}
\item[Acknowledgements] The authors thank A.-L. Barab\'asi, S. Ahnert, J. Park, D.-S. Lee, P.-J. Kim, N. Blumm, D. Wang, M. A. Yildirim and H. Yu. The authors acknowledge the Center for Complex Network Research, supported by the James S. McDonnell Foundation 21st Century Initiative in Studying Complex Systems; the NSF-DDDAS (CNS-0540348), NSF-ITR (DMR-0426737) and NSF-IIS-0513650 programmes; US ONR Award N00014-07-C; the NIH (U01 A1070499-01/Sub \#:111620-2); the DTRA (BRBAA07-J-2-0035); the NS-CTA sponsored by US ARL (W911NF-09-2-0053); and NKTH NAP (KCKHA005). S.L. acknowledges support from the Danish Natural Science Research Council.
 \item[Author Contributions] Y.-Y.A., J.P.B. and S.L. designed and performed the research and wrote the manuscript.
 \item[Author Information] Reprints and permissions information is available at www.nature.com/reprints. The authors declare no competing financial interests. Readers are welcome to comment on the online version of this article at www.nature.com/nature. Correspondence and requests for materials should be addressed to S.L. (email: sune.lehmann@gmail.com).
\end{addendum}

\begin{figure}\centering
\includegraphics[width=\textwidth]{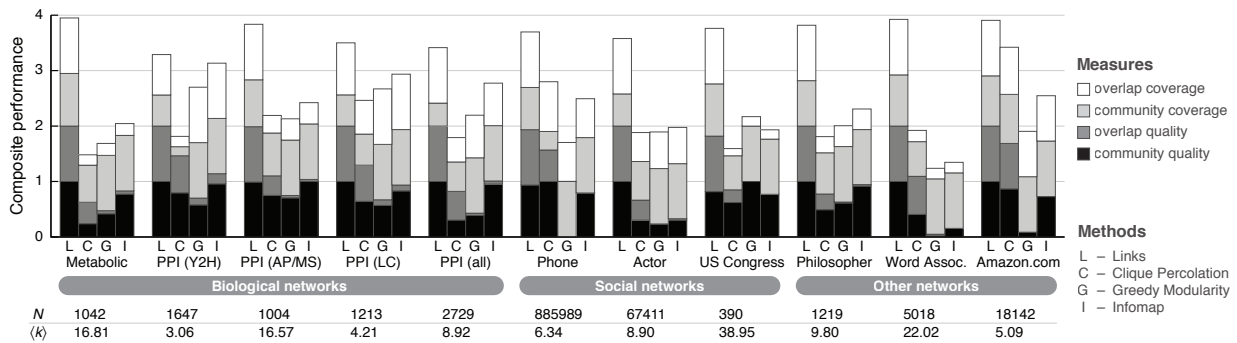}
\caption[]{\textbf{Assessing the relevance of link communities and other methods using real world networks.} Composite performance (Methods and Supplementary Information) is a data-driven measure of the quality (relevance of discovered memberships) and coverage (fraction of network classified) of community and overlap. Tested algorithms are link clustering, introduced here; clique percolation\cite{palla_cpm_2005}; greedy modularity optimization\cite{clauset_2004_finding}; and Infomap\cite{rosvall_infomap_2008}. Test networks were chosen for their varied sizes and topologies and to represent the different domains where network analysis is used. Shown for each are the number of nodes, $N$, and the average number of neighbours per node, $\left<k\right>$. Link clustering finds the most relevant community structure in real-world networks. AP/MS, affinity-purification/mass spectrometry; LC, literature curated; PPI, protein-protein interaction; Y2H, yeast two-hybrid. \label{fig:performance}}
\end{figure}

\begin{figure}\centering
\includegraphics[width=\textwidth]{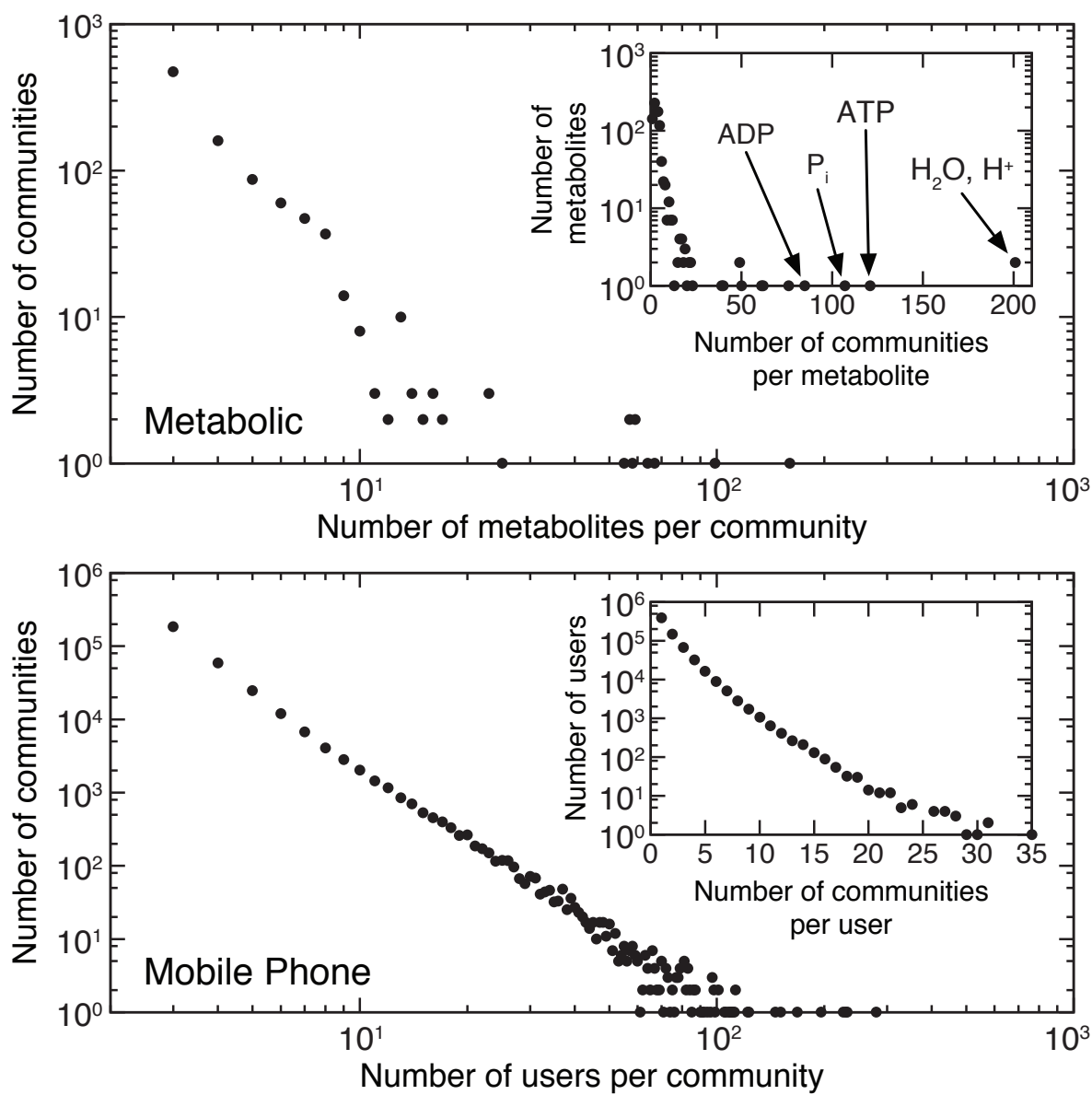}
\caption[]{\textbf{Community and membership distributions for the metabolic and mobile phone networks.}   
The distribution of community sizes and node memberships (insets). Community size shows a heavy tail. The number of memberships per node is reasonable for both networks: we do not observe phone users that belong to large numbers of communities and we correctly identify currency metabolites, such as water, ATP and inorganic phosphate (P${}_\mathsf{i}$), that are prevalently used throughout metabolism. The appearance of currency metabolites in many metabolic reactions is naturally incorporated into link communities, whereas their presence hindered community identification in previous work\cite{ravasz_science_2002,guimera_functional_2005}.\label{fig:stats}}
\end{figure}

\begin{figure}\centering
\includegraphics[width=0.6\textwidth]{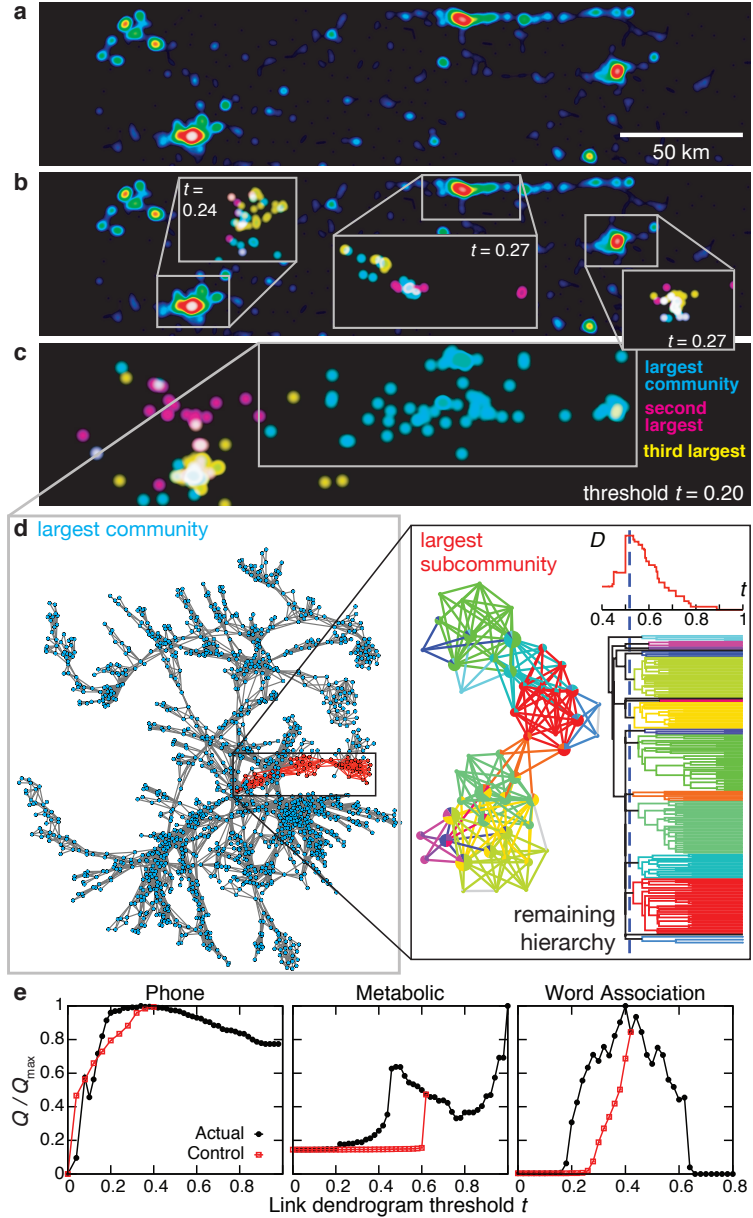}
\caption[]{\textbf{Meaningful communities at multiple levels of the link dendrogram.} \letterD{a}-\letter{c} The social network of mobile phone users displays co-located, overlapping communities on multiple scales. \letter{a} Heat map of the most likely locations of all users in the region, showing several cities. \letter{b} Cutting the dendrogram above the optimum threshold yields small, intra-city communities (insets). \letter{c} Below the optimum threshold, the largest communities become spatially extended but still show correlation. \letter{d} The social network within the largest community in \letterD{c}, with its largest subcommunity highlighted. The highlighted subcommunity is shown along with its link dendrogram and partition density, $D$, as a function of threshold, t. Link colours correspond to dendrogram branches. \letter{e} Community quality, $Q$, as a function of dendrogram level, compared with random control (Methods).\label{fig:heatmap} }
\end{figure}

\clearpage
\setcounter{page}{1}
\setcounter{figure}{0}
\setlength{\parskip}{9pt}

\newcounter{lastnote}
\newenvironment{scilastnote}{%
\setcounter{lastnote}{\value{enumiv}}%
\addtocounter{lastnote}{+1}%
\begin{list}%
{\arabic{lastnote}.}
{\setlength{\leftmargin}{.22in}}
{\setlength{\labelsep}{.5em}}}
{\end{list}}

\renewcommand{\thefigure}{S\arabic{figure}}
\renewcommand{\thesection}{S\arabic{section}}
\renewcommand{\thetable}{S\arabic{table}}
\renewcommand{\theequation}{S\arabic{equation}}

{\huge \noindent Supplementary Information}\\  \\
\baselineskip24pt
\spacing{1}
\noindent \emph{Link Communities Reveal Multi-Scale Complexity in Networks}\\
\noindent by Yong-Yeol Ahn, James P.~Bagrow, Sune Lehmann

\renewcommand*\contentsname{Table of Contents}
\begin{spacing}{0.9}
\tableofcontents
\listoffigures
\listoftables
\end{spacing}
\clearpage

\section{Introduction}

This document is organized as follows. Section~\ref{sec:methods} contains details regarding the implementation of link clustering, as well as the other community detection methods which were used in the main text. In Sec.~\ref{sec:remarks}, we discuss properties of link-partitions and important cases, such as ``what happens when a link should be a member of more than one community?'', and ``what happens in the case of no overlap?''. We show that the link clustering algorithm is able to successfully analyze both cases. Generalizations and extensions of link clustering are discussed in Sec.~\ref{sec:generalizations}.

The final sections of the document focus primarily on our community validation methodology.  To see how meaningful/useful link communities can be, we apply our method to a large corpus of networks, chosen specifically for their diversity and to form a representative sample of common network datasets. First, in Sec.~\ref{sec:testingComms}, we discuss the measures we use to evaluate different community algorithms. Then, details regarding how the chosen networks were collected and curated, and any particular details regarding how to apply the various validation measures are described in Sec.~\ref{sec:realdata}. Section \ref{sec:multi_scale_structure} focuses on studying and validating meaningful communities at multiple levels of the link dendrogram.  The appendix contains raw data regarding the various quality measures.

\section{Methods}\label{sec:methods}
Here we offer a detailed discussion of the different methods we have used in this work.  In particular we offer additional results about our new link communities and we list implementation details for applying other methods, such as parameter choices.  The raw (unnormalized) composite performance scores for all methods are shown in App.~\ref{app:all_measures}.

\subsection{Link clustering}

\subsubsection{Constructing a dendrogram}

The main text has introduced a hierarchical link clustering method to classify links into topologically related groups.  Here we provide further motivation for the suggested pair-wise link similarity measure. For simplicity, we limit ourselves to only \emph{connected} pairs of links (i.e. 
sharing a node) since it is unlikely that a pair of disjoint links are more similar to each 
other than a pair of links that share a node; at the same time this choice is much more 
efficient.  For a connected pair of links $e_{ik}$ and $e_{jk}$, we call the shared node $k$ a 
\emph{keystone} node  and $i$ and $j$ \emph{impost} nodes. 

\begin{figure}[!tbp]\centering
\includegraphics[width=0.9\columnwidth,trim=5 0 0 0,clip=true]{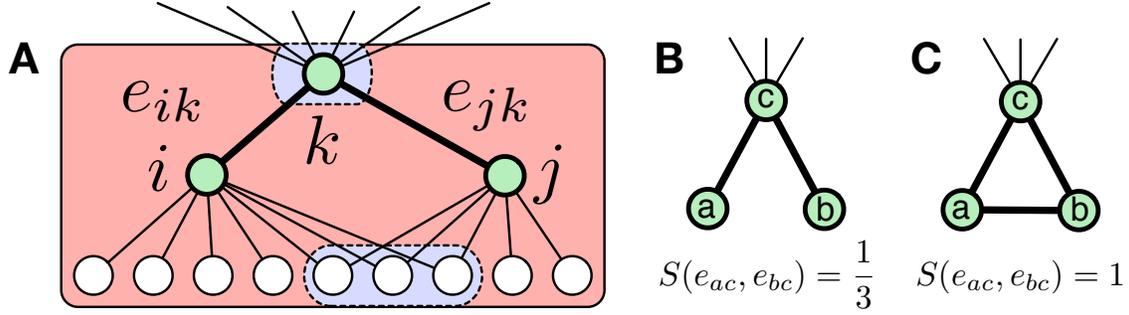}
\caption[Illustration of similarity measure between link pairs]{\letter{A} The similarity measure $S(e_{ik},e_{jk})$ between edges $e_{ik}$ and 
$e_{jk}$ sharing node $k$. For this example, $\left|n_+(i)\cup n_+(j)\right|=12$ and $\left|n_
+(i)\cap n_+(j)\right|=4$, giving $S=1/3$. Two simple cases: \letter{B} an isolated 
($k_a=k_b=1$), connected triple ($a$,$c$,$b$) has $S=1/3$, while \letter{C} an isolated 
triangle has $S=1$.}
\label{sfig:jaccardCartoon}
\end{figure}

If the only available information is the network topology, the most fundamental characteristic 
of a node is its neighbors. Since a link consists of two nodes, it is natural to use the 
neighbor information of the two nodes when we define a similarity between two links. However, 
since the links we are considering already share the keystone node, the neighbors of the 
keystone node provide no useful information. Moreover, if the keystone node is a hub, then the 
similarity is likely to be dominated by the keystone node's neighbors. For instance, if the 
hub's degree increases the similarity between the links connected to the hub also increases. 
This bias due to the keystone node's degree also prohibits us from applying traditional 
methods directly to the \emph{line graph} of the original graph, which is constructed by 
mapping the links into nodes. (Since a hub of degree $k$ becomes a fully connected subgraph of 
size $k$ in the line graph, the community structure can become radically different.) Thus, we 
neglect the neighbors of the keystone. We first define the \emph{inclusive} neighbors of a 
node $i$ as: 
\begin{equation}
	n_+(i) \equiv \{ x \mid d(i,x) \le 1 \}
	\label{eqn:nodeNeighbors}
\end{equation}
where $d(i,x)$ is the length of the shortest path between nodes $i$ and $x$. The set simply 
contains the node itself and its neighbors. From this, the similarity $S$ between links can be 
given by, e.g., the Jaccard index~\cite{jaccard_1901}:
\begin{equation}
	S(e_{ik}, e_{jk}) =   \frac{\left| n_+(i) \cap n_+(j) \right|}{\left| n_+(i) \cup n_+(j) 
\right|}
	\label{eqn:jaccardSim}
\end{equation}
An example illustration of this similarity measure is shown in Fig.~\ref{sfig:jaccardCartoon} 
(see Sec.~\ref{subsec:tanimoto} for generalizations of the similarity). 

\begin{figure}[]
	\begin{minipage}[t]{0.58\textwidth}
		\centering
\includegraphics[width=1\textwidth]{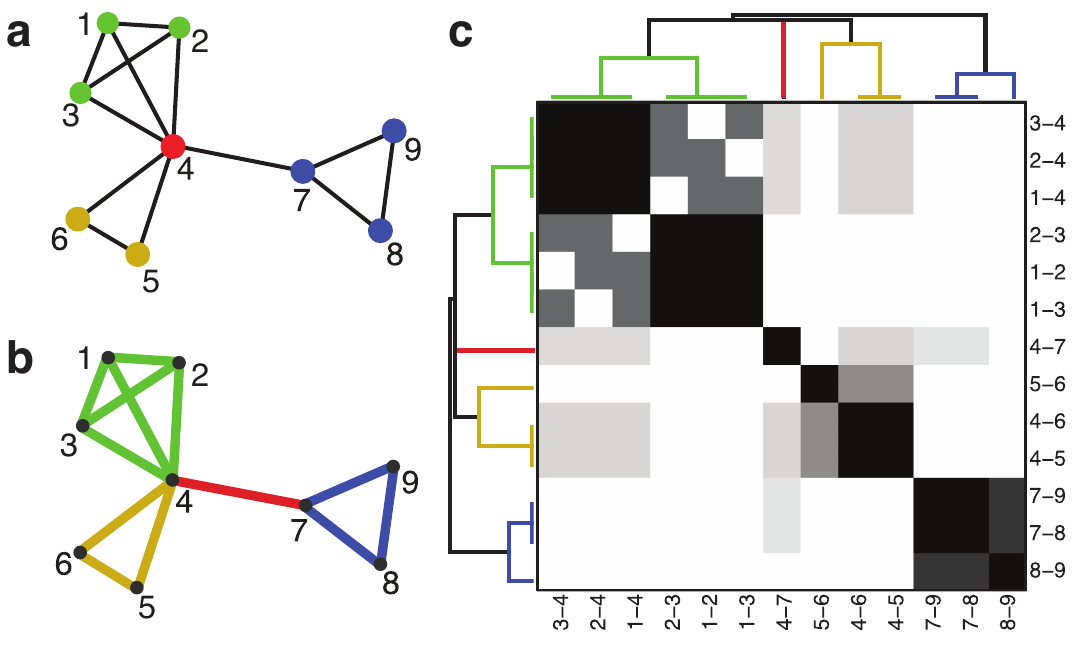}
	\end{minipage}
	\hspace{0.02\textwidth}
	\begin{minipage}[b]{0.35\textwidth}
	\caption[Example node and link communities]{
    An example network with node communities \letter{a} and link communities \letter{b}.  \letter{c} The resulting link similarity matrix and link dendrogram. Compare with main text Fig.~1.\label{sfig:nodelinkdend_example}}
		\vspace{1.9cm}
	\end{minipage}
\end{figure}

With this similarity, we use single-linkage hierarchical clustering to find hierarchical 
community structures. We use single-linkage mainly due to simplicity and efficiency, which 
enables us to apply link clustering to large-scale networks. However, it is also possible to use other options such as complete-linkage or average-linkage clustering. Each link is initially 
assigned to its own community; then, at each time step, the pair of links with the largest 
similarity are chosen and their respective communities are merged. Ties, which are common, are 
agglomerated simultaneously.  This process is repeated until all links belong to a single 
cluster.  The history of the clustering process is then stored in a dendrogram, which contains 
all the information of the hierarchical community organization. The similarity value at which 
two clusters merge is considered as the strength of the merged community, and is encoded as 
the height of the relevant dendrogram branch to provide additional information.  See Fig.~\ref{sfig:nodelinkdend_example} for an example.

\subsubsection{Partitioning the dendrogram: partition density}
\begin{figure}[!t]\centering
\includegraphics[width=0.9\textwidth]{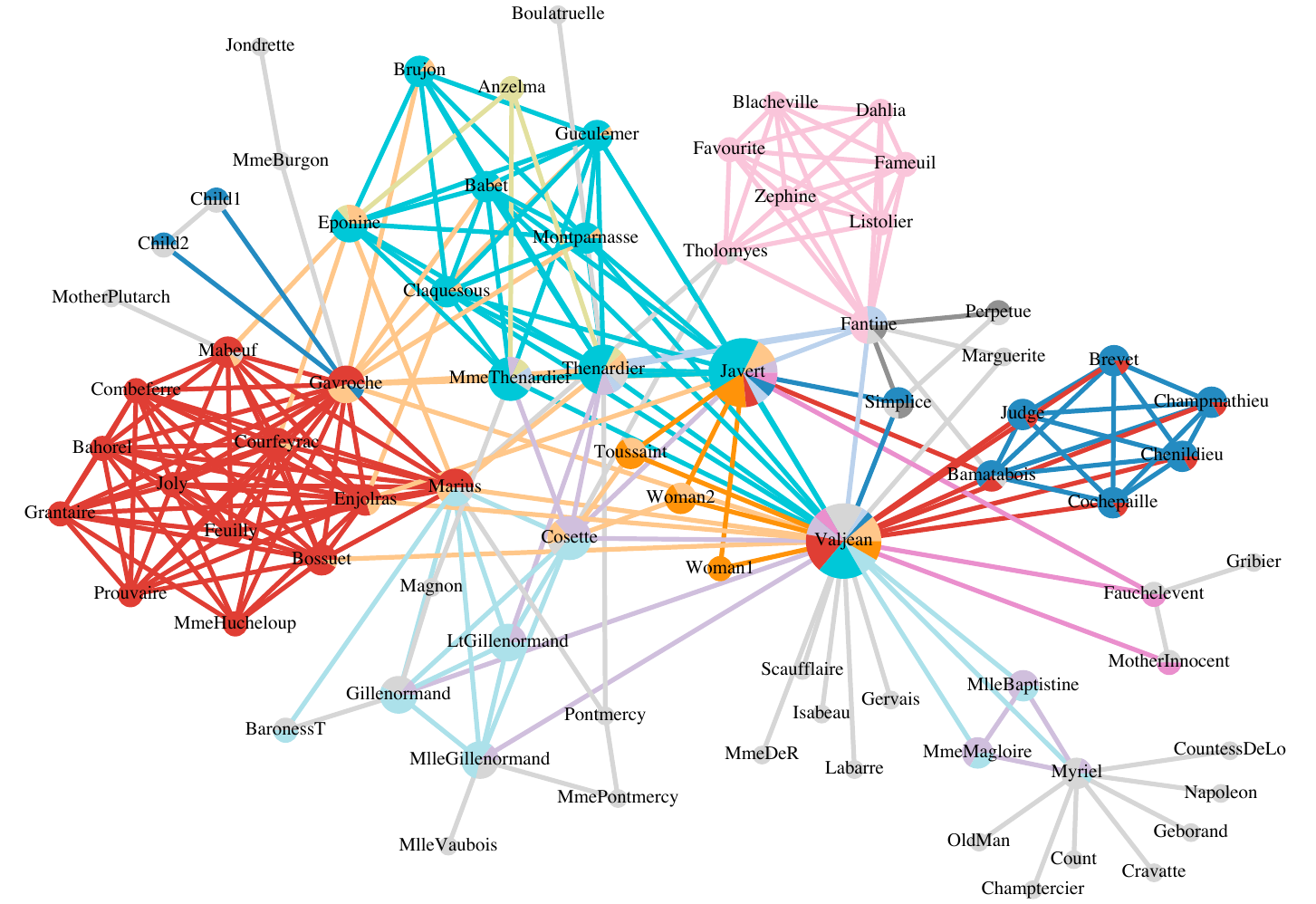}
\includegraphics[width=0.9\textwidth]{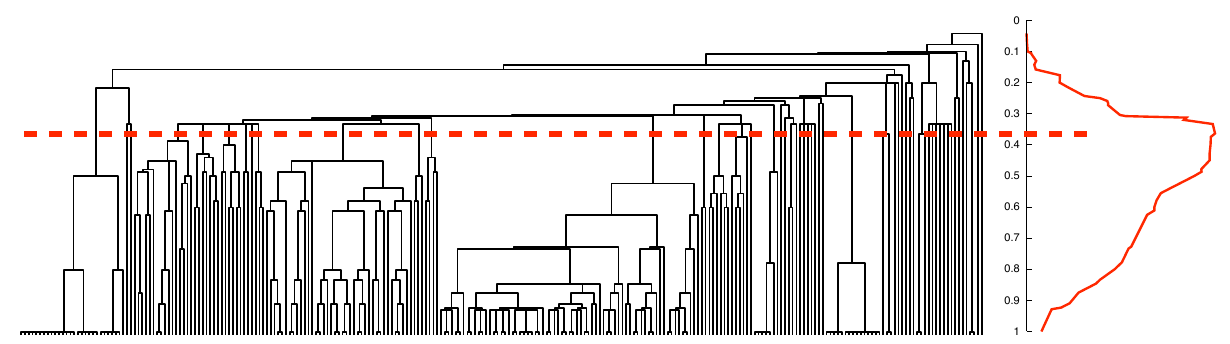}
\caption[Link communities in \emph{Les Mis\'erables}]{Link communities for the coappearance network of 
characters in the novel \emph{Les Mis\'erables} \cite{knuth_stanford_1993}.  \letter{Top} the 
network with link colors indicating the clustering, with grey indicating single-link clusters. 
Each node is depicted as a pie-chart representing its membership distribution.  The main 
characters have more diverse community membership. \letter{Bottom} the full link dendrogram (left)
and partition density (right). Note the internal blue community in the large blue and red clique 
containing Valjean.  Link clustering is able to unveil hierarchical structure even inside of cliques.\label{fig:lesmisDendro}}
\end{figure}

Hierarchical clustering methods repeatedly merge groups until \emph{all} elements are members 
of a single cluster.  This eventually forces highly disparate regions of the network into 
single clusters. To find meaningful communities rather than just the hierarchical organization 
pattern of communities, it is crucial to know where to partition the dendrogram. Modularity 
has been widely used for similar purposes in node-hierarchies~\cite{newman_finding_2004, danon_comparing_2005}, but is not easily defined for overlapping communities.\footnote{Several modifications of modularity that allow for ``fuzzy'' communities with relaxed interfaces (or overlapping nodes) to exist \cite{shenDetect2009,nicosiaExtending2009,PhysRevLett.93.218701,Li:arXiv0807.0521,lancichinetti_detecting_2009} have been suggested. However, in order to avoid the trivial optimum, where all nodes are part of all communities, each of these methods \emph{penalize} overlap, and are therefore not suitable for networks with pervasive overlap. (See Fig.~\ref{fig:overlap} of the main text)} Thus, we introduced a 
new quantity, the \emph{partition density} $D$, that measures the quality of a link partition (see Methods, main text). 
The partition density has a single global maximum along the dendrogram in almost all cases, 
because the value is just the average density at the top of the dendrogram (a single giant 
community with every link and node) and it is very small at the bottom of the dendrogram (most 
communities consists of a single link). This process is illustrated in Fig.~\ref{fig:lesmisDendro}. 

The maximum of $D$ is 1 but it can take values less than zero; $D=1$ when every community is a fully connected clique and $D=0$ when each community is a tree. Essentially, $D$ measures how ``clique-ish'' vs.~``tree-ish'' each link community is.  If a link community is less dense than a tree (when the community subgraph has disconnected components), then that community will give a negative contribution to $D$.  The minimum of $D_c$ is ${}-2/3$, given by one community of two disconnected edges.  Since $D$ is the average of $D_c$, there is a lower bound of $D=-2/3$.

\subsection{Node clustering}
We introduce "node clustering" as a control algorithm to offer a direct comparison to link clustering. In other words, if two algorithms are identical in every possible respect except that one classifies nodes and the other classifies links, how different will their performances be?  The node clustering method is closely related to the method introduced in Ravasz et al. \cite{ravasz_science_2002}. There are many ways to define a similarity between two nodes. We tried four different variations of the node similarity.  The four versions are following:
\begin{itemize}
\item $S(i,j) =  { \left| n(i) \cap n(j) \right| } / {\left| n(i) \cup n(j) \right|}$,
\item $S(i,j) =  { \left| n(i) \cap n(j) \right| } / \min( k_i , k_j )   $,
\item $S(i,j) =  { \left| n_+(i) \cap n_+(j) \right| } / {\left| n_+(i) \cup n_+(j) \right|}$,
\item $S(i,j) =  { \left| n_+(i) \cap n_+(j) \right| } / \min( k_i , k_j ) $,
\end{itemize}
where $n(i)$ means the neighbors, not inclusive neighbors, of the node $i$. Among those, we use the version in Eq.~\eqref{eq:nodesimilarity} since it finds more relevant communities across most networks we used. In addition, it is the definition most similar to link similarity. Thus, the node similarity is chosen to be 
\begin{equation}\label{eq:nodesimilarity}
S(i,j) =  \frac{ \left| n_+(i) \cap n_+(j) \right| }{\left| n_+(i) \cup n_+(j) \right|},
\end{equation}
where, as in the main text, $n_+(i)$ are the inclusive neighbors of node $i$. To determine the node dendrogram,  we use the same single linkage hierarchical clustering as we used for clustering links. This node dendrogram is cut at the point of maximum modularity~\cite{newman_finding_2004}.  Since this method is a nice control, but not necessarily applicable in the real world, we study it only in the SI.

\subsection{Other methods}
In order to evaluate its performance, we compare link clustering to existing, popular community detection methods. We chose three representative algorithms: the clique percolation method (CPM)~\cite{palla_cpm_2005}, which is widely recognized as state-of-the-art for detecting overlapping communities; Infomap \cite{rosvall_infomap_2008} which is the current state-of-the-art algorithm for detecting non-overlapping communities; and a greedy modularity optimization algorithm~\cite{newmanFastAlgorithm}, which is widely used in the literature.

\begin{table}[]\centering
    \begin{minipage}[c]{0.35\linewidth}\centering
\begin{tabular}{lc}
Network       & Modularity $Q$ \\
\hline                         
Metabolic     & 0.360562       \\
PPI (Y2H)     & 0.733042       \\
PPI (AP/MS)   & 0.722658       \\
PPI (LC)      & 0.864972       \\
PPI (all)     & 0.728056       \\
Phone         & 0.652382       \\
Actor         & 0.867364       \\
US Congress   & 0.275167       \\
Philosopher   & 0.454025       \\
Word Assoc.   & 0.343629       \\
Amazon.com    & 0.889058       \\
\end{tabular}
\end{minipage}\hfill
\begin{minipage}[c]{0.65\linewidth}
\caption[Modularity values for the test networks as discovered using greedy optimization]{The modularity values for the test networks studied in the main text, found using greedy modularity optimization~\cite{clauset_2004_finding}.  Many values are very high, indicating that the structure found by the greedy optimization algorithm is highly modular (at least according to the definition of modularity). Good modularity values typically lie between $0.3-0.7$, while higher values are rare \cite{newman_finding_2004}.\label{tbl:modularityBio}}
\end{minipage}
\end{table}

\subsubsection{Clique percolation}
Clique percolation \cite{palla_cpm_2005,palla_directed_2007} provides an elegant and highly useful method to uncover overlapping community structure~\cite{PallaQuantifyingSocialGroupEvolution2007}. It is currently the most popular and most successful tool available for this task.  A particularly interesting feature of this method is that it presents the experimenter with a ``knob'' $k$, the clique size, which can be used to tune the result between high coverage, low community quality (sparse communities) and low coverage, high community quality (dense communities). For some networks, such as the mobile phone network, a precedent exists for the choice of $k$, which we follow. Whenever that is not the case, we have computed the composite performance for a range of $k$'s and chosen the $k$ which results in the optimum overall performance\footnote{For some of the very large or very dense networks, we were not able to run clique percolation for large values of $k$ with the fastest existing software (even on a machine with 32 Gb of RAM), using the fast algorithm developed by Kumpala et al.~{\protect\cite{kumpula2008}}.}.  This weighs coverage and quality equally, however, and it remains at the discretion of the researcher to decide if this is optimal for his or her application. See Appendix~\ref{app:cpm_vals}.

The main drawback of CPM is its somewhat rigid definition of communities. When a network is very dense, it can become super-critical in the sense of clique percolation, which leads to giant clique communities. At the other end of the spectrum, when the network is too sparse, the network is sub-critical and there are not enough connected cliques to find any communities. For example, in the metabolic network, CPM's coverage is largely due to one giant community containing almost all nodes, leading to a minuscule community quality. Removing this giant community increases the enrichment value, but only $\sim 5\%$ of nodes remain. This situation is not unchanged by increasing clique size. For the Y2H network, however, the problem is sparsity: there are not enough cliques to find structure. 

\subsubsection{Modularity optimization}
To study how typical modularity~\cite{newman_finding_2004,newman_detecting_2004,newmanGirvanCommsPNAS} optimization methods perform, we choose the fast/greedy optimization method of Clauset, et al.~\cite{clauset_2004_finding}.  Although this particular modularity algorithm is the most popular one, more accurate methods exist, based on simulated annealing, extremal optimization, and more. (See~\cite{danon_comparing_2005} for additional details.)  However, the modularity values we found are often quite high (good modularity values typically lie between $0.3-0.7$, while higher values are rare \cite{newman_finding_2004}), so the lack of accuracy in our comparison is less likely to be from failing to find partitions near the system's maximum modularity. The modularity values found for the test networks are shown in Table \ref{tbl:modularityBio}.

\subsubsection{Infomap}
The Infomap algorithm~\cite{rosvall_infomap_2008} is becoming accepted as one of the best and most accurate node partitioning methods~\cite{lancichinetti-comparison-2009}.  It exploits deep results from information theory and uses a complex, multi-stage optimization scheme. In our application of this method, we used 100 restarts for the large networks (phone, amazon, etc.) and 1000 restarts for smaller networks.  The final partition that minimized the map length was then used.

\section{Properties of link communities}\label{sec:remarks}

\subsection{Link communities capture multiple memberships between nodes}

\begin{figure}[!ht]	
	\includegraphics[width=0.9\textwidth]{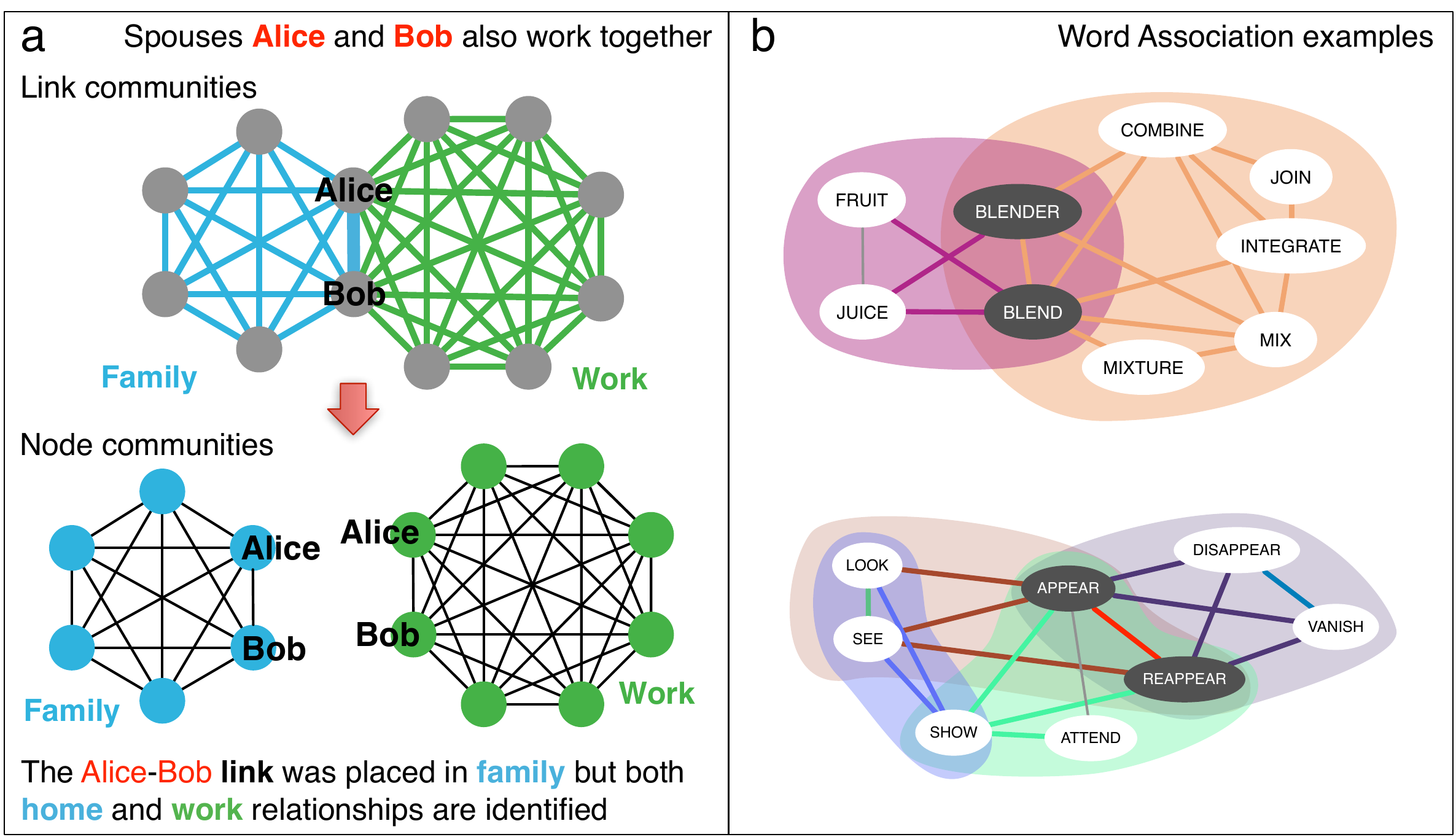}\centering         
\caption[Link communities identify multiple relationships between nodes]{Overlapping links. In the link community framework, a link may be assigned to only one community. By deriving node communities, however, the problem of effectively discovering multiple relationships between nodes is effectively solved. Two nodes can belong to many communities together regardless of the membership of the link between them. Left: illustration of the situation. Right: real examples from word association network. In the upper example, \emph{Blend} and \emph{blender} belong to both `fruit juice' community and `mix' community. In the bottom example, the link between \emph{appear} and \emph{reappear} does not even belong to any of the other communities, but they belong to several communities together. \label{sfig:link_overlap}}
\end{figure}

While clustering links is a much more flexible approach than clustering nodes, one might wonder whether this method is flexible enough---after all, it does not appear to take into account links that appear in multiple contexts (overlapping links). In the main text, we briefly address the issue of multiple relations represented by a single link. Main text Fig.~\ref{fig:overlap}f shows that it is very natural that two nodes of a given link can simultaneously belong to multiple communities even though the link itself belongs to only one community. Here, we let the examples in Fig.~\ref{sfig:link_overlap} provide further illumination of this point.

\begin{figure}[!tbp]  
        \includegraphics[height=0.75\textheight]{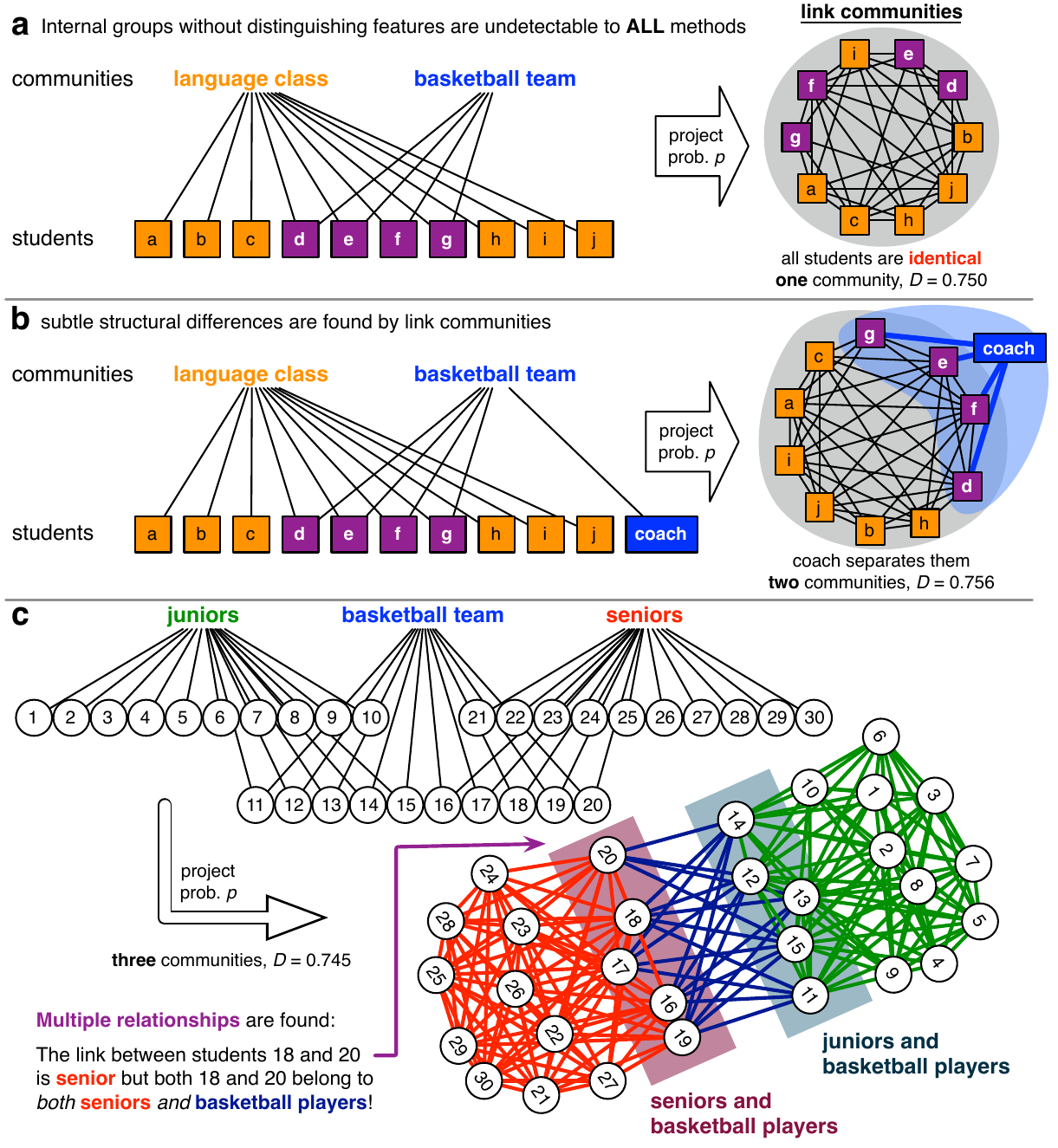}\centering
\caption[Subtle structural differences are detected by link communities]{Some small, illustrative examples of the subtle structural changes that link communities detect, using the bipartite social model of \cite{newman_whysocial_2003} with $p=0.8$, followed by our link communities algorithm.  In (\textbf{a}) there are no distinguishing structural features to separate the ``subsumed'' basketball team from the language class.  Detecting the team is impossible for all methods.  In (\textbf{b}) however, a single change allows for 100\% complete detection.  The entire basketball team is successfully found, even though only the coach-team links are separated. It doesn't take much to achieve the proper node communities.  (\textbf{c}) A more extreme example. Class and team detection are again 100\% accurate. Very subtle patterns are detectable (see, e.g., the word association communities in main text Fig.~\ref{fig:overlap}f and Figs.~\ref{fig:lesmisDendro},~\ref{fig:node_hierarchy}, \ref{sfig:PPI_example_1}, \ref{sfig:PPI_example_2}).\label{sfig:basketball_team2}}
\end{figure}

The simplistic cases in Fig.~\ref{sfig:link_overlap}, however, do not address the complex community structure that arises in real life, where the multiple relationships may include more groups of many nodes and more than one link. Consider a high school with classes of about 30 students. These classes form clusters/communities and are likely to be located by the link community method. Now, students from these classes typically form a number of further communities: Some go to the same class to learn a foreign language, others play on the school's basketball team, etc. Thus, there will be further overlapping communities in such a way that the members in these new communities are in touch with each other in two distinct ways: through going to the same regular class \emph{and} through playing basketball together. Figure~\ref{sfig:basketball_team2} show that the link communities do, in fact, extract these subtle relationships. 

It is true that if a group is completely \emph{subsumed} inside another group, and there are \emph{no structural differences} distinguishing this group, such as different connectivity patterns, then link communities will not find the internal group. \emph{No method} will find it, because it's completely invisible (Fig.~\ref{sfig:basketball_team2}a).  However, if the school's social network is weighted based on the time students spend together, or if basketball players are slightly more likely to become friends with other basketball players than with students not on the team, or if the team has slightly different external connectivity, these will be identified (Fig.~\ref{sfig:basketball_team2}b).  Notice that the link communities shown in Fig.~\ref{sfig:basketball_team2}b only separate the player-coach links.  This is sufficient to completely identify the basketball team. Figure~\ref{sfig:basketball_team2}c shows a further example.  We also identify these sub-communities in practice; note the `clever/wit' community inside the `smart/intelligent' community in main text Fig.~\ref{fig:overlap}f.

\begin{figure}[]%
	\centering%
	\includegraphics[width=\textwidth,trim=0 5 0 0]{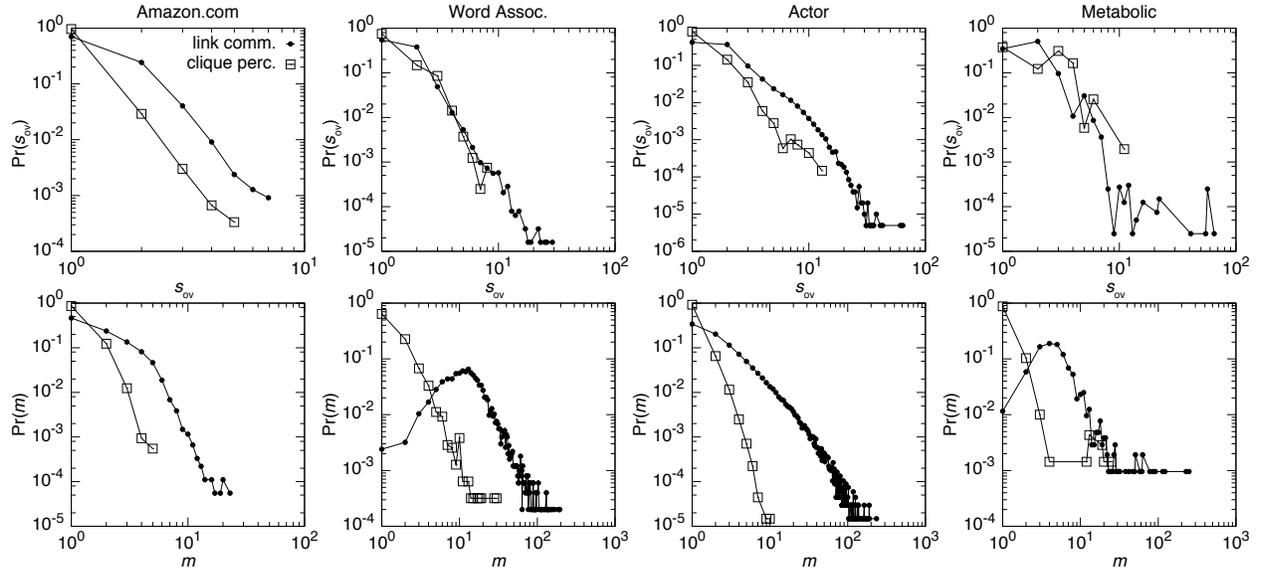}%
		\caption[Membership and overlap statistics for link communities]{Membership and overlap statistics for link communities in sparse (Amazon.com, actor) and dense (word association, metabolic) networks.  Shown are the distributions for overlap size $s_\mathrm{ov}$ (\textbf{top}) and membership number $m$ (\textbf{bottom}), as introduced by Palla \emph{et al.} \cite{palla_cpm_2005}.  Link communities were found at the maximum partition density $D$. We find that link communities extract more highly overlapping communities and a higher average number of overlapping memberships for the denser networks than the sparser ones.  The distribution of $s_\mathrm{ov}$ corresponds to the distribution of weights in the community network. 
	Statistics for clique percolation are shown for comparison (clique size $k$ was chosen from existing literature precedents or else to maximize composite performance).\label{sfig:sparse_dense_comm_stats}}%
\end{figure}

What about in practice? Are multiple relationships between nodes rare or abundant in link communities? To answer this, we study the network of communities, where each node is now a community in the original network, and the weights on each link are the number of shared members.  The distribution of link weights $s_\mathrm{ov}$ in this network, studied by Palla \emph{et al.} \cite{palla_cpm_2005} (we use their notation), explicitly shows how many nodes participate in the same communities together.  (Whenever $s_\mathrm{ov}>1$ we have found multiple relationships between two or more nodes.) The broad distributions of $s_\mathrm{ov}$ in Fig.~\ref{sfig:sparse_dense_comm_stats} (top row) show that link communities successfully capture multiple relationships in practice, for both sparse and dense networks.  Examining the distribution of the number of community memberships per node $m$, also studied by Palla \emph{et al.}, we see (Fig.~\ref{sfig:sparse_dense_comm_stats} bottom row) that link communities capture a great deal of overlap. (See also Fig.~\ref{sfig:distribution_of_community_sizes}.)

\subsection{Link dendrograms, node hierarchy, and overlap}
\begin{figure}[t] 
   \centering
   \includegraphics[width=.9\textwidth]{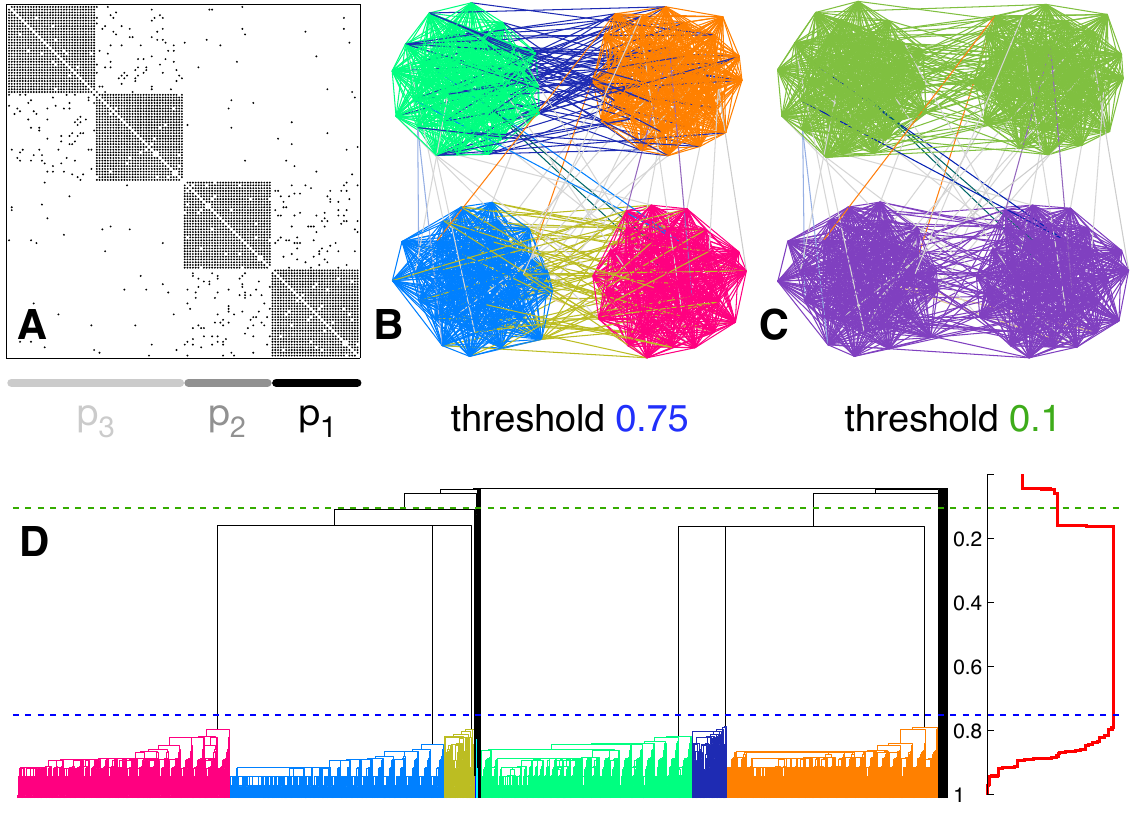} 
   \caption[Building link dendrogram intuition]{Building link dendrogram intuition.  Shown is an example illustrating how hierarchy can be captured at multiple levels of the link dendrogram.  \letter{A} The $128\times128$ adjacency matrix for a network of four densely connected non-overlapping communities (each possible link exists with probability $p_1$), each connected to another community ($p_2$), and finally the two pairs are weakly connected ($p_3$).  For this example, $p_i=\frac{1-\epsilon}{12^{i-1}}, \epsilon = 0.02$. The communities at a high \letter{B} and low \letter{C} threshold, and the full dendrogram \letter{D} are shown. The chosen values of $p_i$ lead to a very ``stretched'' dendrogram and partition density, as expected.  While one expects to identify \textbf{four} communities at the higher threshold, \textbf{six} are actually found, since the inter-community edges are accurately identified by link clustering.   \label{fig:node_hierarchy}}
\end{figure}

A link dendrogram can be very different from a node dendrogram. As an example, consider the graph shown in Fig.~\ref{fig:node_hierarchy}.  Here we have constructed a simple network without overlap, but with two levels of node hierarchy, consisting of four very dense communities, loosely connected into pairs which are then more loosely connected.  At the lower level of the link dendrogram, we find six communities, not the expected four.  \emph{The reason is that link clustering has correctly identified the two sets of cross-community links as structurally related groups}.  

\begin{figure}[!tbp]\centering
\includegraphics[width=0.65\textwidth, trim=10 20 10 5,clip=true]{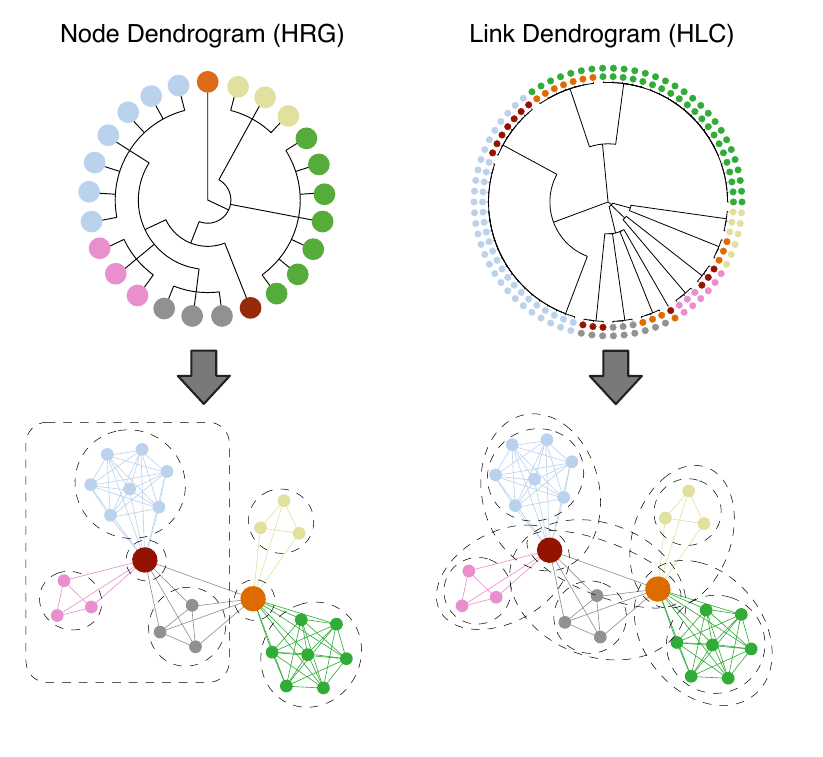}
\caption[Comparison of node and link dendrograms]{Comparison of a node dendrogram and link dendrogram in the presence of overlap. The 
node dendrogram is obtained by using the hierarchical random graph (HRG) method (consensus dendrogram)~
\cite{clauset_nature_2008}, and the link dendrogram is obtained from link clustering. Nodes are colored to distinguish each node or clique and dotted lines represent several hierarchies in the 
dendrogram. In the link dendrogram, two colored circles at each leaf represent the link 
between the nodes with the given colors. Note that HRG isolates the red, orange, and gray 
nodes in the dendrogram, even though they are central to the network and belong to the same clique: one cannot retrieve the full clique communities. In contrast, the link 
dendrogram captures every clique while at the same time constructing a reasonable hierarchical 
tree. Note that the links of the red node are placed in appropriate branches of the dendrogram 
according to their context. Also note the internal hierarchical structures found inside each 
clique. }
\label{fig:node_link_dendrogram}
\end{figure}

\begin{figure}[!tbp]\centering
\includegraphics[width=\textwidth,trim=0 0 0 0,clip=true]{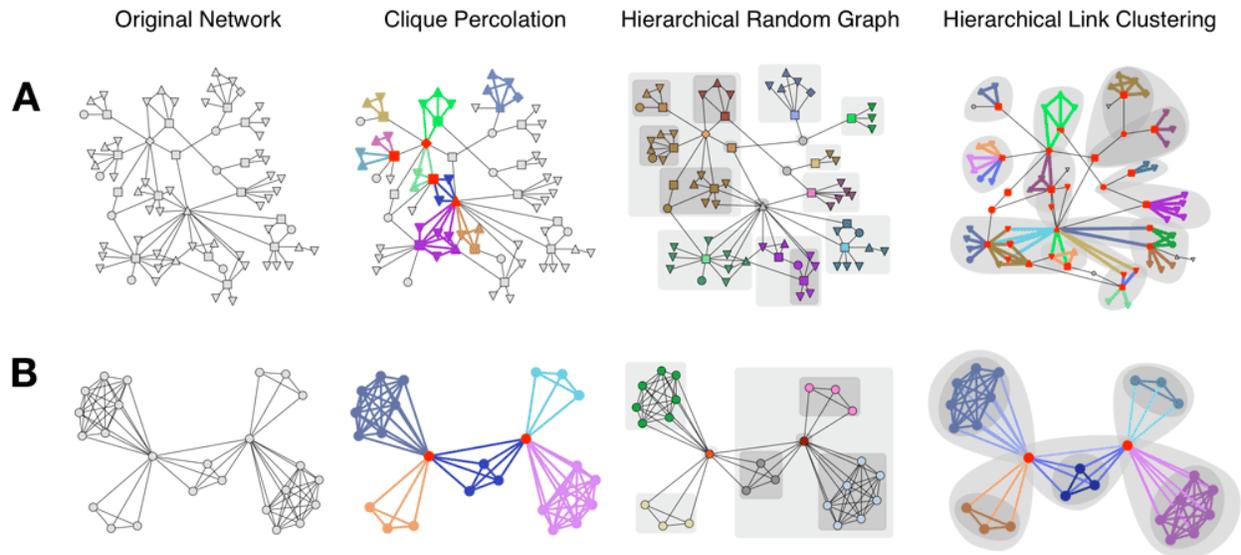}
\caption[Simultaneous overlap and hierarchy in a food web and toy model]{Comparison of methods on a network of UK grassland species interactions~
\cite{martinez_grassland_1999}, which has evident hierarchical structure \letter{A}, and on a 
simple example network with overlapping communities \letter{B}.  Colors and boxes indicate community 
structures while nested boxes illustrate hierarchical information. Red nodes possess multiple 
community memberships. The performance of existing methods depends heavily on the network's 
structural characteristics. CPM fails to detect the structure in sparse, hierarchical networks 
\letter{A}. The HRG model captures the hierarchical structure in \letter{A} but neglects 
overlap, and forces the middle 5-clique in \letter{B} to be arbitrarily spread across 
branches.  In the case of hierarchical link clustering, both hierarchy and overlapping 
structures are well classified.  Again, real social networks possess more overlap than in 
\letter{B}.\label{fig:comparison}}
\end{figure}

Several prominent methods for finding hierarchical organization exist~\cite{salespardo_extracting_2007,clauset_nature_2008}, however, none are able to handle overlap since hierarchical structure always assumes almost disjoint community partitions. For instance, see Fig.~\ref{fig:node_link_dendrogram} for a case where simple overlap prevents node hierarchy from finding true hierarchical structure. Structurally, the red and orange node should be members of the full cliques to which they are connected, but node clustering assigns them to their own community. The situation is more severe than it appears since in a network with pervasive overlap, \emph{all nodes} are in a situation similar to that of the orange and red node.
Clique percolation finds overlapping community structure (cliques) in the example network very easily, while the hierarchical random graph model fails to find all of them. Figure \ref{fig:comparison} illustrates a similar situation.

\subsection{Partition density}\label{ssubsec:PDrules}
\begin{figure}[]\centering
\includegraphics[width=0.8\textwidth,trim=0 0 0 0,clip=true]{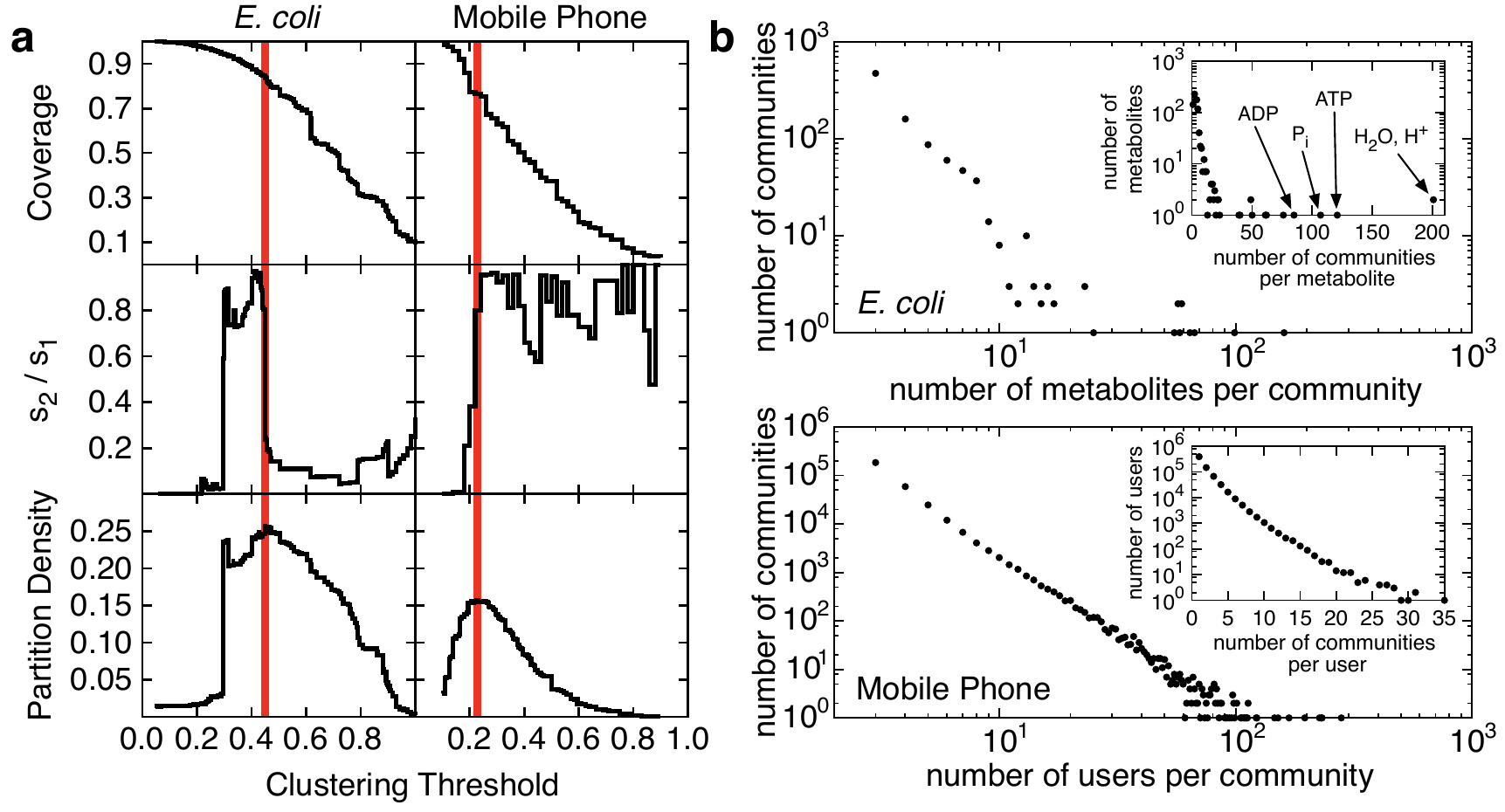}
\caption[Partition density is meaningful]{
Statistics for the E. coli metabolic and mobile phone networks. \letter{a} Community coverage, the ratio of the number of links in the two largest communities, and the partition density $D$, respectively. In both networks, peaks in $D$ align with $s_2/s_1 \to 1/2$, implying that the maximum of $D$ corresponds to the percolation transition point where community size exhibits a power-law distribution. \letter{b} The distribution of community sizes and node memberships (insets). The distribution of community size shows a heavy tail. The number of memberships per node is reasonable for both networks: we do not observe phone users that belong to large numbers of communities and we correctly identify currency metabolites, such as water and ATP, that are prevalently used throughout metabolism. The appearance of currency metabolites in many metabolic reactions is naturally incorporated into link communities, whereas their presence hindered community identification in previous work.
\label{sfig:partitionDensityMetaBile}}
\end{figure}

\begin{figure}[!tbp]\centering
\includegraphics[]{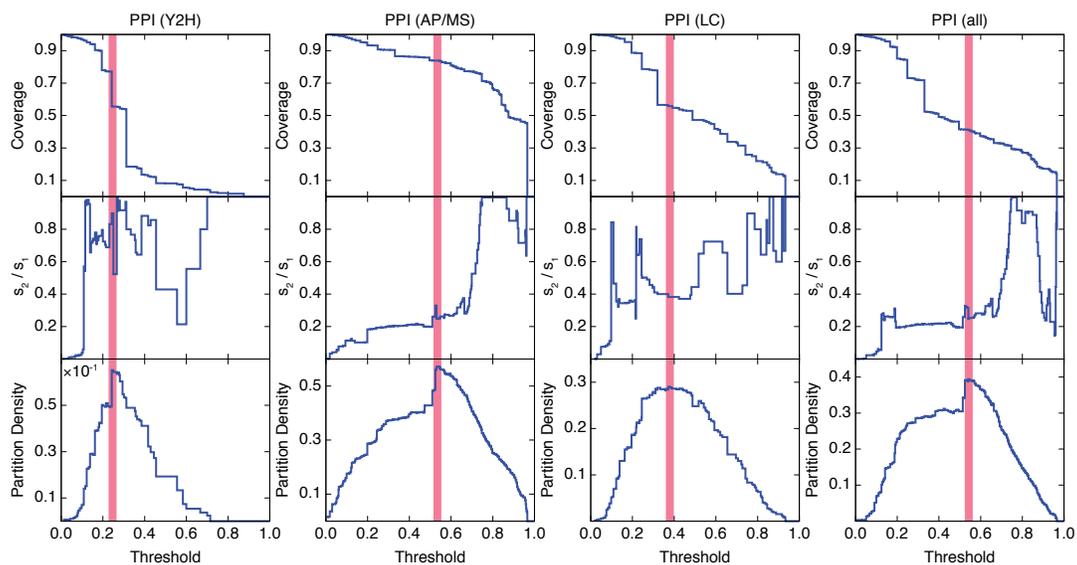}
\caption[Statistics for protein-protein interaction networks]{Several statistics for the \textbf{protein-protein interaction} networks, as a function of the link dendrogram cut threshold.  Compare with Fig.~\ref{sfig:partitionDensityMetaBile}a.\label{sfig:PPI_stats}}
\end{figure}

\begin{figure}[!tbp]\centering
	\noindent\makebox[\textwidth]{%
	\includegraphics[]{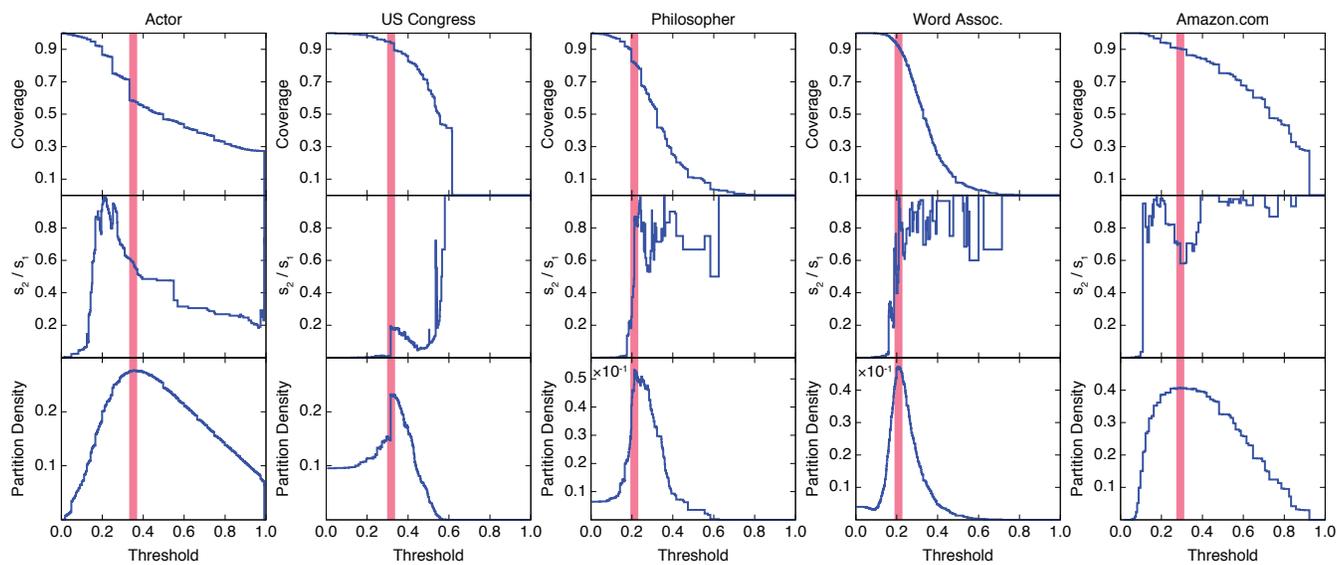}}
\caption[Statistics for other networks]{Several statistics for the remaining corpus networks.  Compare with Fig.~\ref{sfig:PPI_stats} and Fig.~\ref{sfig:partitionDensityMetaBile}a.\label{sfig:otherCorpus_stats}}
\end{figure}

To support the relevance of the structure found at the optimum partition density, we examine the link communities of the metabolic and mobile phone networks, presented in Fig.~\ref{sfig:partitionDensityMetaBile}. Here we show community coverage, the ratio of the number of links within the second largest to largest communities $s_2/s_1$, and partition density $D$, as a function of the dendrogram cut threshold (Fig.~\ref{sfig:partitionDensityMetaBile}a). That maxima in $D$ coincide with $s_2/s_1 \to 1/2$ indicates that discovered link communities are well structured~\cite{palla_cpm_2005,vicsek_CP_PRL}. Likewise, the community size distribution at the optimum $D$ is heavy tailed for both networks (Fig.~\ref{sfig:partitionDensityMetaBile}b). These properties suggest that the optimum $D$ is related to a critical point where the link communities are neither fragmented nor gelated. These statistics for the remaining test corpus are shown in Figs.~\ref{sfig:PPI_stats} and \ref{sfig:otherCorpus_stats}.

\subsection{Link communities and fuzzy membership weights}\label{subsec:fuzzymems}
Most fuzzy community methods require membership weights quantifying how strongly a node belongs to a particular community, such that the sum of every node's weights is 1.  Link communities can be mapped into fuzzy community memberships simply by counting the number of link membership a node has. If node $i$ with 8 total links has 5 links to community $A$ and 3 links to community $B$ then its membership weights are $w_{iA} = 5/8$ and $w_{iB} = 3/8$. 

It is, however, often more natural to consider each node as a full member of its communities. A person's family would be disappointed if anyone proclaimed that he or she was only $1/5$th of a member of it; in the metabolic network, it would also be strange to say that H${}_2$O was only $1/200$th a member of a given pathway.

\subsection{Filtering weighted networks}\label{subsubsec:filtering}

While the networks composing our test corpus are considered unweighted, it may happen that a researcher is presented with a weighted network. A common pre-processing step is filtering the network, deleting all edges below some defined weight threshold.  This was done in ~\cite{palla_cpm_2005}, where the clique percolation method was applied to networks after removing links below some weight $w_*$.  This approach may not be ideal, however, as useful information may be lost.

\begin{figure}[!tbp]%
	{\begin{minipage}[c]{0.35\linewidth}%
		\vspace{0pt}%
		\includegraphics[]{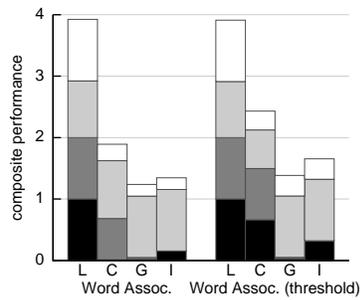}%
	\end{minipage}}%
	{\begin{minipage}[c]{0.65\linewidth}%
		\vspace{0pt}%
		\centering
		\caption[Filtering dense networks is not necessary for link communities]{Thresholding or filtering a weighted network is not critical for link communities, whereas other methods benefit from this procedure. (Symbols and colors as per main text Fig.~\ref{fig:performance}. Here we show the composite performance for the original word association network (\textbf{left}) and the same network after thresholding weak links (\textbf{right}).  For the thresholding we use $w_*=0.025$, the same value used in~\cite{palla_cpm_2005}, as well as $k=4$ for clique percolation. Clique  percolation, particularly its community quality (black), greatly improves.  We see that the link community procedure is robust to ``noisy'' links, unlike other approaches, and actually benefits from all available information.\\ \label{sfig:wordassoc_threshold}}%
	\end{minipage}}%
\end{figure}

Since this technique is common, it is important to see how link communities are affected by such filtering.  The word association network (Sec.~\ref{subsub:wordassoc}) possesses such weights, and was filtered with $w_*=0.025$ in Palla \emph{et al.} \cite{palla_cpm_2005} (using clique size $k=4$).  In Fig.~\ref{sfig:wordassoc_threshold} we show the composite performance for the tested methods on the original unfiltered word association network and the thresholded network.  Several methods benefit a great deal, but the link communities remain the leader both overall and in community quality.  This is strong evidence that link communities are better at dealing with dense networks than other methods, and at exploiting all available information. 

\subsection{Examples of link community structure}\label{sec:examples}
This section contains additional examples of link communities in various networks, all intended to illustrate that link clustering finds meaningful and relevant structure.

\subsubsection{Biological networks} 
\begin{figure}[!tbp]\centering
\includegraphics[width=0.8\textwidth]{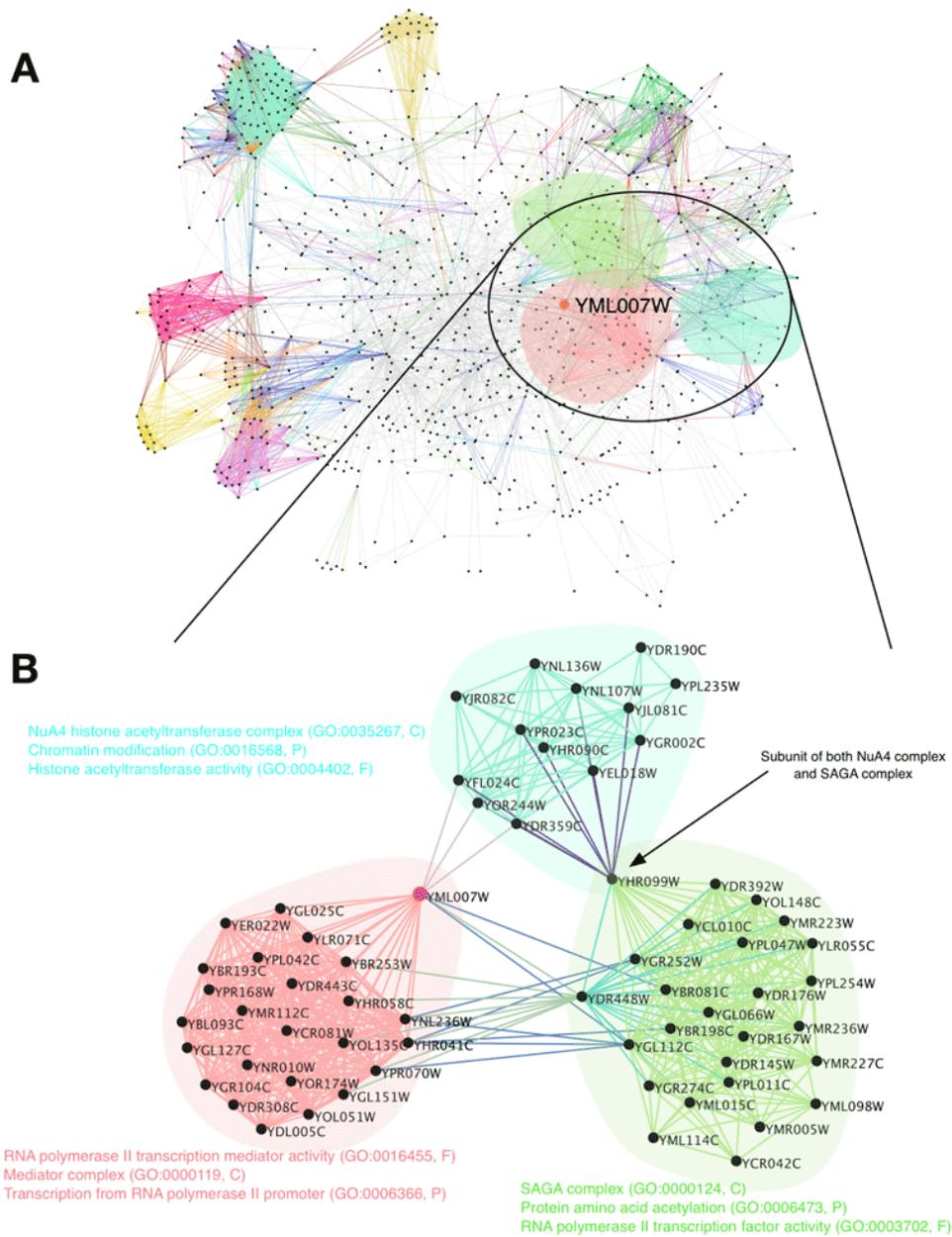}
\caption[Link community structure for the PPI network]{An example of overlapping community structure in the PPI compendium network. \letter{A} The subnetwork surrounding protein YML007W (snowball sampled out to three steps). \letter{B} The communities around YML007W.   Only GO terms with $p$-value smaller than $10^{-10}$ are displayed (colors correspond to communities).\label{sfig:PPI_example_1}}
\end{figure}

\begin{figure}[!tbp]\centering
\includegraphics[width=0.8\textwidth,trim=4 30 2 0,clip=true]{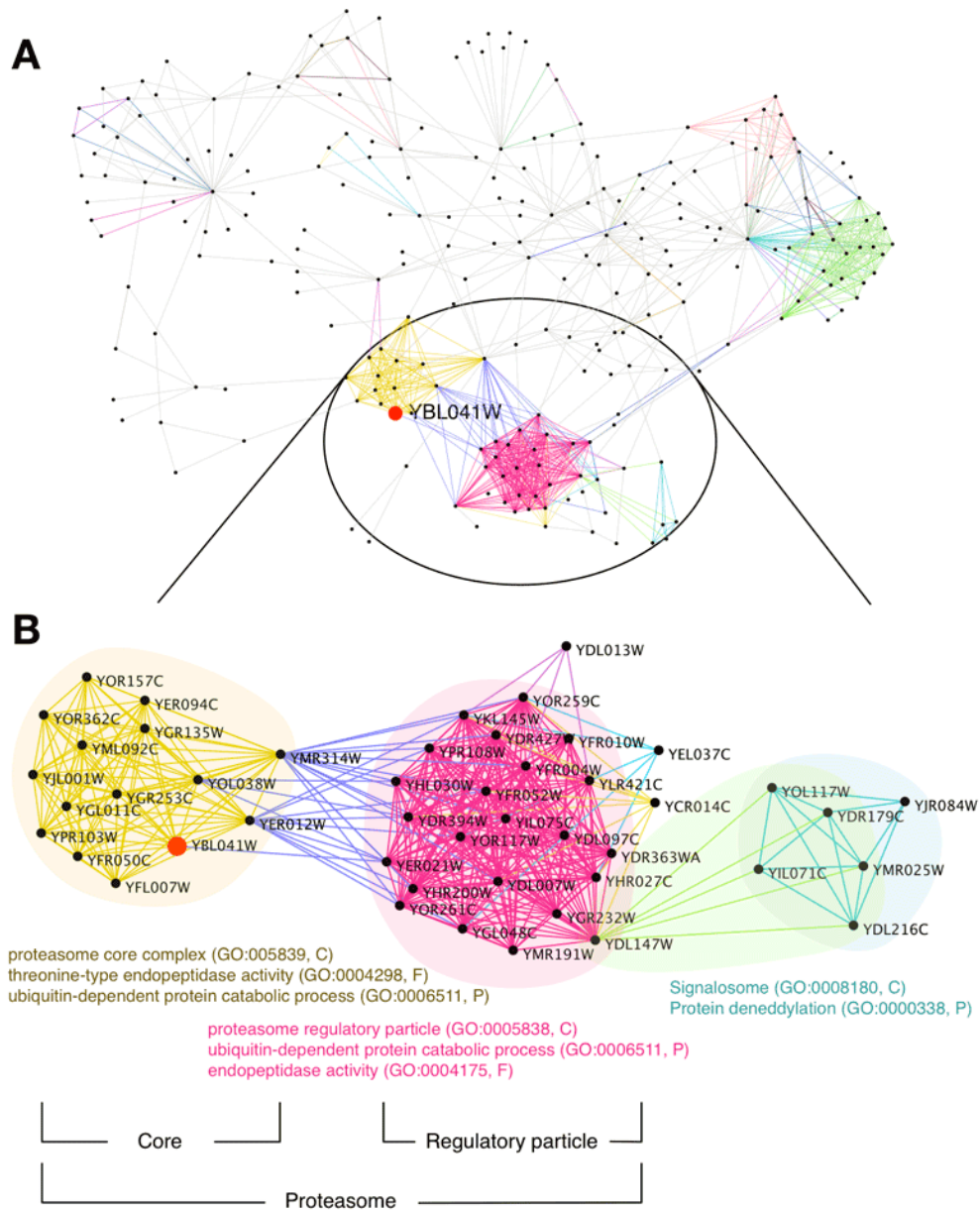}
\caption[More link communities in the PPI network]{Another example of overlapping community structure. \letter{A} The subnetwork surrounding protein YBL041W (snowball sampled out to three steps). \letter{B} The communities surrounding YBL041W. Only GO terms with $p$-value smaller than $10^{-10}$ are displayed (colors indicate communities). These communities correspond to the core and the regulatory particles of the proteasome complex and a community connecting the two.\label{sfig:PPI_example_2}}
\end{figure}

\begin{figure}[!tbp]\centering
\includegraphics[width=0.8\textwidth]{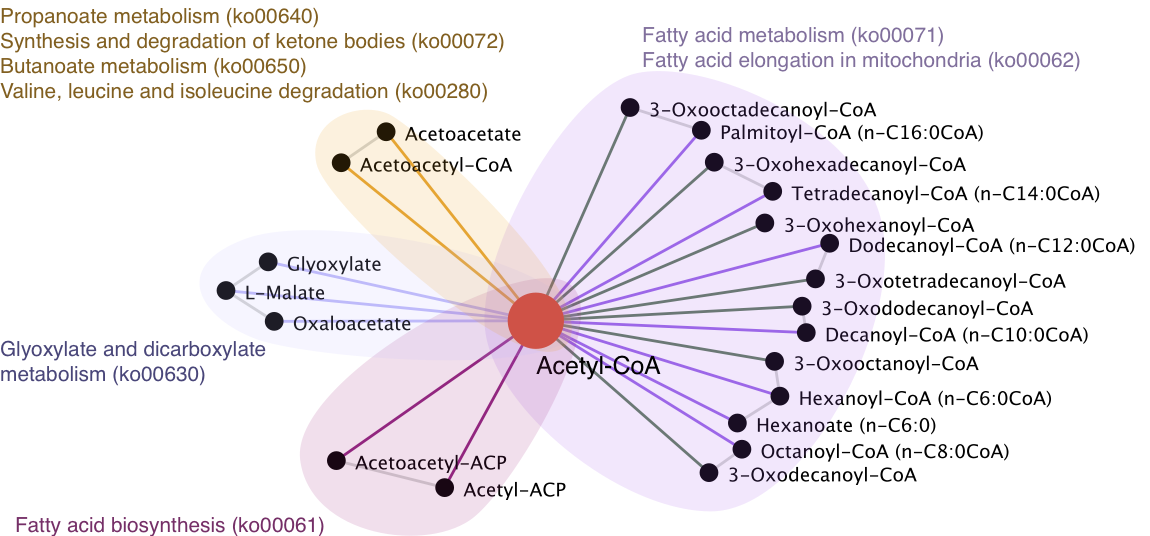}
\caption[Example link communities around Acetyl-CoA in the metabolic network]{Overlapping community structure around Acetyl-CoA in the \emph{E. coli} metabolic network. Acetyl-CoA plays several different and important roles in metabolism. Shown are only communities with homogeneity score equal to 1 (all compounds inside each community share at least one pathway annotation); all other links, including those that contribute to community structure, are omitted. Pathway annotations shared by all community members are displayed with corresponding colors. The two communities to the right of Acetyl-CoA are grouped since they share the same exact pathway annotations.}\label{sfig:metabolic_example}
\end{figure}

Figure \ref{sfig:PPI_example_1} shows the community structure around protein YML007W. There are three major communities, all three are related to the transcription process, identified as the mediator complex, NuA4 HAT complex, and SAGA complex~\cite{doyonHat2004, Dotson12192000, Wu2004199}, respectively. Note the overlapping membership of protein YHR099W, which is already known as a subunit of both the NuA4 complex and the SAGA complex~\cite{AymanSaleh10091998, Brown06222001,Bhaumik02012004}. Figure~\ref{sfig:PPI_example_2} shows three major communities around the protein YBL041W, which belongs to the core of the proteasome complex~\cite{Baumeister1998367}. We can directly observe that the proteasome consists of two parts: the core and the regulatory particle, and link clustering finds two corresponding communities plus a community connecting the two. As expected from the structure of the proteasome, the core is less exposed to other communities, while the regulatory particle has several connected communities. Likewise, Fig.~\ref{sfig:metabolic_example} shows the community structure around Acetyl-CoA, illustrating several roles that Acetyl-CoA plays in the metabolic network. 

In addition, we supply in Supplementary Table 1 the list of all communities found by link clustering along with its most relevant GO terms or pathway annotations. For the PPI networks, we use GO-TermFinder \cite{Boyle12122004} version 0.82 to find enriched GO terms and estimate the $p$-values for each GO term. First, we find all GO terms with $p$-value less than $0.05$, then we pick up only the most significant term for each aspect (biological process, cellular component, molecular function). These terms and $p$-values are listed along with the community members in Supplementary Table 1. This table shows that more than 80\% of communities have at least one enriched GO-term with $p$-value lower than $0.0001$ and more than 30\% of communities have at least one enriched GO-term with $p$-value lower than $10^{-10}$.  

For the metabolic network, we first filter out communities where less than three members possess pathway annotations. Then, we calculate the enriched pathway annotations shared by the largest number of community members. We compile this information in Supplementary Table 2.

\subsubsection{Word association networks}

\begin{figure}[!tbp]\centering
\includegraphics[width=0.5\textwidth]{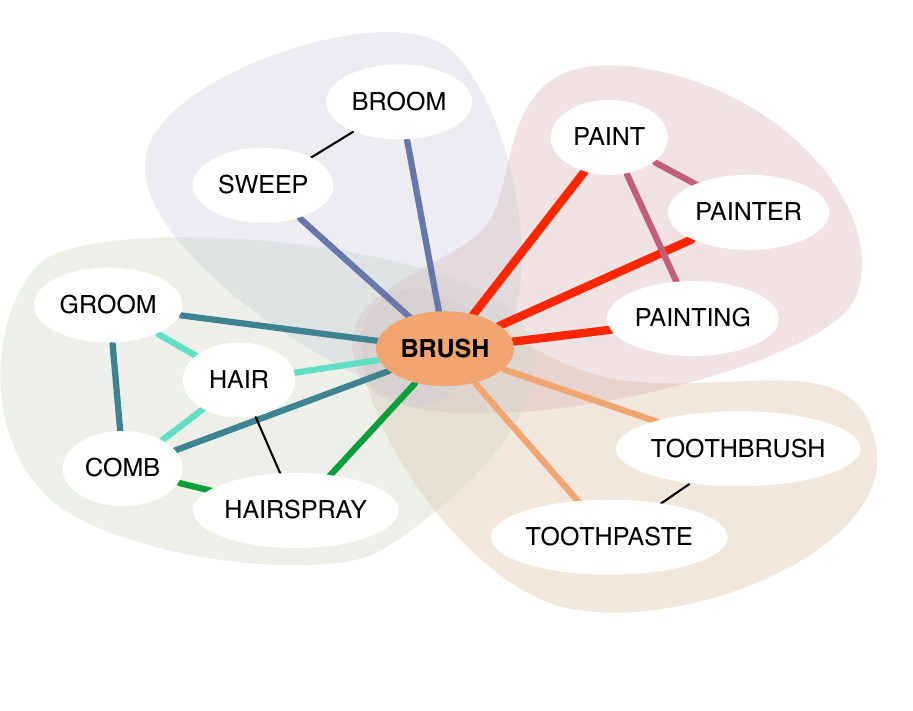}
\includegraphics[width=0.7\textwidth]{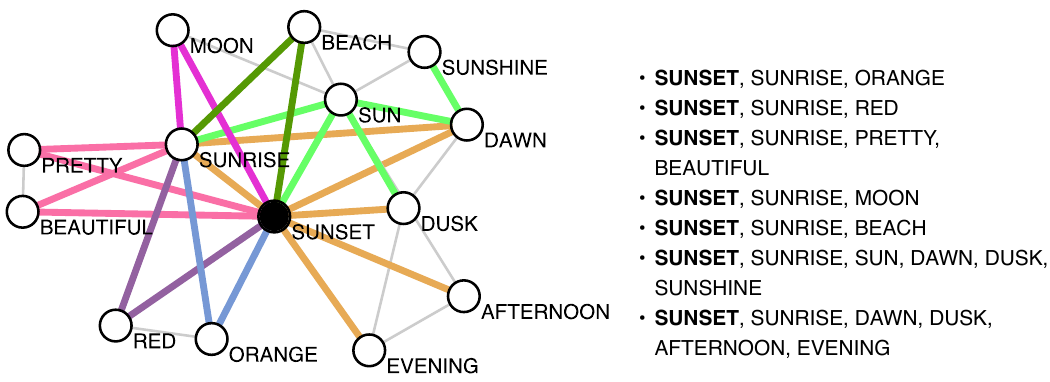}
\caption[Example link communities in the word association network]{More link community examples in the word association network. Top: link communities successfully captures various meanings of the word \textsc{brush}. Bottom: Link communities captures diverse associations of the word pair \textsc{sunrise}-\textsc{sunset}. The translated node communities are listed. }\label{sfig:word_examples}
\end{figure}

We present more examples of link communities in the word association network in Fig.~\ref{sfig:word_examples}. We also attach the list of all link communities found by link clustering at the maximum $D$ in the word association network as Supplementary Table 3. 

\section{Generalizations and extensions of link communities}\label{sec:generalizations}
\subsection{Networks with weighted, directed, or signed links}\label{subsec:tanimoto}
The similarity between links can be easily extended to networks with weighted, directed, or 
signed links (without self-loops), since the Jaccard index generalizes to the Tanimoto 
coefficient~\cite{tanimoto_elementary_1958}. Consider a vector $\mathbf{ a}_i = 
\left( \widetilde{A}_{i1}, \ldots ,  \widetilde{A}_{iN} \right)$ with 
\begin{equation}
\widetilde{A}_{ij} = \frac{1}{k_i}\sum_{i' \in n(i)} w_{i i'}\delta_{ij} + w_{ij}
\end{equation}
where $w_{ij}$ is the weight on edge $e_{ij}$, $n(i) = \left\{j | w_{ij} > 0\right\}$ is the 
set of all neighbors of node $i$, $k_i = \left| n(i) \right|$, and $\delta_{ij}=1$ if $i=j$ 
and zero otherwise. The similarity between edges $e_{ik}$ and $e_{jk}$, analogous to Eq.~\eqref{eqn:jaccardSim}, is now:
\begin{equation}
	S(e_{ik}, e_{jk}) =  \frac{ \mathbf{ a}_i \cdot  \mathbf{ a}_j}{\left|  \mathbf{ a}_i 
\right|^2 + \left|  \mathbf{ a}_j \right|^2 -  \mathbf{ a}_i \cdot  \mathbf{ a}_j}
	\label{eqn:weightedSim}
\end{equation}

\subsection{Multi-partite networks}
A multi-partite network is a network in which the nodes can be divided into  $K$ disjoint sets 
and all links must terminate in two distinct sets.  This creates additional constraints on the 
existence of certain edges which must be accounted for in both the link similarity and the 
partition density.

\textbf{Link similarity}:
The similarity measures, Eqs.~\eqref{eqn:jaccardSim} and \eqref{eqn:weightedSim}, depend only 
upon connectivity, and therefore automatically account for multi-partite structure. The one
change necessary is incorporating the forbidden connections between the same kind of nodes, 
which can be achieved by using the set of neighbors instead of the inclusive neighbor set when 
calculating the similarity. 

\textbf{Partition density}:
We must modify the definition of partition density since a fully connected $K$-partite 
clique is much sparser than a clique in a unipartite network. In general, the $K$-partite 
partition density of a subset $c$ can be written as
\begin{equation}
D_c^{(K)} = \frac{m_c+1-\sum_k n_c^{(k)}}{\sum_k \!\! \left( n_c^{(k)}\sum_{k' \ne k} 
n_c^{(k')}\right)-2\left[\left(\sum_k  n_c^{(k)}\right) - 1 \right]},
\end{equation}
where the index $k$ runs over the $K$ node types and the notation $n_c^{(k)}$ refers to nodes 
of type $k$. The full partition density is achieved by summing over individual communities, 
$D^{(K)} = 2 M^{-1}\sum_c m_c D_c^{(K)}$.

\subsection{Local methods}
Since our definition of similarity between links only uses local information, a local version~
\cite{bagrowbollt:lcd,bagrow:EvalLocalMethods,clauset:localcomm} of link clustering can be trivially 
obtained.  One can simply choose a starting \emph{link}, compute its similarity $S$ with all 
adjacent links, agglomerate the one with the largest $S$ into the community, compute any new 
similarities between edges inside the community and bordering it, and repeat. A stopping 
criteria to determine when the community has been fully agglomerated is still necessary~
\cite{bagrow:EvalLocalMethods}. For instance, one can monitor the partition density as links 
are agglomerated, in order to establish a reasonable community boundary. Another, simpler 
approach is to fix the similarity threshold and agglomerate only links with similarity larger 
than that threshold. To find all the overlapping communities of a node one can simply begin 
the above methods with each of that starting node's links or start from one link, find its 
community (which may end up including another starting node link), then pick another 
unassigned link from the starting node, find that community, and repeat until all the starting 
node's links are contained within communities.

\subsection{Partition density optimization}
Since the partition density is a quality function of link community structures in networks, it 
is possible to find link communities by direct optimization. Begin by assigning links to 
communities at random, then use, e.g. simulated annealing. The fact that link communities are disjoint partitions enables us to apply many traditional optimization techniques to find overlapping 
communities.

\section{Testing community methods}\label{sec:testingComms}
\subsection{Methodology}
Our goal is to provide a fair evaluation of all the community methods we test.  Unfortunately, evaluation of community structure in real networks is akin to a ``chicken and egg'' problem: since we don't know what the actual communities are, we must use algorithms to try and discover them. But if we don't know the real communities, how can we determine if the found communities are any good?

While common in the biological sciences, where enrichment analysis or similarity analysis using annotations (e.g. GO terms) is the standard method to assess computational predictions about a group of proteins, quantitative validation using real-world networks has not been a common practice in community research. Even the most widely cited, state-of-the-art papers about community identification do not provide quantitative validation, but only provide qualitative arguments with one or two small networks that are small enough to draw and look at the structure \cite{rosvall_infomap_2008,newman_mixture_2007,guimera_functional_2005,palla_cpm_2005}. A recent survey paper~\cite{fortunato_community_2009} about community structure, although very extensive, does not contain even a single section regarding quantitative validation using real-world networks.

Some literature has answered the problem of validating community detection methods using model graphs (\emph{benchmarks}) designed to generate a random, pre-programmed community structure as ``ground truth''. However, since the community structure in these graphs reflects the \emph{conceptual model} of communities held by their creator, there is no guarantee that the results can be extrapolated to real networks. Worse, this approach introduces serious biases towards the algorithms that conform with the same conceptual model as the benchmark graphs and are directly biased \emph{against} other theories of community structure. 

For instance, every existing benchmark graph has the underlying principle that a community should have more intra-community links than outgoing links, which is not true in networks with pervasive overlap. Furthermore, no existing benchmark graph takes into account the highly non-random abundance of triangles, one of the most important and fundamental characteristics of real world networks, and one of the earliest discoveries of the complex networks field~\cite{watts_wsmodel_1998}. The randomized nature of current benchmark graphs shows evident bias against algorithms such as clique percolation~\cite{palla_cpm_2005}, which exploits these triangles (and cliques) and is based on a different community definition than modularity~\cite{newman_finding_2004}, which is the conceptual model behind current benchmark graphs.

To avoid requiring the hidden ``ground truth'' communities, we have focused on networks that possess descriptive \emph{metadata}.  This information does not directly contribute to the construction of the network, but it allows us to understand what the nodes in the network do, how similar they are to one another, and how many contexts or roles each node has.  An example of a network and its metadata is presented in Fig.~\ref{fig:metadata_amazon}.  Using these metadata to describe how similar nodes are within communities (community quality, see Sec.~\ref{sec:measures}), we can compare and contrast the results of different methods, relating how much each method's results tell us about the relevant (hidden) metadata.

\subsection{Measures}\label{sec:measures}
There are some subtle aspects to consider when comparing disparate community algorithms.  Some methods find excellent communities (high quality) but only for a very small fraction of the network (low coverage).  Others find medium-quality communities but classify the majority of the network.  Some methods find overlapping memberships, others do not.  Since it is difficult and unfair to compare all methods along any one of these directions, we have introduced a simple \emph{composite performance} measure to fairly account for these differences while also allowing a researcher to focus on the individual aspects.  

We study four distinct aspects of the quality and coverage of the communities found---the quality measures are based on metadata and the measures of coverage focus on the amount of information extracted from the network.

\begin{figure}[!tbf]
 \centering
 \includegraphics[width=\hsize]{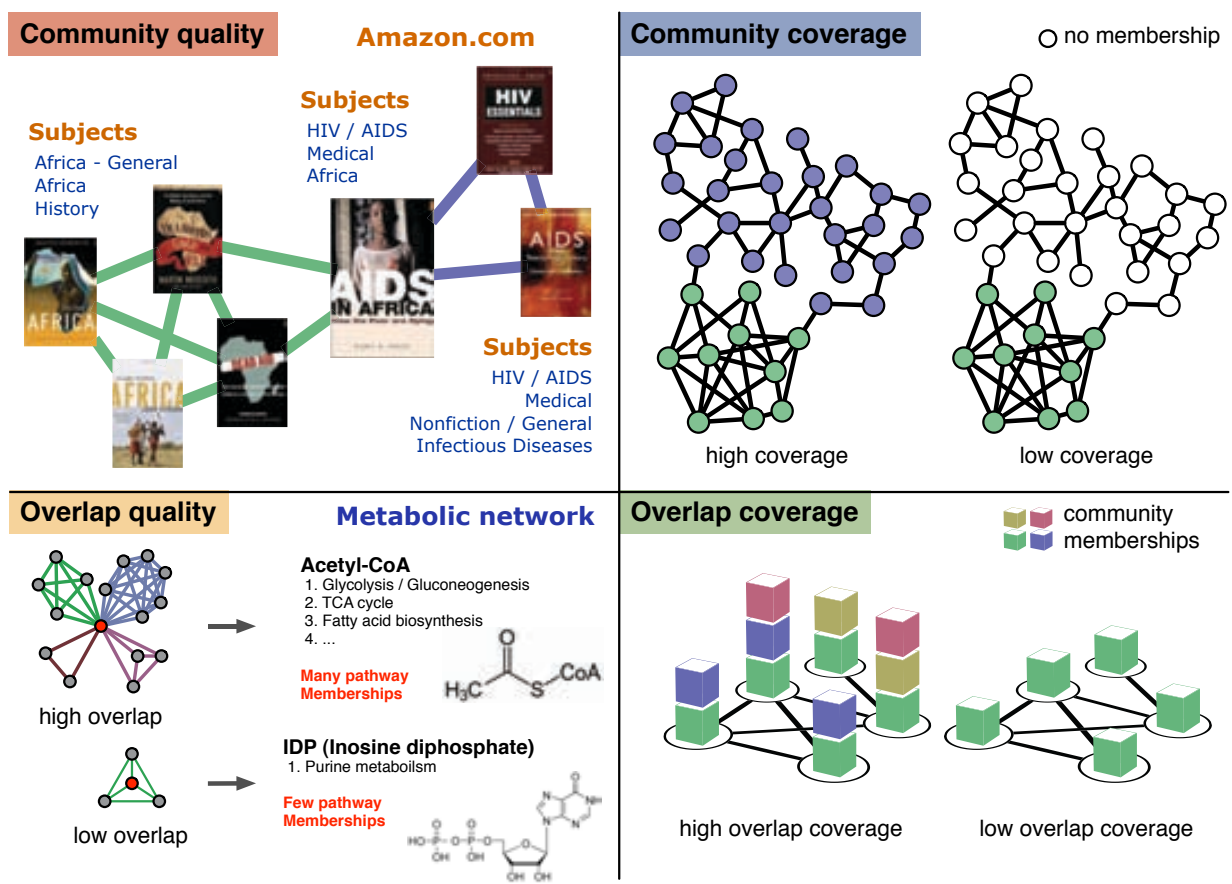}
 \caption[The elements of composite performance]{The elements of composite performance. (\textbf{top left}) \textit{Community quality} measures the similarity between nodes within each community compared to a null model, based on metadata. (\textbf{bottom left}) \textit{Overlap quality} compares the amount of overlap found for each node with a measure of real-wold overlap, based on metadata. (\textbf{top right}) \textit{Community coverage} is simply the fraction of nodes categorized by the algorithm. (\textbf{bottom right}) Two methods may have the same community coverage but one may extract many more overlapping memberships and will yield more information about the network.  Thus we introduce \textit{Overlap coverage}, the average number of memberships per node.  This is equivalent to community coverage for non-overlapping methods. \label{fig:composite_performance}}
\end{figure}

\begin{description}
 \item[Community Quality.] Many of the networks studied here possess metadata that attaches a small set of \emph{annotations} or \emph{tags} to each node.  For example, in the Amazon.com network, each product is categorized into several subjects (see Figs.~\ref{fig:composite_performance}, \ref{fig:metadata_amazon}); each actor's career in the Actor collaboration network can be described by a set of plot keywords; each protein in the Protein-Protein Interaction networks is given a set of GO-terms, which describe the biological process that the protein participates in. Assuming that these metadata form a description of the node, beyond the network itself, we can reasonably state that ``similar'' nodes share more metadata than dissimilar nodes.  To quantify this, we compute, e.g., the \emph{enrichment of node pair similarity}:
\begin{equation}
	\mathrm{Enrichment} = \frac{\Big<\mu(i,j)\Big>_{\substack{\mathrm{all}~i,j~\mathrm{within}\\\mathrm{same~community}}} }{ \Big<\mu(i,j)\Big>_{\substack{\mathrm{all~pairs}~i,j}} }, \label{eqn:CommSimilarity}
\end{equation}
where $\mu(i,j)$  is a metadata-based similarity between node $i$ and $j$ whose exact definition depends on the particular network (each similarity is discussed in detail in Sec.~\ref{sec:realdata}). In other words, enrichment is the average metadata similarity between all pairs of nodes that share a community, divided by the average metadata similarity between all pairs of nodes\footnote{For very large networks or very large communities, we may not be able to test every possible pair of nodes.  In this case, if the network is more than around 1M nodes, we compute the baseline from $10^7$ randomly chosen pairs of nodes.  Likewise, for communities of more than 1000 nodes, we chose $10^5$ random pairs to compute the numerator in Eq.~\eqref{eqn:CommSimilarity}.}. The denominator serves as a baseline similarity and larger values of enrichment show that the communities are ``tighter,'' according to the metadata. Note that it is important to compare all pairs of nodes, not just links, since links themselves are often enriched beyond average, depending on the properties of the metadata. See Fig.~\ref{fig:composite_performance}, top left.

This approach is very similar to that used in~\cite{yu_ppi_2008} to quantify the relevance of interactions. 

 \item[Overlap Quality.] For each node $i$ in the network, we extract from the metadata a scalar quantity (call this the overlap metadata) that we expect to be closely related to the number of true communities that node $i$ participates in. For example, in the word association network, each community corresponds to a set of words that share the same general topic.  The more definitions a word has, the more topics the word is expected to belong to.  In the metabolic network, the number of reaction pathways that a metabolite participates in corresponds to the number of communities (contexts or roles) of the metabolite. 

To rigorously quantify the amount of information gained by community algorithms, we use \emph{mutual information} to relate the number of memberships and the overlap metadata.  This quantity tells us how much information about the true overlap of a node is gained by knowing or learning the number of communities that a particular method has assigned to the node. Mutual information works well since detected relationships need not be linear or obey a predisposed functional form.  By running multiple algorithms and computing this mutual information, we can see which methods let us know the most about the overlap metadata. Note that even non-overlapping methods may learn information about the overlap metadata, since some nodes may be placed within zero communities. See Fig.~\ref{fig:composite_performance}, bottom left.

 \item[Community Coverage.] To measure community coverage, we simply count the fraction of nodes that belong to at least one community of three or more nodes.  A size of three was chosen since it is the smallest \emph{nontrivial} community. This measure provides a sense of how much of the network is analyzed. See Fig.~\ref{fig:composite_performance}, top right.

 \item[Overlap Coverage.] Two algorithms may both completely classify a network, giving complete coverage, but one method may extract more information by finding many more densely overlapping communities than the other. It is therefore important to consider  overlap coverage as well as community coverage.  To do so, we count the average number of memberships in nontrivial communities that nodes are given. For non-overlapping community methods, both coverage measures are identical. This measure shows how much information is extracted from that portion of the network that the particular algorithm was able to analyze. See Fig.~\ref{fig:composite_performance}, bottom right.
\end{description}
Note that the evaluation of the community and overlap quality include neither trivial communities nor singleton nodes, since their absence is considered by the coverage measures.  

For many networks, these measures do not necessarily fall between 0 and 1.  For example, in the Amazon.com product network and the word association network, link communities find enrichments 80--100 times higher than the global baseline.  Therefore, we renormalize all community and overlap quality values such that the maximum value is 1 for the best performing method\footnote{If a method happens to yield a negative value for a particular measure, all the methods are subsequently scaled such that the minimum value is 0.}. This allows us to directly compare performance across networks whose metadata similarities may cover vastly different ranges of values. Likewise, overlap coverage is often greater than 1 for overlapping methods; these values are likewise rescaled.  Community coverage is also renormalized, although there is typically always one algorithm that yields complete coverage and the values are already constrained to $[0,1]$.

\begin{figure}[t]
\noindent\makebox[\textwidth]{%
 \includegraphics[]{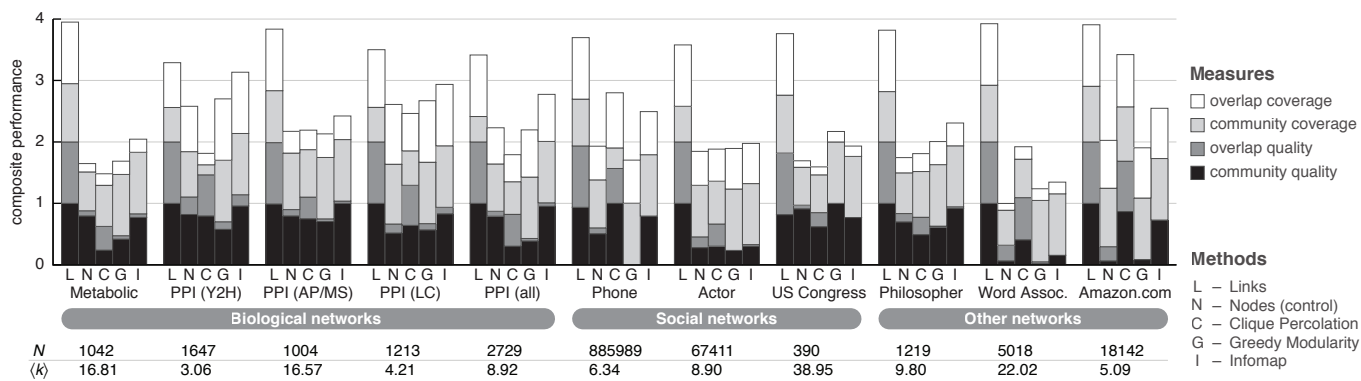}}
 \caption[Network validation results with node clustering]{Data-driven evaluation of community algorithms over a large corpus of real networks. (Compare with main text Fig.~\ref{fig:performance}, which lacks the node clustering control algorithm.) Each column represents an algorithm's \textbf{composite performance}, measuring community/overlap accuracy and sensitivity. Also shown for each network is the number of nodes $N$ and the average degree $\left<k\right>$.  Link communities achieve the best performance in every network. \label{fig:composite_performance_withNodes}}
\end{figure}

We are now left with four measures quantifying the performance of each algorithm. In order to provide a clean, simple representation of each algorithm's performance, we show a stacked bar chart summing all four measures.  Since each measure is normalized to have values between 0 and 1, so that the best method for each measure has a value of 1, the maximum composite performance will be 4. Note that this composite performance measure weighs each of the four aspects equally, while providing a simple and easily understood bar chart that nevertheless allows the researcher to evaluate the individual merits of each performance criterion.  We find this stacked representation simpler to understand than multiple bar charts while still presenting sufficient information to be fair to all aspects of the problem.  Results are shown in Fig.~\ref{fig:composite_performance_withNodes} (compare with main text Fig.~\ref{fig:performance}).

\section{Network datasets}\label{sec:realdata}

\subsection{Overview}

Here we discuss the network datasets used throughout this work, including properties of their metadata, how they were collected, and how the metadata was used to compute the composite performance.  Table \ref{tab:tableNetworks} summarizes all the networks used in this study.

We have chosen eleven networks to test (one is the union of three other networks).  This test set contains some of the most relevant networks in recent network research: protein-protein interaction networks for \emph{S.~cerevisiae}~\cite{yu_ppi_2008}, the metabolic network reconstruction of \emph{E.~coli}~\cite{feist_ecoli_2007}, and a large, dynamic social network derived from mobile phone telecommunication records~\cite{onnela_structure_2007,onnela_analysis_2007,pallaQuantifying_2008,gonzales_uncovering_2008}. A variety of other networks were also chosen to serve as diverse test topologies, representative of the diverse datasets used in complex networks research, and to enable the comprehensive validation procedure of Sec.~\ref{sec:testingComms}, due to their rich metadata.  Table~\ref{tab:tableNetworks} includes brief descriptions of this associated metadata.

\begin{table}[f]
	\centering
	\footnotesize
	\rowcolors{4}{MyGray}{white}
  \begin{minipage}{15cm}
	\centering \ra{1.3} 
	\begin{tabular}
		{lp{3.75cm}llp{3.3cm}p{2.4cm}}
		        &             &     &                  &  \multicolumn{2}{c}{metadata}\tabularnewline
		\cmidrule{5-6}
		network & description & $N$ & $\left<k\right>$ &  \multicolumn{1}{c}{community} & \multicolumn{1}{c}{overlap}\tabularnewline
		\midrule
			
		PPI (Y2H) & \rr PPI network of \emph{S. cerevisiae} obtained by yeast two-hybrid (Y2H) experiment~\cite{yu_ppi_2008}
		            & 1647  & 3.06 &  \rr 	Set of each protein's known functions (GO terms)\footnote{GO terms are ``structured, controlled vocabularies (ontologies) that describe gene products in terms of their associated biological processes, cellular components and molecular functions in a species-independent manner.''  See \url{http://wiki.geneontology.org/index.php/GO_FAQ}}  & \rr The number of GO terms \tabularnewline
		PPI (AP/MS) & \rr Affinity purification mass spectrometry (AP/MS) experiment
		             & 1004 & 16.57 &  \rr  GO terms & \rr GO terms \tabularnewline
		PPI (LC) & \rr Literature curated (LC)
		             & 1213  & 4.21 &  \rr 	GO terms & \rr GO terms \tabularnewline
		PPI (all) & \rr Union of Y2H, AP/MS, and LC  PPI networks
		 	& 2729 & 8.92 & 
			\rr GO terms & 
			\rr GO-terms \tabularnewline
		Metabolic & \rr Metabolic network (metabolites connected by reactions) of \emph{E. coli}
			& 1042 & 16.81 &
			\rr Set of each metabolite's pathway annotations (KEGG)\footnote{KEGG database provide metabolic pathway annotations for metabolites. See \url{http://www.genome.jp/kegg/}} &
			\rr The number of KEGG pathway annotations \tabularnewline
		Phone & \rr Social contacts between mobile phone users~\cite{onnela_analysis_2007,pallaQuantifying_2008,gonzales_uncovering_2008} & 885989 & 6.34 & 
			\rr Each user's most likely geographic location & 
			\rr Call activity (number of phone calls)  \tabularnewline %
		Actor & \rr Film actors that appear in the same movies during 2000--2009~\cite{IMDB}& 67411 & 8.90 &
			\rr Set of plot keywords for all of the actor's films &
			\rr Length of career (year of first role) \tabularnewline
		US Congress & \rr Congressmen who co-sponsor bills during the 108th US Congress~\cite{fowlerConnectingCongress2006,fowlerCosponsorNets2006} & 390 & 38.95 &
			\rr Political ideology, from the common space score~\cite{pooleRecovering1998,PooleBook2005} &
			\rr Seniority (number of congresses served) \tabularnewline
		Philosopher & \rr Philosophers and their philosophical influences, from the English Wikipedia\footnote{These influences are treated independently from the global wikipedia hyperlink structure and are particularly easy to extract for philosopher biographies.}
			& 1219 & 9.80 &
			\rr Set of (wikipedia) hyperlinks exiting in the philosopher's page &
			\rr Number of wikipedia subject categories \tabularnewline
		Word Assoc. & \rr English words that are often mentally associated~\cite{fellbaum:wordnet} & 5018 & 22.02 &
			\rr Set of each word's \emph{senses}, as documented by WordNet\footnote{See \url{http://wordnet.princeton.edu/wordnet/man/wngloss.7WN.html}}   &
			\rr Number of senses \tabularnewline
		Amazon.com & \rr Products that users frequently buy together & 18142 & 5.09\footnote{Amazon.com's XML Service only returns the five most co-purchased products, though considering the network as undirected will boost some node degrees.  This artificial constraint makes the network to have very narrow degree distribution, and serves as a unique test set.} &
			\rr Set of each product's user tags (annotations) &
			\rr Number of product categories \tabularnewline
	\end{tabular}
\end{minipage}
	\caption[Summary of the 11 network test corpus]{A brief description of the networks used in the paper.  Shown are the number of nodes $N$, the average degree $\left<k\right>$, and brief descriptions of the metadata available to study node similarity and the expected amount of overlap.  Full details in Sec.~\ref{sec:realdata}. \label{tab:tableNetworks}}
\end{table}

\subsection{Biological networks}\label{sec:results_bio}

\subsubsection{Protein-protein interaction}\label{subsec:ppi_metadata}
We analyzed the protein-protein interaction (PPI) network of \emph{S.~cerevisiae}, the most studied PPI network. 

\begin{description}

\item[Construction] We use a recently published dataset of PPI networks compiled into three genome-scale networks: yeast two-hybrid (Y2H), affinity purification followed by mass spectrometry (AP/MS), and literature curated (LC)~\cite{yu_ppi_2008}. We also use the union of these three networks (PPI (all)). We use only the largest component of each network. 

\item[Metadata] We use the Gene Ontology (GO) terms as metadata for the PPI network. The GO project is \emph{``a major bioinformatics initiative with the aim of standardizing the representation of gene and gene product attributes across species and databases.''}~\cite{go} And it provides controlled vocabulary (GO terms) which describes certain aspects of protein characteristics (function, location, etc). We choose GO terms as the most reasonable metadata for PPI networks, since they are the most elaborate protein annotations available, provide structured information along with statistical information for each term, and there are established methods to calculate the functional similarity between proteins. 

\item[Community quality] We adopt the same measure as the paper that published the datasets~\cite{yu_ppi_2008}. First, a $p$-value that two proteins share similar GO terms by chance is calculated using GO biological process terms and the total ancestry measure~\cite{yu_tam_2007}. The similarity between two proteins $\mu(i,j)$ is defined as either one (if $p < 10^{-3}$) or zero (if $p \ge 10^{-3}$). Then, the enrichment of functionally similar pairs is calculated using Eq.~\eqref{eqn:CommSimilarity}:

\item[Overlap quality] We use the total number of GO terms as a proxy for the amount of overlap, since it is likely that a protein with many GO terms functions in more diverse contexts. We compute the mutual information between the number of GO terms and the number of discovered memberships as overlap quality.

\end{description}

\subsubsection{Metabolic}
We use a metabolic network reconstruction of \emph{E. coli} K-12 MG1655 strain (iAF1260), one of the most elaborate metabolic network reconstructions currently available~\cite{feist_ecoli_2007}. 

\begin{description}
\item[Construction] From the metabolic network reconstruction iAF1260, we retain only cellular reactions, ignore information regarding the compartments (cytoplasm and periplasm), and project the network into metabolite space (two metabolites are connected if they share a reaction). For instance, if an enzyme catalyzes the reaction where metabolites $A$ and $B$ are transformed into $C$ and $D$, the resulting network would contain a clique of $A$, $B$, $C$, and $D$. 

\item[Metadata] We use the pathway annotations from KEGG database~\cite{kegg}, which is one of the most widely used metabolic network databases. Each metabolite has zero or more metabolic pathway annotations. For instance, Acetyl-CoA is annotated with 38 pathways including Glycolysis, citrate cycle, and fatty acid biosynthesis. 

\item[Community quality] To measure the similarity between a pair of metabolites $a$ and $b$, we calculate the Jaccard index between their pathway sets, i.e. $\mu(a,b) = \left| P_a \cap  P_b \right| / \left| P_a \cup  P_b \right|$, where $P_m$ is the set 
of pathways that contain metabolite $m$. With this similarity, the community quality is then calculated using Eq.~\eqref{eqn:CommSimilarity}.

\item[Overlap quality] The number of pathways represents the number of contexts that a given metabolite participates in. We measure the mutual information between the number of pathways and the number of community memberships found by the algorithms. 

\end{description}

\subsection{Social networks}

\subsubsection{Mobile phone}
This dataset catalogs approximately 8 million users, all calls among these users, and the 
locations of users when they initiate a phone call (the tower from which the call originated). 
Self-reported demographic information such as age and gender is also available for some users.

\begin{description}
	\item[Construction]  We generate the social network by constraining the location to a 350 km by 80 km region and two nodes in the region are connected only if they each call the other person at least once during a 30-week period. We assign to each user a single location, that of the tower they most frequently used.  The final network contains approximately 2.8 million links.
	
	\item[Community quality]  Unlike most other networks, we do not possess tags for each node, but instead the nodes are embedded spatially, using each phone user's most likely location.  To compute the similarity between nodes, we use the euclidean distance between their most likely locations, hypothesizing that social contact is more frequent for users that are geographically related.  Since nodes with higher similarity have \emph{smaller} distance, we do not use Eq.~\eqref{eqn:CommSimilarity}, but instead:
	\begin{equation}
		\mathrm{Community~quality} = 1 - \frac{\Big<d(i,j)\Big>_{\substack{\mathrm{all}~i,j~\mathrm{within}\\ \mathrm{same~community}}} }{ \Big<d(i,j)\Big>_{\substack{\mathrm{all~pairs}~i,j}} },\label{eqn:spatialEnrichment}
	\end{equation}
where $d(i,j)$ is the euclidean distance between the most likely locations of nodes $i$ and $j$.
	\item[Overlap quality] To quantify how much information was discovered about the amount of overlap, we use the total number of phone calls each user made during the observation window.  This operates under the assumption that frequent phone users may fulfill broader roles in their social networks. 
\end{description}

\subsubsection{Actor}
For this network, we use the Internet Movie Database (IMDb) to find working collaborations between film actors.  We focus on actors who star in at least one movie during the years 2000 and 2009, and at least two movies during their entire career.  Television shows, video games, and other performances were not used.
	\begin{description}
		\item[Construction] The raw IMDb files were downloaded from \url{http://us.imdb.com/interfaces} on 2009-12-08.  From this data, we construct a bipartite network of movies and actors.  We remove films and actors who do not satisfy the above criteria and then project the bipartite network onto the actors, creating a network where two actors $i$ and $j$ are linked with a weight $w_{ij}$ if they co-star in $w_{ij}$ films.  Finally, we remove projected links with weights $w<2$ and keep only the largest connected component. By ensuring that the actors have appeared together in at least two films, we increase the likelihood that they developed a working relationship.
		\item[Community quality] Associated with each film is a set of plot keywords.  We can roughly summarize each actor's career during 2000--2009 by taking the union of all the keywords of the movies that actor appeared in.  Since many keywords are very finely grained, we consider only those that label at least 100 films (over the entire IMDb dataset).  The Jaccard index between these sets is then used as the node-node similarity in Eq.~\eqref{eqn:CommSimilarity} to compute the ``keyword enrichment'' of each community algorithm.
		\item[Overlap quality] One option for overlap metadata is to use the \emph{seniority} of the actor, defined as the year of his or her first film role (not necessarily during 2000--2009).  We expect actors with longer careers to be professionally capable of participating in more collaborative groups.  The mutual information between the number of communities an actor belongs to and the first year of his or her career is then used to quantify this relationship.
	\end{description}

\subsubsection{US Congress}
\begin{figure}[t!]
	\begin{minipage}[t]{0.58\textwidth}
		\centering
		\includegraphics[width=1\textwidth,trim=0 0 0 0,clip=true]{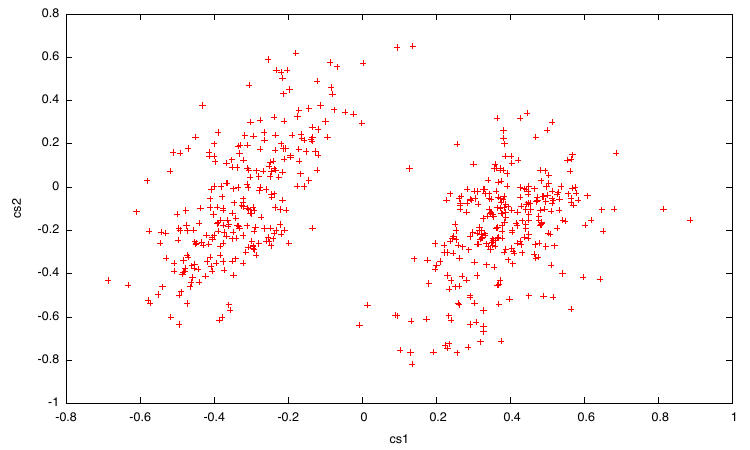}
	\end{minipage}
	\hspace{0.02\textwidth}
	\begin{minipage}[b]{0.4\textwidth}
	\caption[Political ideology metadata in the US Congress network]{Scatter plot of the Common Space Scores for the 108th US Congress (House and Senate). The ideological and political breakdown is visible in the clustering of the points, which closely follow party lines (republicans and democrats).\label{fig:UScongress_CSS}}
		\vspace{1.5cm}
	\end{minipage}
\end{figure}

	The network of legislative collaborations between US congressional representatives (not senators) during the 108th US congress (2003-2005).  
	\begin{description}
		\item[Construction] Using the dataset of \cite{fowlerConnectingCongress2006,fowlerCosponsorNets2006}\footnote{Downloaded from \url{http://jhfowler.ucsd.edu/cosponsorship.htm}.}, we construct a bipartite network $B$ of representatives and the legislative bills they (co)-sponsored.  Many bills are co-sponsored by the majority of representatives and there were many bills introduced (7765 total), so projecting this bipartite network onto the representatives results in a very dense, nearly complete graph.  To avoid this, we filter out edges to capture only the tightest working relationships.  To do this, we apply two filtering criteria.  First, we remove all introduced bills that contain more than 10 (co)-sponsors total.  This network is then projected onto the representatives to form network $G_1$.  Meanwhile, we also project the unfiltered $B$ onto the representatives and then delete all links with weights less than 75, forming network $G_2$.  The final network $G$ that we feed to the community detection algorithms is then the intersection of $G_1$ and $G_2$, i.e., each link in $G$ must exist in both $G_1$ and $G_2$.  This network is still fairly dense but was disconnected, so we focus on only the giant connected component.  This is why there are only 390 representatives.
		\item[Community quality] Associated with each representative are two values between -1 and 1 known as the \emph{common space score}~\cite{pooleRecovering1998,PooleBook2005}.  These values form a two-dimensional space where distances capture political and ideological similarity (Fig.~\ref{fig:UScongress_CSS}).  The first dimension generally represents liberal/conservative bias while the second is related to women's rights and abortion issues.  We simply compute the euclidean distance between pairs of points as the node-node similarity measure, and compute the overall ``enrichment'' of an algorithm's communities using Eq.~\eqref{eqn:spatialEnrichment}.
		\item[Overlap quality] For the overlap metadata we use the seniority of each congressional representative, measured as the number of elected terms that person has served.  We roughly expect that longer-serving representatives will more easily participate in multiple collaborations than those who are newly elected.  The mutual information between the number of community memberships and the number of elected terms is then used to quantify this relationship. 
	\end{description}

\subsection{Other networks}
\begin{figure}[t!]
	\begin{minipage}[b]{0.72\textwidth}
		\centering
		\frame{\includegraphics[width=1\textwidth,trim=10 530 75 170,clip=true]{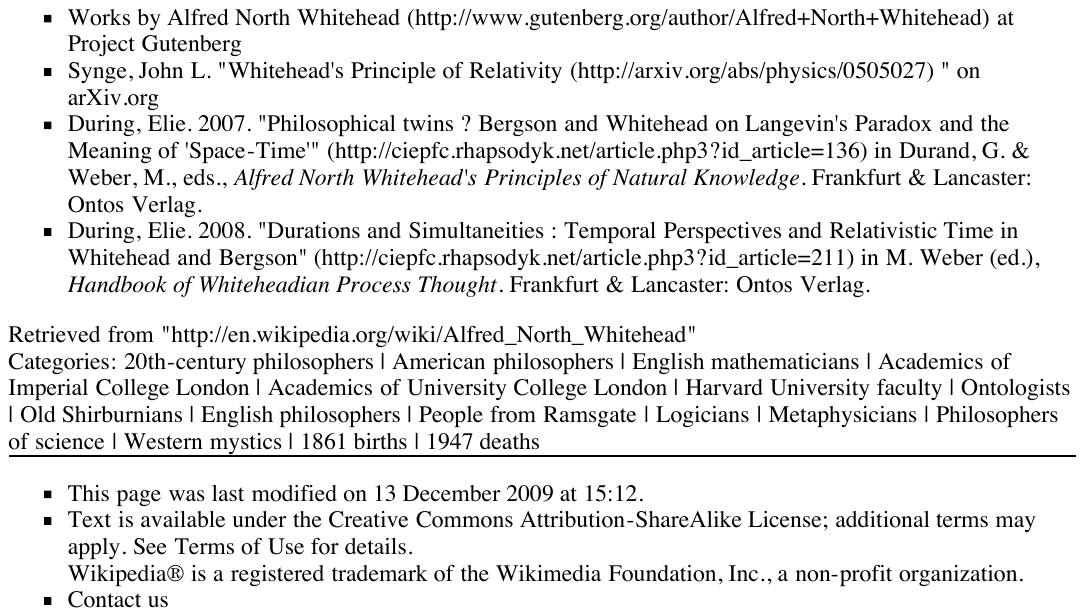}}
		\vspace{0.2cm}
		\caption[Example metadata for philosopher network]{The network of \textbf{philosopher}'s and their philosophical influences, as captured by Wikipedia.  Here we show the \emph{infobox} for mathematician and philosopher A.~N.~Whitehead (\textbf{right}), and the categories that his page is grouped into (\textbf{top}), many of which represent his chosen profession.  The bottom of the infobox lists the other philosophers who influenced his work and the philosophers who were later influenced by him.   
		The page also has a collection of hyperlinks to other wikipedia pages, which we use to quantify the similarity between pairs of philosophers. \label{fig:philosopher_meta}}
		\vspace{3cm}
	\end{minipage}
	\hspace{0.02\textwidth}
	\begin{minipage}[b]{0.25\textwidth}
		\centering
		\frame{\includegraphics[width=1\textwidth,trim=5 5 5 5,clip=true]{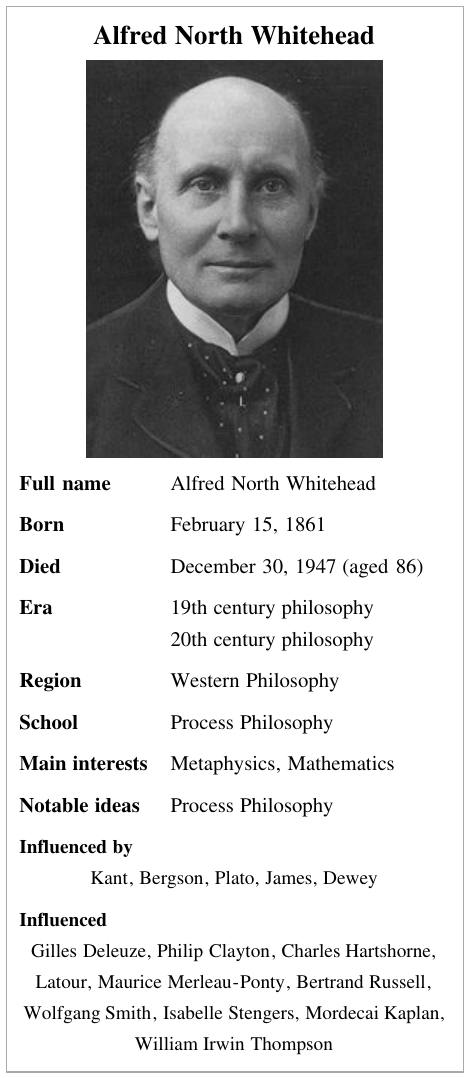}}
	\end{minipage}
\end{figure}

\subsubsection{Philosopher}
Network of famous philosophers and their philosophical influences, as recorded by users of the english-language Wikipedia\footnote{\url{http://en.wikipedia.org}}. 
	\begin{description}
		\item[Construction] The raw data consists of the file \texttt{enwiki-latest-pages-articles.xml} containing all articles in Wikipedia per 2009-12-02, 22:35:45, which was obtained from the site's download section\footnote{\url{http://download.wikimedia.org/enwiki/latest/}}. Wikipedia maintains a list of all philosophers, sorted by name\footnote{\url{http://en.wikipedia.org/wiki/Lists_of_philosophers}}. This set of names forms the nodes of the philosopher network; an example is shown in Fig.~\ref{fig:philosopher_meta}. Internal Wikipedia hyperlinks between philosophers form the network links\footnote{Another choice of links between philosophers would have been the set of links listed under \emph{Influenced by} and \emph{Influenced} in the philosopher `infobox' (see Fig.~\ref{fig:philosopher_meta}), However, most of the articles describing lesser known philosophers do not have infoboxes, so in order to work with the largest possible dataset, we chose to use all internal hyperlinks.}.
		\item[Community quality] Associated with each philosopher's webpage is the set of all (internal) Wikipedia hyperlinks. Besides links to other philosophers, used to build the network, each page has many hyperlinks to philosophical concepts, philosophical schools of thought, time periods, geographical areas, and so on.  We expect more similar philosophers to have more Wikipedia pages in common, so we use the Jaccard index between these sets as the node-node similarity measure in Eq.~\eqref{eqn:CommSimilarity}.
		\item[Overlap quality] Each philosopher is placed into a number of categories (see Fig.~\ref{fig:philosopher_meta} top).  We expect that philosophers that belong to more categories will participate in more communities, due to their broader interests, etc., though the relationship is not necessarily linear.  The mutual information between the number of community memberships and the number of categories is then used to quantify this relationship.  
	\end{description}

\subsubsection{Word association}\label{subsub:wordassoc}

This network is constructed from existing datasets about free association of word pairs~\cite{nelson1998}. This dataset is not only interesting as is, but also acts as a nice testbed for community identification: Since nodes are plain english words, we can qualitatively evaluate how reasonable each community is just by looking at the members of a community.  This network is quite dense and possesses pervasive overlap.

\begin{description}
	\item[Construction] The dataset was created at the University of South Florida and University of Kansas~\cite{nelson1998}. They presented 5,019 stimulus words to more than 6,000 participants and asked them to write the first word that came to mind. For instance, if you hear the word \emph{cheddar}, you will almost certainly think about the word \emph{cheese}. They gathered all of these word pairs and assigned a weight that represents how frequently two given words are associated. This data itself is a weighted, directed network between words. We reduce this network into an undirected, unweighted network by ignoring weight and direction (cf. Palla \emph{et al}.~\cite{palla_cpm_2005}). 
	
	\item[Metadata] We use the WordNet database for the metadata~\cite{fellbaum:wordnet}, assigning a set of meanings/definitions or \emph{senses} to each word (known as \emph{synsets}). Since this database was specifically built for semantic analysis, each detailed meaning of a word has a unique ID, which enables quantitative analysis.
	
	\item[Community quality] We define a pair of words to be similar when they share at least one meaning ID, i.e. $\mu(i,j) = 1$ if $i$ and $j$ share at least one meaning, $0$ otherwise. Then the community quality is defined using Eq.~\eqref{eqn:CommSimilarity}.
	
	\item[Overlap quality] We calculate the mutual information between the number of meanings for each word and the number of non-trivial community memberships for the node.
\end{description}

This network was previously studied using clique percolation in \cite{palla_cpm_2005}.  They used clique size $k=4$ but first removed all edges with weights less than $w_* = 0.025$.  Here we consider the unweighted, unfiltered network and so instead use $k=5$, which gives much higher quality $k$-clique communities and improved composite performance.  In Sec.~\ref{subsubsec:filtering} we discuss this filtering, and show results for $k=4$ with and without weight thresholding (Fig.~\ref{sfig:wordassoc_threshold}).

\subsubsection{Amazon.com products}
\begin{figure}[t]
		\begin{minipage}[b]{0.5\linewidth}
		\raisebox{0ex}{\framebox{\includegraphics[width=0.8\textwidth]{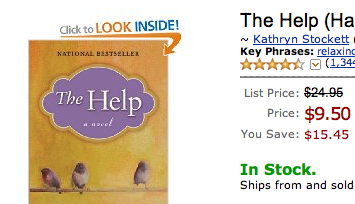}}}\\[1ex]
		\framebox{\includegraphics[width=0.95\textwidth,trim=0 0 250 0,clip=true]{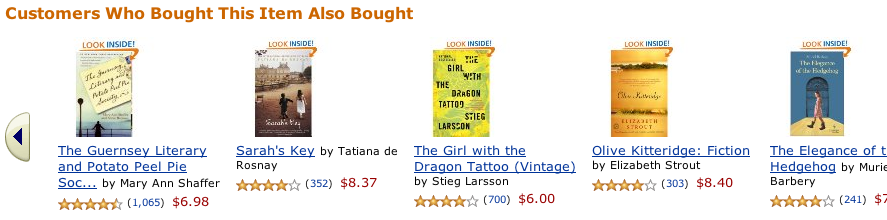}}
		\end{minipage}
		\hspace{2ex} 
		\begin{minipage}[b]{0.5\linewidth}
		\framebox{\includegraphics[width=0.75\textwidth]{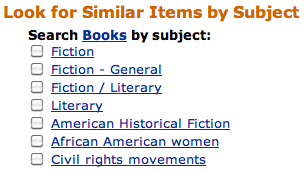}}\\[1ex]
		\framebox{\includegraphics[width=0.88\textwidth,trim=0 0 250 0,clip=true]{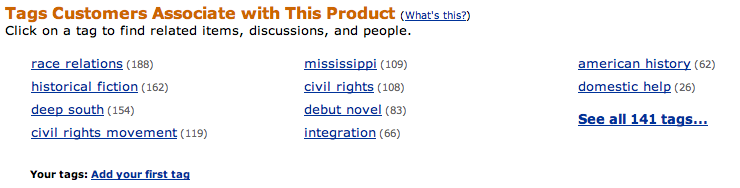}}
	\end{minipage}%
		\caption[Example network and metadata for Amazon.com]{Example of the network and available metadata for the \textbf{Amazon.com} product co-purchases network.  Here we show a \href{http://www.amazon.com/dp/0399155341/}{particular book}, some of the books it is often bought with, the set of subjects it is classified into by Amazon.com, and the set of popular ``tags'' Amazon.com users have chosen to describe or annotate the book's content.  We can use shared tags to quantify how similar pairs of books are, and the more subjects a book has, the more communities it might be expected to belong to. Other combinations of metadata are certainly possible.  Other networks have similar quantities.\label{fig:metadata_amazon}}
\end{figure}

	Products that are frequently purchased at the same time by customers at Amazon.com.  The Amazon Web Service (\url{http://aws.amazon.com/}) provides a tool to programmatically access information about any given product sold on their website.  For a particular product, we retrieve the top five most frequently co-purchased products, the set of tags or annotations that users have applied to describe the product, and the list of subjects the product is sold under.  The former is used to construct the network while the latter two are used for metadata.  See Fig.~\ref{fig:metadata_amazon} for an example product.
	\begin{description}
		\item[Construction] Using Amazon.com's XML web service, on 2009-12-24 we performed a breadth-first search (BFS) crawl (or snowball sample) of co-purchased products by repeatedly retrieving encountered products' top five co-purchases (along with relevant metadata), starting from the number one bestselling book at the time, \textsc{The Help} by Kathryn Stockett.  This crawl continued out to depth $d=12$.  At the final layer of the BFS snowball, many nodes may point to unexplored products at the next step.  These unexplored products are removed from the network, since we do not know their connectivity, resulting in a final network of $N=18142$ nodes.  This network is interesting not only because of the rich metadata that is available but also because this snowball sampling technique does not completely capture the network yet is a common approach when sampling dynamic web data.  Likewise, since Amazon.com only returns the top five most co-purchased products, the network's degree distribution is not accurate (we treat the final network as being undirected).  This provides an interesting test to see how reliant or customized a community method is to the broader degree distributions that are commonly encountered.
		\item[Community quality] Each product is associated with a set of keywords or annotations known as tags.  These tags were applied by users of the website and describe the product, e.g., the plot or characters of a book.  The Jaccard index between the sets of tags was used as the node-node similarity in Eq.~\eqref{eqn:CommSimilarity} to compute the overall ``tag enrichment'' for each algorithm.
		\item[Overlap quality] Similar to user tags, each product is associated with a set of subjects categorizing it.  We expect that products with more subjects will belong to more communities due to the broader nature of the product, as well as user purchasing interests.  Thus we use the number of subjects as the overlap metadata and compute the mutual information between the number of communities and number of subjects.  This tells us how much we have ``learned'' about the subjects a product belongs to merely by learning the number of communities the algorithm has placed the product into. 
\end{description}

	Reversing this metadata choice (using subjects for community quality and number of tags for overlap quality) does not qualitatively alter our composite performance results, indicating that our test procedure is not reliant on particular metadata.

\clearpage

	\section{Validating hierarchical organization} \label{sec:multi_scale_structure}
	The main text shows that link communities present an excellent way to reconcile the apparently disparate notions of hierarchy and overlap, something which has not been accomplished before. As illustrated in main text Fig.~\ref{fig:overlap} and Fig.~\ref{fig:node_link_dendrogram}, it is impossible to find a node hierarchy that captures any pervasively overlapping community structure, even in a very simple case. In this sense, the current approach contrasts with all other hierarchical community methods, because our approach---link communities---is a straightforward way to unify hierarchy and overlap.

	In most of the examples used in the main text, we pick out a scale (determined by the maximal partition density $D$), resulting in the set of `best' communities to study. However, we believe that the choice of a best level of communities is often made because the tools to analyze hierarchy are not as advanced as the tools for communities and that the full structure is currently more difficult to deal with, and not because the best level is the only level worth exploring. 

	Here, we elaborate on the part of the main text showing that the \emph{best} level of a hierarchy is not the \emph{only} level worth exploring. This is true in many domains: For example, faculty, staff, and students at a university may organize at multiple scales, from schools (school of science, school of business, etc.), down to the departmental level (physics department, chemistry department, etc.) and then further down to research groups and small-scale collaborations.  The most modular structure may form at, say, the departmental level, but the structures of both smaller research groups and larger school-wide organizations are still relevant.

	In the main text, the evidence for this point is contained in main text Fig.~\ref{fig:heatmap}.  Below, we present additional evidence for the presence of meaningful, multi-scale structure represented in the link dendrogram, as well as results for the full network corpus. A small number of networks possess metadata about the hierarchy itself, so we also provide alternative evidence for the existence of such structure in those networks. 
	
	No previous methods have captured pervasively overlapping structures across multiple system levels; the combination of pervasive overlap and meaningful community structure on multiple levels of the dendrogram is the multi-scale complexity to which we refer throughout the text. 

	\subsection{Examples of hierarchical structure}

		Before we begin a quantitative analysis, it is useful to qualitatively inspect samples of the detected hierarchical organization.  Here we choose the word association network to illustrate the multi-scale hierarchical structures; in other networks, it is more difficult to appreciate the meanings of communities and their hierarchical organization since we are less familiar with the node labels.

\begin{figure}
		 \centering
		 \includegraphics[width=\textwidth]{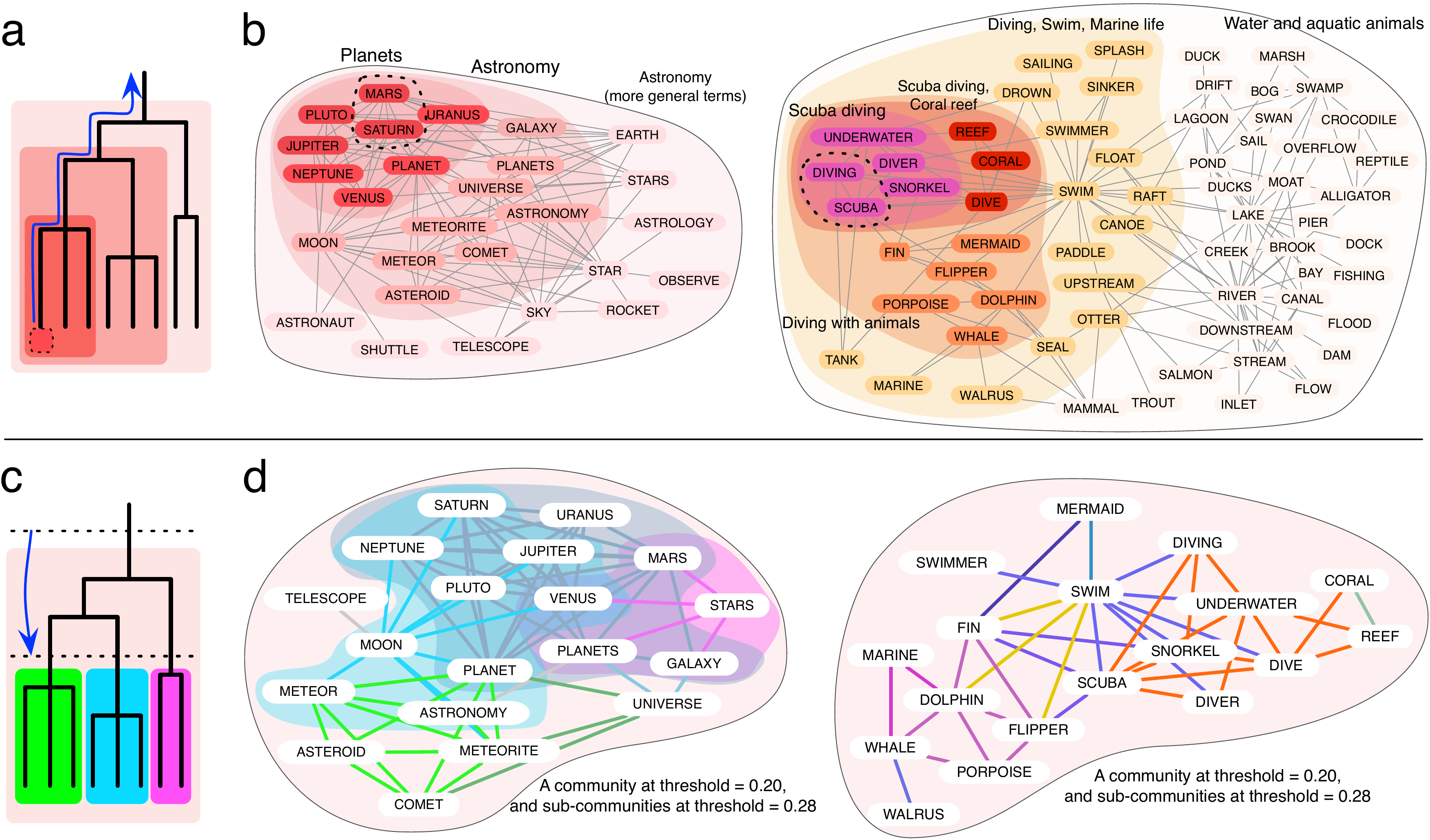}
		 \caption[Examples of hierarchical structure in the word association network]{Examples of hierarchical structure in the word association network. The word association network is a nice example for this purpose, since it is easy to appreciate the meanings and contexts of the individual words and communities. \letter{a} Here we pick a link and follow how the link merges with others as we climb the hierarchical tree. \letter{b} We start from the link \textsc{mars}--\textsc{saturn} on the left, and the link \textsc{scuba}--\textsc{diving} on the right. As we move towards the root of the hierarchical tree, the link \textsc{mars}--\textsc{saturn} forms a `planet' community, an `astronomy' community, and then a more general `astronomy' community. The link \textsc{scuba}--\textsc{diving} results in richer hierarchical structure: the link's community becomes more and more general until we reach a large community of water-related words. \letter{c} Here we delve into the hierarchical structure from a high level community into its sub-communities at a lower level. \letter{d} We pick a sub-community from the example in (b) at threshold 0.20. We then identify its sub-communities at threshold 0.28. These sub-communities are represented by links with different colors. The sub-communities split into meaningful groups of similar words. Note that many links are not shown here because we are only drawing the link communities from these branches. \label{sfig:hierarchy_word_examples}}
\end{figure}

		We use two approaches to decipher complex, hierarchical structure. One is tracking how a single link forms larger and larger super-communities (bottom-up) and the other is drilling down into the sub-communities of a large community (top-down). As shown in Fig.~\ref{sfig:hierarchy_word_examples}, both perspectives clearly (but qualitatively) illustrate the success of the link dendrogram in capturing the network's meaningful communities at multiple levels.

\begin{figure}
		 \centering
		 \includegraphics[width=\textwidth]{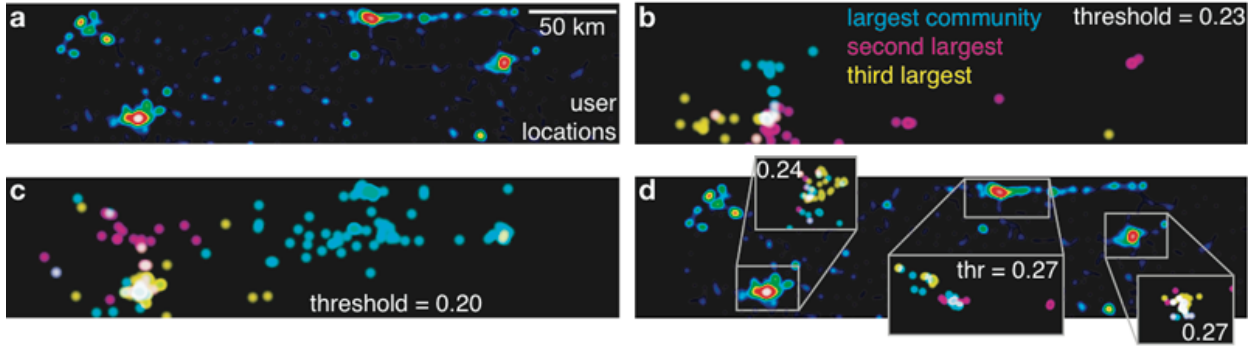}
		 \caption[Spatial hierarchy of mobile phone users]{A spatial hierarchy of link communities amongst mobile phone users. \letter{a} A heatmap showing the most likely geographic locations of all users in the network, several cities are present.  \letter{b} The three largest communities at the link dendrogram threshold with maximum partition density cluster around a single city. \letter{c} Cutting the dendrogram at a lower threshold reveals regional but still spatially correlated communities.  \letter{d} At thresholds above that shown in \textbf{b} we see smaller, intra-city communities. Compare with main text Fig.~\ref{fig:heatmap}.\label{sfig:hierarchy_phone_maps}}
\end{figure}

Figure~\ref{sfig:hierarchy_phone_maps} presents a further example of the spatial hierarchy of link communities within the mobile phone network, expanding on that shown in main text Fig.~\ref{fig:heatmap}.

	\subsection{When is hierarchical structure meaningful?}\label{subsec:what_is_multiscale}
	We begin by noting that finding a hierarchical tree does not necessarily imply the discovery of meaningful structure; one can always build a random tree, for example. The hierarchical tree is only meaningful when the encoded structure is relevant to the system being studied. 
	
	To show that the link dendrogram contains meaningful structure at multiple levels, we now investigate the following:
	\renewcommand\theenumi{(\roman{enumi})}
	\begin{enumerate}
	\item \textbf{Structural changes across the dendrogram.} We show that dendrogram structure is `dynamic' in the sense that when we cut the dendrogram at different thresholds, the community structure changes significantly.  This means that there is not one optimal structure frozen into the dendrogram across a wide range of thresholds.

	\item \textbf{Meaningful communities.} We have already established the partition density $D$ as a measure of the \emph{structural} quality of a given partition of the dendrogram. At the optimal value of $D$, our algorithm finds high quality communities (see main text Fig.~\ref{fig:performance}). As discussed in Sec.~\ref{ssubsec:PDrules}, the partition density $D$ may take on a variety of shapes as a function of the dendrogram cut. The fact that $D$ is sharply peaked does not necessarily imply that multiple,  meaningful levels of community structure do not exist. This is both because a large amount of very different structure may be captured in a very narrow band of the dendrogram and because the partition density is an averaged quantity such that there may be many high quality communities alongside less dense groups.

	While structural quality is important---in particular to community detection algorithms---the network structure \emph{a priori} does not reveal information about how `meaningful' the structure is. In order to quantitatively show that structures at multiple scales are `meaningful' we use metadata to study community quality (see Sec.~\ref{sec:testingComms}) as a function of the link dendrogram cut threshold.
	\end{enumerate}
	The remainder of Sec.~\ref{sec:multi_scale_structure} is devoted to exploring these two aspects in further detail.

	\subsection{Dynamic dendrogram structure}\label{subsec:dynamicdendrogramstructure}
	To begin, we now explore the rate of change of the overlapping community structures encoded in the link dendrograms.  One possible concern is that the number of mergers could potentially drop over a range of the dendrogram, resulting in large gaps where the structure is fixed (e.g., Fig.~\ref{fig:node_hierarchy}d).  In this section, we present evidence that the dendrogram structures for networks in our test corpus are indeed dynamic over a large range of thresholds. 

	\subsubsection{Branching probability}\label{subsubsec:branchingProb}
\begin{figure}
		\includegraphics[width=\hsize]{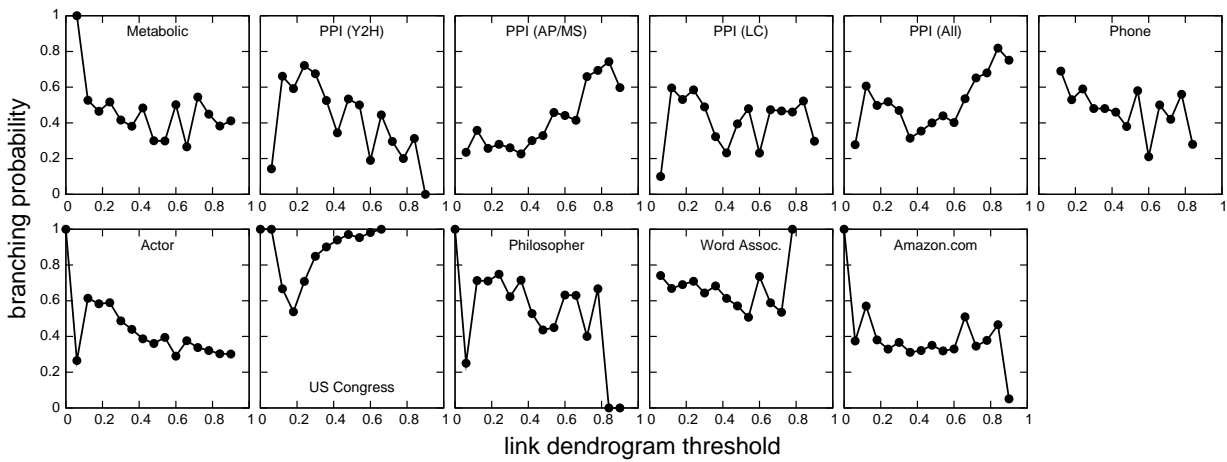}
		\caption[Link dendrogram branching probabilities]{Branching probabilities for the link dendrograms of the networks studied in our test corpus.  In all networks, the branching probability is high over a large range of thresholds, indicating that the structures encoded by the dendrograms are constantly changing. \label{sfig:branching_prob}}
	\end{figure}

	One straightforward way to illustrate the dynamic nature of the link dendrogram is to compute the \emph{branching probability}, the fraction of communities at some threshold $t$ that subsequently split into multiple communities slightly farther down the dendrogram, at threshold $t+\Delta t$.  Low branching probability means that few communities are changing in that level of the dendrogram; conversely, the dendrogram's structure is rapidly changing when the branching probability is high.  As shown in Fig.~\ref{sfig:branching_prob}, all networks in our test corpus possess significant and steady branching probabilities over a wide range of thresholds.
	
\begin{figure}
		{\begin{minipage}[c]{0.45\linewidth}%
		\centering
			\vspace{0pt}%
			\includegraphics[]{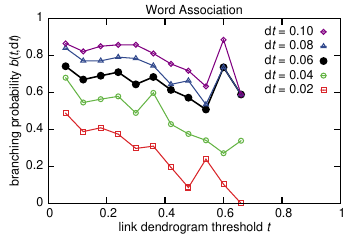}
		\end{minipage}}%
		{\begin{minipage}[c]{0.5\linewidth}%
			\vspace{0pt}%
			\centering
		\caption[Branching probability as a function of window size]{Studying the dependence of the branching probability $b(t,\Delta t)$ on the threshold window $\Delta t$.  Since $b\to0$ when $\Delta t\to0$ and $b\to1$ when $\Delta t\to1$, we must demonstrate that there is a range where $\Delta t$ is small but $b$ is still large.  To do so, we plot $b$ versus $t$ for several small values of $\Delta t$.  We see that even for the lowest value, $b$ is substantial for a wide range of $t$.  Here we show the word association network, but this fact is generic over the test corpus. (We start the curves at $t = 0.7$ because the dense word association network does not begin clustering until $t\approx0.8$, see  Fig.~\ref{sfig:quality_vs_threshold}.)\label{sfig:branching_prob_vs_dt}}
		\end{minipage}}%
		
	\end{figure}

	Here we use $\Delta t=0.06$, but we have tested the dependence of the branching probability on $\Delta t$ in Fig.~\ref{sfig:branching_prob_vs_dt} and find high probabilities over a wide range of values.

	\subsubsection{Distributions of community sizes and node memberships}\label{subsubsec:DistributionsVsThreshold}
	In addition to the branching probabilities, we also examine the distribution of community sizes (nodes per community) and memberships (communities per node) at multiple cuts of the link dendrogram.  These distributions tell us the scales of the detected communities for each threshold, and how those communities overlap.
	
\begin{figure}
		\centering
			\includegraphics[width=0.64\hsize]{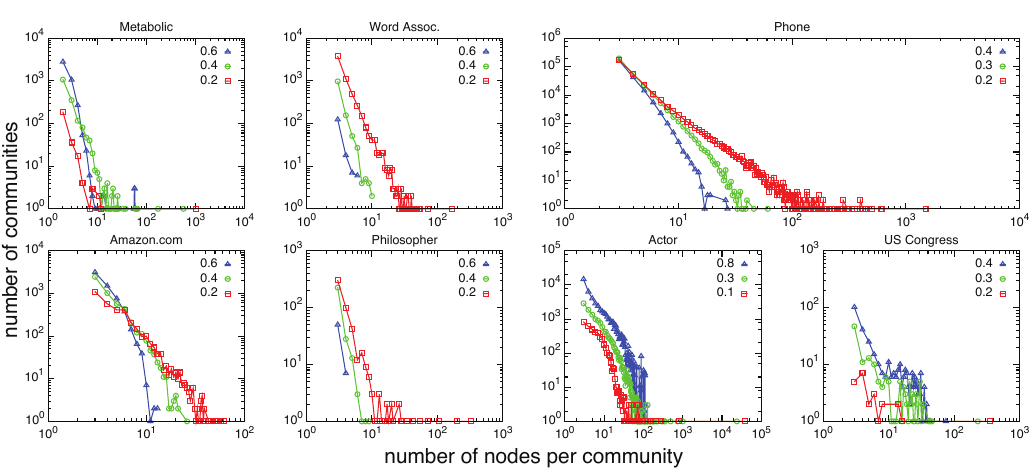}
			\includegraphics[width=0.32\hsize]{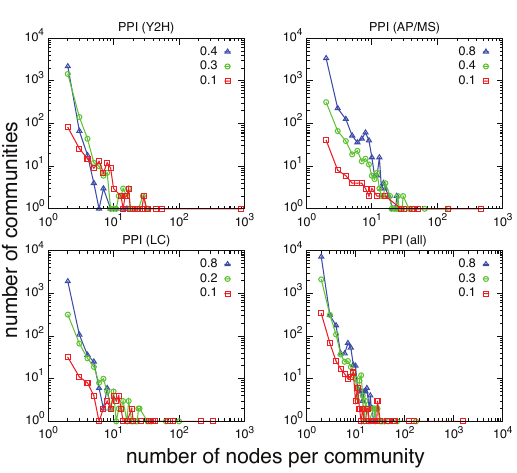}
			\includegraphics[width=0.64\hsize]{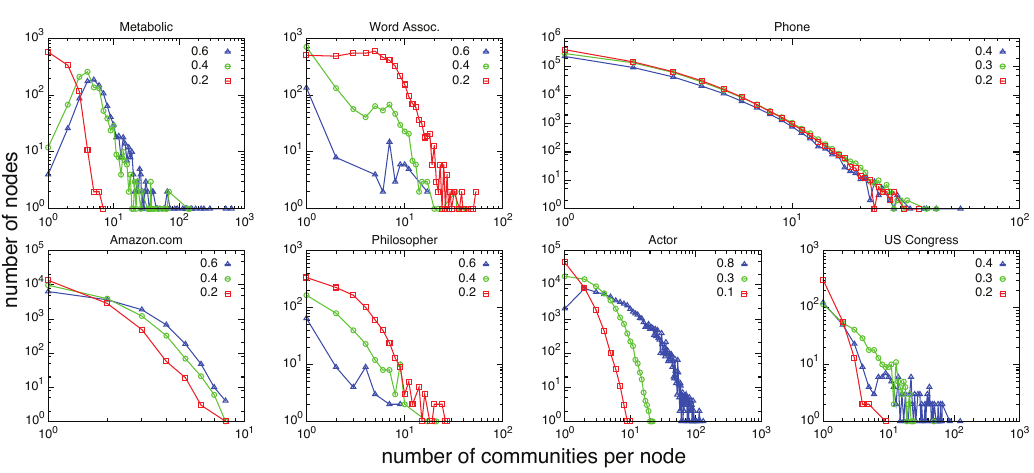}
			\includegraphics[width=0.32\hsize]{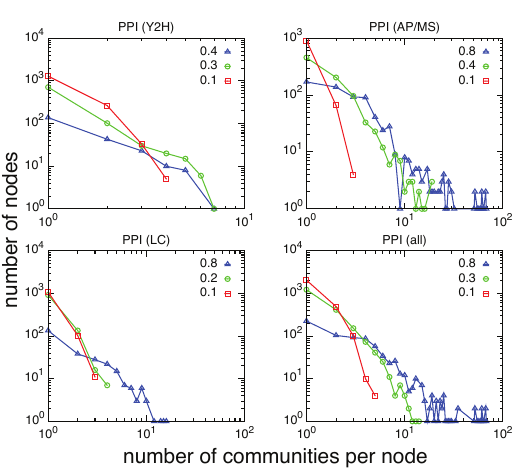}
		\caption[Community and membership distributions for various thresholds]{Overlapping community structure is very different when cutting the link dendrograms at different thresholds.  Shown are the distributions of community sizes and memberships for the networks in our test corpus, each at three different link dendrogram thresholds. A broad, heavy-tailed distribution of community sizes arises at high thresholds in most networks and then persists over a wide range of the link dendrogram, indicating that the link dendrogram does not suddenly collapse but changes smoothly over much of its range.  Meanwhile, the distributions of community memberships per node remain broad over the same region of the dendrogram (this effect is particularly striking in the phone network), indicating that overlapping structure is maintained throughout the dendrograms in nearly all networks.  These results show that the community structures contained in the link dendrograms cover a wide range of scales while maintaining significant overlap.\label{sfig:distribution_of_community_sizes}}
	\end{figure}

	In Fig.~\ref{sfig:distribution_of_community_sizes}, we show these distributions at three different levels of each network's link dendrogram.  We observe that many networks possess broad distributions of community sizes, indicating that a variety of size scales are encoded at each level of the dendrogram.  The broad membership distributions simultaneously indicate that the amount of overlap remains significant at those same levels.
	These results mean that the structures encoded in the link dendrogram do not suddenly collapse but vary smoothy as a function of dendrogram threshold.   We also observe that in some networks the community scales change while the amount of overlap remains steady (particularly the phone network), whereas in other networks the distributions of sizes vary less but the amount of overlap changes drastically (particularly the metabolic and PPI (all) networks).  In conjunction with the branching probability, these properties highlight how the link dendrogram can reveal multiple aspects of the network's levels of hierarchical community structure.

\subsection{Revealing meaningful communities at multiple scales}\label{subsec:revealingMultiple}
	Now that we have shown that very different scale structures are contained throughout the link dendrograms, we must also demonstrate that these structures are meaningful.

	\subsubsection{Community quality as a function of cut-level}\label{subsubsecQualityVsThreshold}
	As we move from the leaves of the dendrogram (where each link is isolated) towards the root (where all links are merged into a giant community) communities must grow in size. Due to the construction of the community quality measures (see Sec.~\ref{sec:testingComms} for details about specific types of metadata), the community quality is likely to drop whenever two communities are joined---since a larger community is likely to be more diverse. For example, while `Physics' and `Chemistry' may be subsumed under the heading `Natural Science', each field on its own is more homogeneous than the merger of the two.

	Thus, it is likely that, relative to the optimal communities, the community quality will decay as the dendrogram cut approaches the root of the dendrogram. For this reason, meaningful communities are expressed as a \emph{slow} decay of community quality, compared to a properly randomized control dendrogram. We now show that all link dendrograms for our test corpus exhibit such slow decay, compared with the following control.
	
\begin{figure}[!htbf]
		\centering
		\includegraphics{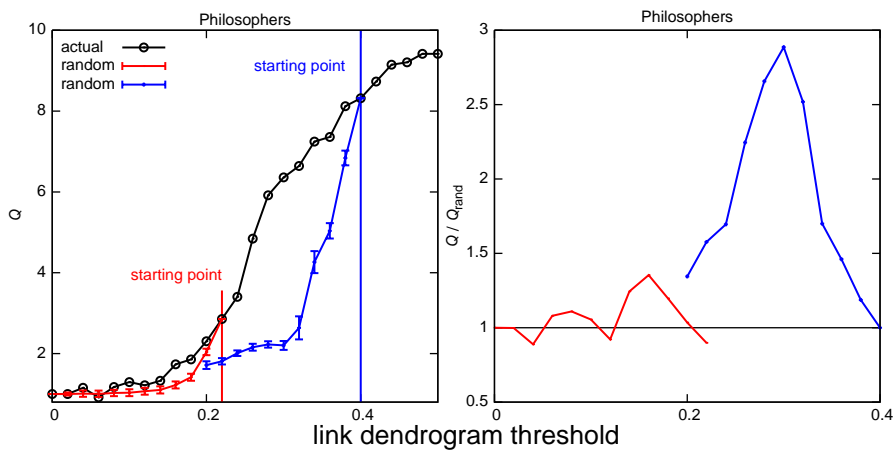}
		\caption[Example of the hierarchical control]{An illustration of the link dendrogram control, using the Philosopher network.  We wish to test whether the hierarchical structure is valid beyond some threshold $t_*$.  To do so, we first compute the edge-pair similarities of all ``cluster-able'' edges.  We then cluster edges according to their similarity (as normal) until we have reached $t_*$.  Afterwards, we then cluster the remaining edge-pairs at random.  This control is much stronger than, e.g., clustering random pairs of edges, since the exact same edge-pairs are being clustered together, only the ordering of the clustering is changed.  If there is significant, meaningful structure for $t<t_*$, we should expect the actual community's quality $Q$ to decay slower than the control's quality $Q_\mathrm{rand}$.  In this example, we choose two values of $t_*$ (vertical lines) and show that the philosopher network's communities possess significant structure beyond $t_*=0.4$, but little structure beyond $0.22$. \label{sfig:demo_linkDendroNullPhilo}}
	\end{figure}

	\begin{description}
		\item[Randomized control dendrogram]  We wish to test whether the hierarchical structure is valid beyond some threshold $t_*$, e.g., that with maximum partition density.  To do this, we introduce the following control:  first, compute the similarities $S(e_{ik},e_{jk})$ for all connected edge-pairs $(e_{ik},e_{jk})$, as normal.  Then perform our standard single-linkage hierarchical clustering, merging all edge-pairs in descending order of $S$ while $S \geq t_*$, fixing the community structure at $t = t_*$.

		Below $t_*$, randomly shuffle similarities amongst the remaining edge-pairs with $S<t_*$, then proceed with the merging process as before.  This randomization only alters merging order, and ensures that the \emph{rate} of edge-pair merging is preserved, since the same similarities are clustered.  This strictly controls not only the merging rate, but also the similarity distributions and the high-quality community structure found at $t_*$. This procedure ensures that the dendrogram is properly randomized while other salient features are conserved. See Fig.~\ref{sfig:demo_linkDendroNullPhilo}.
	\end{description}

\begin{figure}[!htbf]
		\centering
		\includegraphics[width=1\hsize]{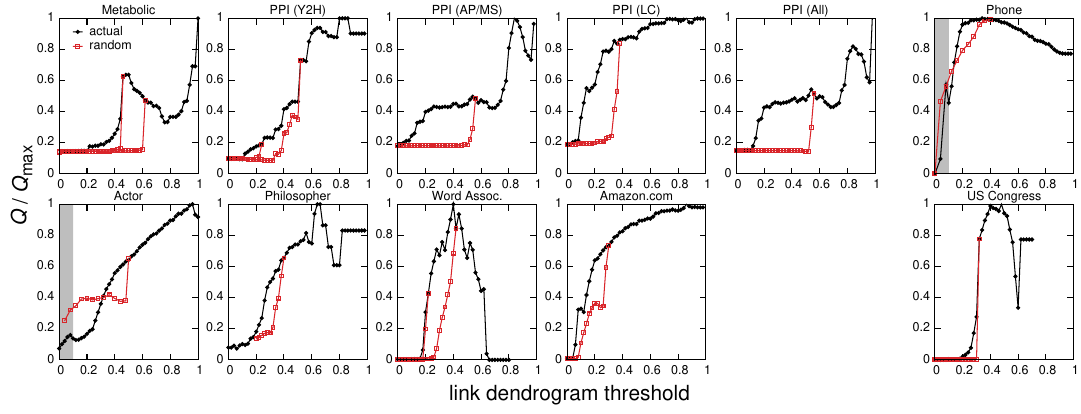}
		\includegraphics[width=1\hsize]{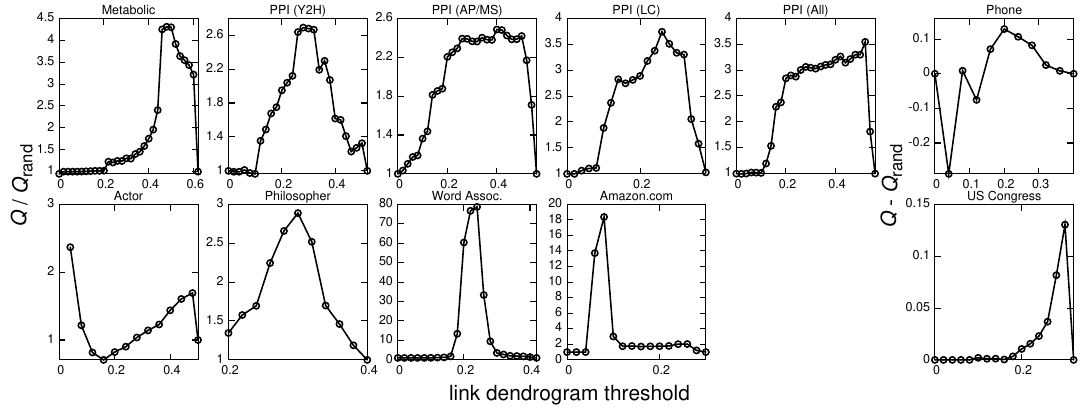}
		\caption[Community quality across the link dendrograms]{(\textbf{top}) The community quality $Q$ (see Sec.~\ref{sec:measures}) as a function of dendrogram threshold for the corpus networks.  We see that most networks possess very slow decay of quality across a wide range of the dendrogram.  This is particularly true for PPI (AP/MS), PPI (LC), PPI (All), Phone, word association, and Amazon.com networks.  The control, shown in red, indicates that all networks possess meaningful hierarchical structure beyond the examined threshold. (For metabolic, PPI (Y2H), and the word association networks, we test multiple thresholds.) Notice in the Actor network we see that the very large link communities appear \emph{worse} than the control.  The IMDb data is known to strongly split at very large scales, according to language groups~\cite{LeskobitchLarge}.  Since our quality measure is based on plot keywords and not languages, the dendrogram may capture the true, large scale structure but this is not reflected in the metadata.   We plot $Q / Q_\mathrm{max}$, normalizing the enrichments (dispersions in the case of the Phone and US Congress networks) by their maximal value.  For the large Phone and Actor networks, we sample communities to speed up the calculation of the quality of the null partitions.  This may introduce a small positive bias in the shaded regions. (\textbf{bottom}) The relative quality $Q / Q_\mathrm{rand}$ (the ratio of the two curves), highlighting the validity of each link dendrogram's hierarchy.  For the Phone and US Congress networks we instead plot $Q - Q_\mathrm{rand}$ as the difference is more meaningful than the ratio for dispersive measures.\label{sfig:quality_vs_threshold}}
\end{figure}

If there is significant, meaningful structure for $t<t_*$, we expect the actual community's quality $Q$ to decay slower than the randomized control quality $Q_\mathrm{rand}$.  As shown in Fig.~\ref{sfig:quality_vs_threshold}, this is the exact behavior we find across the entire network corpus\footnote{Notice in the Actor network we see that the very large link communities appear \emph{worse} than the control.  The IMDb data is known to strongly split at very large scales, according to language groups~\cite{LeskobitchLarge}.  Since our quality measure is based on plot keywords and not languages, the dendrogram may capture the true, large scale structure but this is not reflected in the metadata.}.

	\subsubsection{Hierarchical metadata}\label{subsecHierarchicalMetadata}
\begin{figure}
	 \centering
	 \includegraphics[width=0.6\textwidth]{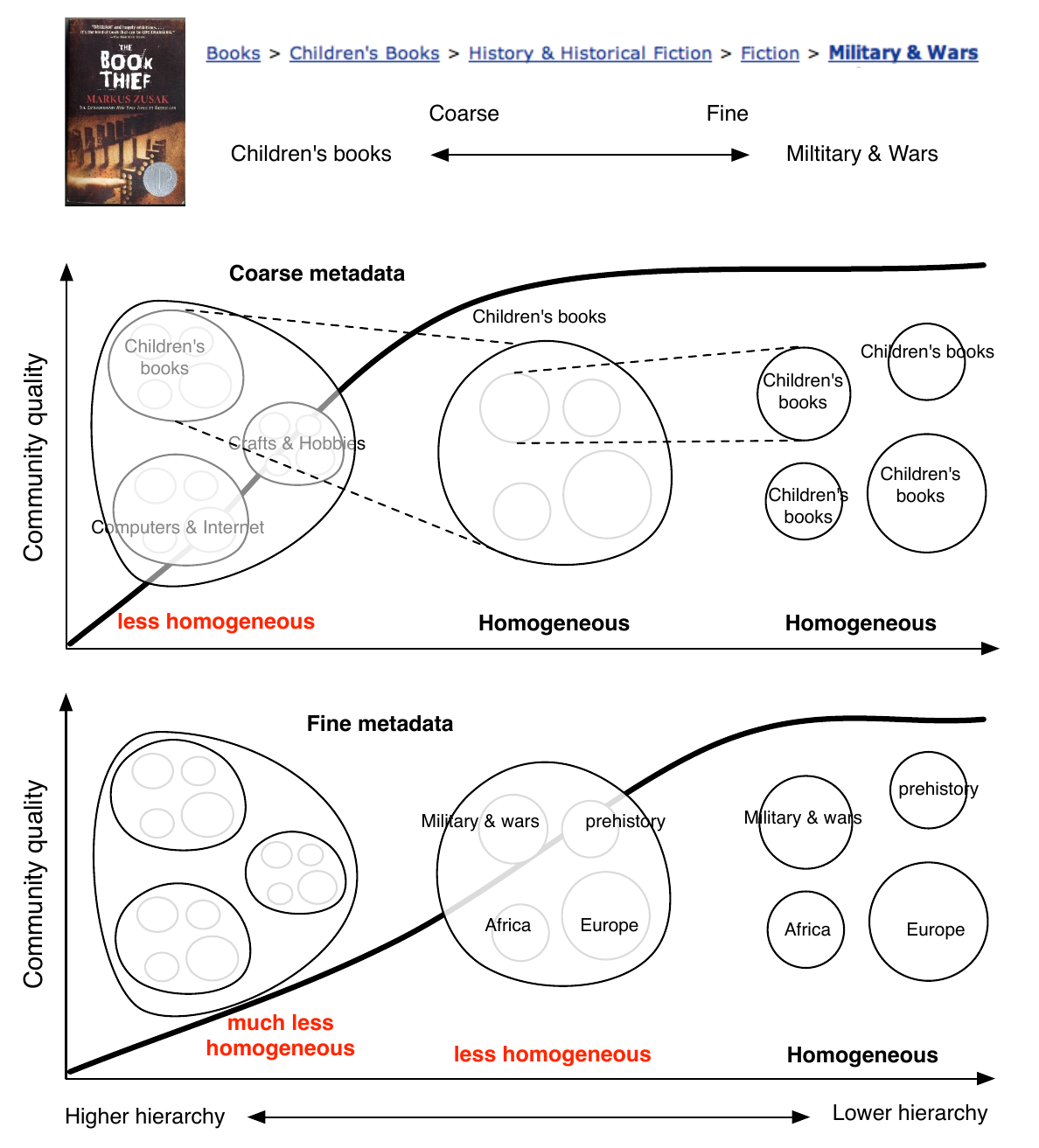}
	 \caption[Multi-scale metadata]{A cartoon explaining multi-scale metadata. \letter{top} Nodes in some of our test networks have metadata that are organized hierarchically. We can use these data to study the hierarchical organization of the communities we detect. This schematic figure illustrates the case where community structure at multiple levels is successfully revealed. \letter{middle} If we use coarse metadata to evaluate the community quality, it will remain high until we reach the point where the scale of communities is larger than the scale described by the coarse metadata. \letter{bottom} Meanwhile, if we use fine metadata, the quality will remain high until the point where the scale of communities is larger than the scale described by the fine metadata. That is, a clear distinction between the two curves of community quality versus threshold will emerge: one with coarse metadata and the other with fine metadata. The difference will vanish if one fails to capture the hierarchical structure between the two scales that are described by coarse and fine metadata. See Figs.~\ref{sfig:hierarchy_multi_metadata_PPI} and \ref{sfig:hierarchy_multi_metadata_amazon} for results. \label{sfig:coarse_fine_cartoon}}
	\end{figure}

	Finally, the Amazon.com and PPI networks in our test corpus possess multi-level metadata.  For these networks, we can construct a direct test of whether there are meaningful communities at different levels of the link dendrogram. For instance, a book in the Amazon.com network has category information at multiple levels of granularity, see Fig.~\ref{sfig:coarse_fine_cartoon} (top) for an example. The PPI networks also contain hierarchical information: GO terms (see Sec.~\ref{subsec:ppi_metadata}) are organized hierarchically, forming a directed acyclic graph; the MIPS functional catalog also provides a hierarchical categorization of each protein. 

	From these hierarchical metadata, we now extract two sets of metadata: \emph{coarse} and \emph{fine}. If our method is able to find meaningful structures at multiple scales, we expect that the community quality based on the fine metadata will have high values at cuts near the leaves of the dendrogram, and the community quality based on coarse metadata will high values for lower thresholds (higher than those using the fine metadata). That is, coarse-grained communities at the lower threshold will conform well with the coarse metadata while detailed, fine communities at higher thresholds will conform well with the fine metadata, as illustrated in Fig.~\ref{sfig:coarse_fine_cartoon}. 

	For the Amazon.com network, we use the available subject categories given for each book, stored as lists, each of which are ordered by level of granularity (one list for \textsc{The Book Thief} is shown at the top of Fig.~\ref{sfig:coarse_fine_cartoon}).  Broad categories such as `General' are removed.  The coarse metadata for each book is then the set of first elements of that book's category lists, and the fine metadata are the last elements.

	In the PPI network, we use the MIPS functional catalog annotations since they provide a clearly defined set of hierarchical metadata: For instance, \emph{metabolism} is labeled `01', \emph{amino acid metabolism} is `01.01', \emph{assimilation of ammonia} is `01.01.03', and so on. Each level is separated by a period, and each level is represented by two digits. The coarse metadata is obtained by reducing every annotation to its first hierarchical level. For instance, if a protein has an annotation `01.01.03', we can represent it by `01'. These metadata constitute the coarse metadata for the protein. The fine metadata is obtained by removing all metadata that have two or less levels of information, and reducing longer metadata to three levels. For example, `01.01' or `01' will be removed from the annotation, and `01.01.01.01.01' becomes `01.01.01'. We choose the third level as the fine metadata because there are only a few proteins that have finer levels of annotations, and thus these finer levels are too noisy. 
	
\begin{figure}
	{\begin{minipage}[c]{0.5\linewidth}%
	\vspace{0pt}%
	\includegraphics[width=1\textwidth]{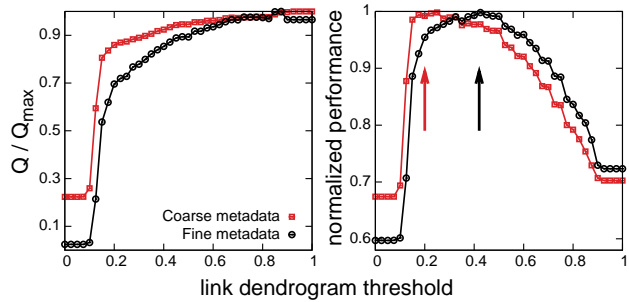}
	\end{minipage}}%
	\hfill
	{\begin{minipage}[c]{0.475\linewidth}%
	\vspace{0pt}%
	\centering
	\caption[Amazon.com network with hierarchical metadata]{Hierarchically organized product category metadata for the Amazon.com network confirms the validity of the discovered link dendrogram.  \letter{left} Community quality remains high for the coarse metadata for longer than the fine metadata, although both decay quite slowly.  Note that controlling for the global baseline enrichment by normalizing with $\left(Q-Q_\mathrm{min}\right) / \left(Q_\mathrm{max}-Q_\mathrm{min}\right)$ does not change this effect. \letter{right} Normalized performance, the normalized sum of community quality and coverage, reveals that the fine metadata peaks earlier (threshold $\sim 0.4$) than the coarse metadata (threshold $\sim 0.2$), indicating that the community partitions at multiple levels of the link dendrogram are meaningful according to the hierarchical metadata.   \label{sfig:hierarchy_multi_metadata_amazon}}
	\end{minipage}}%
\end{figure}

\begin{figure}[htbp]%
{\begin{minipage}[c]{0.6\linewidth}%
	\vspace{0pt}%
	\includegraphics[width=0.85\textwidth]{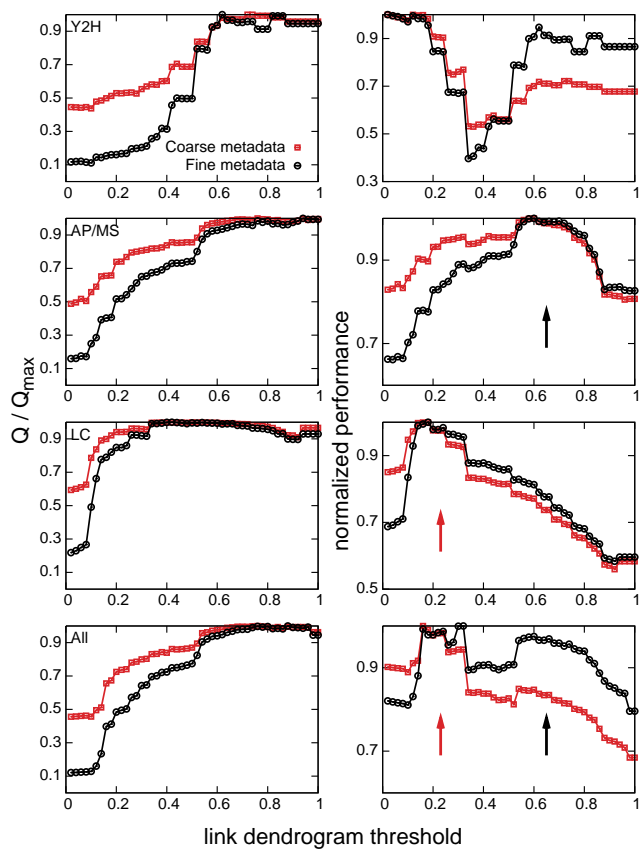}
\end{minipage}}%
\hfill
{\begin{minipage}[c]{0.4\linewidth}%
	\vspace{0pt}%
	\centering
	\caption[Yeast PPI networks with hierarchical metadata]{ Hierarchical metadata confirms that distinct structures are visible throughout the link dendrograms of the PPI networks.  Here we compute community quality (left column) and normalized performance, the normalized sum of quality and coverage (right column) for all four networks.  As with the Amazon.com network, the quality decays more rapidly for the fine metadata than for the coarse (see Fig.~\ref{sfig:coarse_fine_cartoon}), indicating that each link dendrogram's structures correspond well with the networks' existing metadata.
	Using normalized performance, the highly clustered AP/MS network shows a distinct pattern in the link dendrogram compared to the LC network.  The dense AP/MS protein co-complex clusters give that network a clear optimum at higher thresholds ($\sim 0.6$, black arrow) than the LC network, which peaks at $\sim0.2$ (red arrow).  The PPI (all) network, which contains AP/MS and LC, shows \emph{two} distinct peaks in performance, one corresponding to the AP/MS structure and one corresponding to LC.  Thus the link dendrogram for the PPI (all) network captures AP/MS-specific structure at one level and LC-specific structure at another.\label{sfig:hierarchy_multi_metadata_PPI}}
\end{minipage}}%
\end{figure}

	With these two sets of metadata, we calculate community quality and coverage for the different networks.   Figures \ref{sfig:hierarchy_multi_metadata_amazon} and \ref{sfig:hierarchy_multi_metadata_PPI} clearly show the difference between coarse and fine metadata. In every case, the coarse metadata remains relatively more important at lower thresholds (near the root of the dendrogram) and the fine metadata becomes less important. This confirms our hypothesis shown in Fig.~\ref{sfig:coarse_fine_cartoon} and indicates that the structures throughout the link dendrogram correspond well to the hierarchical metadata.
	
	Finally, it is interesting to note that the highly clustered AP/MS network shows a distinct pattern in the link dendrogram compared to the LC network.  By calculating `normalized performance,' the normalized sum of community quality $Q/Q_\mathrm{max}$ and coverage, we see that the dense AP/MS protein co-complex clusters give that network a clear optimum at higher thresholds ($\sim 0.6$) than the LC network, which peaks at $\sim0.2$.  Meanwhile, the PPI (all) network, which contains all other PPI networks, shows \emph{two} distinct peaks in performance, one corresponding to the AP/MS structure and one corresponding to LC.  Thus the link dendrogram for the PPI (all) network captures AP/MS-specific structure at one level and LC-specific structure at another.  The sparse Y2H network does not exhibit as much community structure as LC and AP/MS, and thus has little impact on the community structures of PPI (all), compared with the other constituent networks.

\clearpage

\appendix
\section{Tables of measures}\label{app:all_measures}
Here we list the raw (unnormalized) values for the four calculated measures, the networks and the algorithms that were shown in main text Fig.~\ref{fig:performance} and Fig.~\ref{fig:composite_performance_withNodes}.  For clique percolation we have chosen the value of $k$ that gives the best overall composite score (see Appendix \ref{app:cpm_vals}), unless there is an existing precedent in the literature.  Note that this weighs coverage and quality equally, but an experimenter may wish to prioritize coverage for quality, or vice versa.

\subsection{Overall methods}
{\tiny 
\begin{verbatim}
#1### metabolic #############
#     comm. quality   comm. coverage  overlap quality overlap coverage
L     4.77233482      0.95009597      0.41809907      4.65642994  
N     3.79668962      0.63339731      0.03529230      0.63339731  
C     1.14228117      0.66890595      0.16156926      0.87523992  
G     1.99859878      0.99808061      0.02294501      0.99808061  
I     3.69066921      1.00000000      0.02419719      1.00000000  
# overall winner: L

#2### PPI (Y2H) #############
#     comm. quality   comm. coverage  overlap quality overlap coverage
L     2.35223830      0.55555556      0.08653618      0.72374013  
N     1.92594816      0.73102611      0.02482696      0.73102611  
C     1.87169405      0.16393443      0.05782404      0.18397086  
G     1.36153975      0.99149970      0.01060934      0.99149970  
I     2.25216779      0.98785671      0.01590075      0.98785671  
# overall winner: L

#3### PPI (AP/MS) #############
#     comm. quality   comm. coverage  overlap quality overlap coverage
L     2.73145231      0.83864542      0.38406704      2.57669323  
N     2.19482167      0.91135458      0.03996271      0.91135458  
C     2.07359450      0.76792829      0.13443620      0.81673307  
G     1.94785560      0.99103586      0.01632051      0.99103586  
I     2.75864056      0.99203187      0.01480369      0.99203187  
# overall winner: L

#4### PPI (LC) #############
#     comm. quality   comm. coverage  overlap quality overlap coverage
L     4.52990197      0.55812036      0.17366791      0.93075021  
N     2.34665560      0.96537510      0.02541182      0.96537510  
C     2.91313090      0.55647156      0.11309138      0.60428689  
G     2.58449740      0.99175598      0.01737294      0.99175598  
I     3.76173052      0.99175598      0.01857447      0.99175598  
# overall winner: L

#5### PPI (All) #############
#     comm. quality   comm. coverage  overlap quality overlap coverage
L     3.51593751      0.41260535      0.19629188      1.29754489  
N     2.78187442      0.76511543      0.01589616      0.76511543  
C     1.07221941      0.52876512      0.10141995      0.57053866  
G     1.36531606      0.99523635      0.00797972      0.99523635  
I     3.35047694      0.99340418      0.01124843      0.99340418  
# overall winner: L

#6### phone #############
#     comm. quality   comm. coverage  overlap quality overlap coverage
L     0.75761102      0.76180404      0.13760916      1.42556059  
N     0.33284757      0.78113498      0.01301070      0.78113498  
C     0.82114799      0.33514186      0.07811141      1.27819838  
G    -0.17040443      0.99970880      0.00029690      0.99970880  
I     0.61369550      0.99967268      0.00031225      0.99967268  
# overall winner: L

#7### actor #############
#     comm. quality   comm. coverage  overlap quality overlap coverage
L     6.65974811      0.57986085      0.04076675      1.51764549  
N     1.87424645      0.83947724      0.00706428      0.83947724  
C     2.03239313      0.69482725      0.01468963      0.79485544  
G     1.56548814      1.00000000      0.00000000      1.00000000  
I     2.01709951      0.99273116      0.00106879      0.99273116  
# overall winner: L

#8### congress #############
#     comm. quality   comm. coverage  overlap quality overlap coverage
L     0.34647780      0.94358974      0.68222751      5.89743590  
N     0.38692427      0.61794872      0.03855387      0.61794872  
C     0.26286049      0.61282051      0.15720036      0.77435897  
G     0.42350813      1.00000000      0.00000000      1.00000000  
I     0.32595601      0.99487179      0.00000000      0.99487179  
# overall winner: L

#9### philosopher #############
#     comm. quality   comm. coverage  overlap quality overlap coverage
L     2.40739272      0.81788351      0.45773225      2.66119770  
N     1.68405991      0.66037736      0.06243616      0.66037736  
C     1.18575668      0.74405250      0.12858942      0.77276456  
G     1.47736530      0.99835931      0.00791000      0.99835931  
I     2.20936130      0.99015587      0.01235264      0.99015587  
# overall winner: L

#10### word assoc. #############
#     comm. quality   comm. coverage  overlap quality overlap coverage
L    83.16274063      0.92447190      0.09459306      5.23455560  
N     5.31083477      0.56954962      0.02424692      0.56954962  
C    33.94752060      0.62554803      0.06495803      1.05579912 
G     1.69216772      0.99820646      0.00275916      0.99820646  
I    12.98083645      1.00000000     -0.00000000      1.00000000  
# overall winner: L

#11### amazon #############
#     comm. quality   comm. coverage  overlap quality overlap coverage
L   102.81247272      0.90629479      0.01281968      1.22103406  
N     6.71393780      0.95022599      0.00296223      0.95022599  
C    89.11665793      0.88562452      0.01051039      1.03836402  
G     8.95745118      1.00000000      0.00000000      1.00000000  
I    75.04521188      1.00000000      0.00000000      1.00000000  
# overall winner: L
\end{verbatim}}

\subsection{Clique Percolation}\label{app:cpm_vals}
When applying clique percolation we picked the value of clique size $k$ that gave the best overall (normalized) composite score.  Here we list the raw values for multiple $k$ (shown as \texttt{cp3}, \texttt{cp4}, etc.).  The overall winner lists the chosen value of $k$ used in the main text and in Appendix \ref{app:all_measures}.  If there is an existing precedent for which value of $k$ to use, such as with the mobile phone data \cite{PallaQuantifyingSocialGroupEvolution2007}, we follow the original work.

It is important to note that choosing the $k$ to maximize the composite performance score weighs coverage and quality equally, whereas a researcher may wish to sacrifice coverage for quality.  Higher values of $k$ tend to find very high quality communities; it is up to the researcher's discretion if such a choice is appropriate to his or her particular application.

{\tiny 
\begin{verbatim}
#1### metabolic #############
#                 cp3          cp4          cp5          cp6          cp7          cp8          cp9
comm. quality     1.05405749   1.08502531   1.14092849   1.14228117   1.24034244   1.31172522   1.29127564
comm. coverage    0.99328215   0.97696737   0.88291747   0.66890595   0.46065259   0.32053743   0.17850288
over. quality     0.02817960   0.05334991   0.10411422   0.16156926   0.14877168   0.17817169   0.15137653
over. coverage    0.99328215   1.01919386   0.97312860   0.87523992   0.54798464   0.44913628   0.19673704
# overall winner: cp6

#2### PPI (Y2H) #############
#                 cp3          cp4          cp5
comm. quality     1.87169405   8.85655602   0.00000000
comm. coverage    0.16393443   0.01700061   0.00000000
over. quality     0.05782404   0.01462019   0.00000000
over. coverage    0.18397086   0.01700061   0.00000000
# overall winner: cp3

#3### PPI (AP/MS) #############
#                 cp3          cp4          cp5
comm. quality     1.84479625   2.07359450   1.98196162
comm. coverage    0.87250996   0.76792829   0.67430279
over. quality     0.08580265   0.13443620   0.11253619
over. coverage    0.91235060   0.81673307   0.70019920
# overall winner: cp4

#4### PPI (LC) #############
#                 cp3          cp4          cp5
comm. quality     2.91313090   3.59607890   3.98440063
comm. coverage    0.55647156   0.30502885   0.19043693
over. quality     0.11309138   0.09132684   0.06699107
over. coverage    0.60428689   0.32976092   0.20857378
# overall winner: cp3

#5### PPI (All) #############
#                 cp3          cp4          cp5
comm. quality     1.07221941   2.43214030   2.36350203
comm. coverage    0.52876512   0.35507512   0.28325394
over. quality     0.10141995   0.09320351   0.07553584
over. coverage    0.57053866   0.38402345   0.29754489
# overall winner: cp3

#6### phone #############
#                 cp4
comm. quality     0.82114799
comm. coverage    0.33514186
over. quality     0.07811141
over. coverage    1.27819838
# overall winner: cp4

#7### actor #############
#                 cp3          cp4          cp5
comm. quality     2.03239313   2.16441181   2.65452628
comm. coverage    0.69482725   0.49747074   0.36856003
over. quality     0.01468963   0.01664214   0.01300756
over. coverage    0.79485544   0.60769014   0.45345715
# overall winner: cp3

#8### congress #############
#                 cp3          cp4          cp5          cp6          cp7          cp8          cp9         cp10        cp11        cp12
comm. quality     0.00062736   0.00447642   0.00983444   0.00973442   0.02589494   0.05050994   0.26286049  0.29205756  0.31483181  0.31804022 
comm. coverage    0.94358974   0.88717949   0.83846154   0.78461538   0.71794872   0.66410256   0.61282051  0.55384615  0.48717949  0.40512821 
over. quality     0.04074278   0.02453166   0.03971995   0.07602430   0.07531725   0.06842514   0.15720036  0.04223622  0.05535796  0.0
over. coverage    0.94358974   0.88717949   0.84871795   0.81538462   0.75641026   0.68717949   0.77435897  1.11574074  1.04736842  1.0
# overall winner: cp9

#9### philosopher #############
#                 cp3          cp4          cp5          cp6          cp7
comm. quality     1.18575668   1.42170714   1.76620515   3.78843554   4.61831912
comm. coverage    0.74405250   0.49056604   0.26579163   0.11812961   0.05004102
over. quality     0.12858942   0.19299861   0.20831356   0.16526745   0.11187828
over. coverage    0.77276456   0.55865463   0.34536505   0.18375718   0.07957342
# overall winner: cp3

#10### word assoc. #############
#                 cp3          cp4          cp5          cp6          cp7          cp8          cp9
comm. quality     1.00181181   1.16419232  33.94752060  61.96886046  63.47129301 101.26453476  47.66309121
comm. coverage    0.99860502   0.93941809   0.62554803   0.25308888   0.05759267   0.01215624   0.00378637
over. quality     0.00787718   0.06339547   0.06495803   0.03193241   0.01055262   0.00234616   0.00188072
over. coverage    1.03786369   1.41072140   1.05579912   0.34436030   0.06695895   0.01335193   0.00378637
# overall winner: cp4 (note: we use k=5 because the k=4 comm. quality was too low.  These results are unfiltered)

#11### amazon #############
#                 cp3          cp4          cp5          cp6
comm. quality    89.11665793 123.14041107 132.90590155 138.69567284
comm. coverage    0.88562452   0.60577665   0.30729798   0.07871238
over. quality     0.01051039   0.01587309   0.01210945   0.00709472
over. coverage    1.03836402   0.66563775   0.32526734   0.08020064
# overall winner: cp3
\end{verbatim}}

\bibliographystyle{naturemag}
\spacing{1}

{\Large\noindent References}

\begin{thebibliography}{10}
\expandafter\ifx\csname url\endcsname\relax
  \def\url#1{\texttt{#1}}\fi
\expandafter\ifx\csname urlprefix\endcsname\relax\def\urlprefix{URL }\fi
\providecommand{\bibinfo}[2]{#2}
\providecommand{\eprint}[2][]{\url{#2}}

\bibitem{newman_structure_2006}
\bibinfo{author}{Newman, M. E.~J.}, \bibinfo{author}{Barab\'asi, A.-L.} \&
  \bibinfo{author}{Watts, D.~J.}
\newblock \emph{\bibinfo{title}{The Structure and Dynamics of Networks:}}
  (\bibinfo{publisher}{Princeton University Press}, \bibinfo{year}{2006}),
  \bibinfo{edition}{1} edn.

\bibitem{caldarelli_scale-free_2007}
\bibinfo{author}{Caldarelli, G.}
\newblock \emph{\bibinfo{title}{Scale-Free Networks: Complex Webs in Nature and
  Technology}} (\bibinfo{publisher}{Oxford University Press, USA},
  \bibinfo{year}{2007}).

\bibitem{dorogovtsev_critical_2008}
\bibinfo{author}{Dorogovtsev, S.~N.}, \bibinfo{author}{Goltsev, A.~V.} \&
  \bibinfo{author}{Mendes, J. F.~F.}
\newblock \bibinfo{title}{Critical phenomena in complex networks}.
\newblock \emph{\bibinfo{journal}{Reviews of Modern Physics}}
  \textbf{\bibinfo{volume}{80}}, \bibinfo{pages}{1275--61}
  (\bibinfo{year}{2008}).

\bibitem{newmanGirvanCommsPNAS}
\bibinfo{author}{Girvan, M.} \& \bibinfo{author}{Newman, M. E.~J.}
\newblock \bibinfo{title}{Community structure in social and biological
  networks.}
\newblock \emph{\bibinfo{journal}{Proceedings of the National Academy of
  Sciences}} \textbf{\bibinfo{volume}{99}}, \bibinfo{pages}{7821--7826}
  (\bibinfo{year}{2002}).

\bibitem{Fortunato201075}
\bibinfo{author}{Fortunato, S.}
\newblock \bibinfo{title}{Community detection in graphs}.
\newblock \emph{\bibinfo{journal}{Physics Reports}}
  \textbf{\bibinfo{volume}{486}}, \bibinfo{pages}{75--174}
  (\bibinfo{year}{2010}).

\bibitem{krogan_APMS_2006}
\bibinfo{author}{Krogan, N.~J.} \emph{et~al.}
\newblock \bibinfo{title}{Global landscape of protein complexes in the yeast
  saccharomyces cerevisiae}.
\newblock \emph{\bibinfo{journal}{Nature}} \textbf{\bibinfo{volume}{440}},
  \bibinfo{pages}{637--643} (\bibinfo{year}{2006}).

\bibitem{gavin_APMS_2006}
\bibinfo{author}{Gavin, A.-C.} \emph{et~al.}
\newblock \bibinfo{title}{Proteome survey reveals modularity of the yeast cell
  machinery}.
\newblock \emph{\bibinfo{journal}{Nature}} \textbf{\bibinfo{volume}{440}},
  \bibinfo{pages}{631--636} (\bibinfo{year}{2006}).

\bibitem{wassermanFaustBookSocNetAnalysis}
\bibinfo{author}{Wasserman, S.} \& \bibinfo{author}{Faust, K.}
\newblock \emph{\bibinfo{title}{Social Network Analysis: Methods and
  Applications}}.
\newblock Structural analysis in the social sciences
  (\bibinfo{publisher}{Cambridge University Press}, \bibinfo{year}{1994}).

\bibitem{palla_cpm_2005}
\bibinfo{author}{Palla, G.}, \bibinfo{author}{Der\'eny, I.},
  \bibinfo{author}{Farkas, I.} \& \bibinfo{author}{Vicsek, T.}
\newblock \bibinfo{title}{Uncovering the overlapping community structure of
  complex networks in nature and society}.
\newblock \emph{\bibinfo{journal}{Nature}} \textbf{\bibinfo{volume}{435}},
  \bibinfo{pages}{814} (\bibinfo{year}{2005}).

\bibitem{pallaQuantifying_2008}
\bibinfo{author}{Palla, G.}, \bibinfo{author}{Barab\'asi, A.} \&
  \bibinfo{author}{Vicsek, T.}
\newblock \bibinfo{title}{Quantifying social group evolution}.
\newblock \emph{\bibinfo{journal}{Nature}} \textbf{\bibinfo{volume}{446}},
  \bibinfo{pages}{664--667} (\bibinfo{year}{2007}).

\bibitem{ravasz_science_2002}
\bibinfo{author}{Ravasz, E.}, \bibinfo{author}{Somera, A.~L.},
  \bibinfo{author}{Mongru, D.~A.}, \bibinfo{author}{Oltvai, Z.~N.} \&
  \bibinfo{author}{Barab\'asi, A.-L.}
\newblock \bibinfo{title}{Hierarchical organization of modularity in metabolic
  networks}.
\newblock \emph{\bibinfo{journal}{Science}} \textbf{\bibinfo{volume}{297}},
  \bibinfo{pages}{1551--1555} (\bibinfo{year}{2002}).

\bibitem{salespardo_extracting_2007}
\bibinfo{author}{Sales-Pardo, M.}, \bibinfo{author}{Guimera, R.},
  \bibinfo{author}{Moreira, A.} \& \bibinfo{author}{Amaral, L.}
\newblock \bibinfo{title}{Extracting the hierarchical organization of complex
  systems}.
\newblock \emph{\bibinfo{journal}{PNAS}} \textbf{\bibinfo{volume}{104}},
  \bibinfo{pages}{15224--15229} (\bibinfo{year}{2007}).

\bibitem{clauset_nature_2008}
\bibinfo{author}{Clauset, A.}, \bibinfo{author}{Moore, C.} \&
  \bibinfo{author}{Newman, M. E.~J.}
\newblock \bibinfo{title}{Hierarchical structure and the prediction of missing
  links in networks}.
\newblock \emph{\bibinfo{journal}{Nature}} \textbf{\bibinfo{volume}{453}},
  \bibinfo{pages}{98} (\bibinfo{year}{2008}).

\bibitem{yu_ppi_2008}
\bibinfo{author}{Yu, H.} \emph{et~al.}
\newblock \bibinfo{title}{{High-Quality Binary Protein Interaction Map of the
  Yeast Interactome Network}}.
\newblock \emph{\bibinfo{journal}{Science}} \textbf{\bibinfo{volume}{322}},
  \bibinfo{pages}{104--110} (\bibinfo{year}{2008}).

\bibitem{guimera_functional_2005}
\bibinfo{author}{Guimer\`a, R.} \& \bibinfo{author}{Amaral, L. A.~N.}
\newblock \bibinfo{title}{Functional cartography of complex metabolic
  networks}.
\newblock \emph{\bibinfo{journal}{Nature}} \textbf{\bibinfo{volume}{433}},
  \bibinfo{pages}{895--900} (\bibinfo{year}{2005}).

\bibitem{feist_ecoli_2007}
\bibinfo{author}{Feist, A.~M.} \emph{et~al.}
\newblock \bibinfo{title}{A genome-scale metabolic reconstruction for
  \emph{Escherichia coli} k-12 mg1655 that accounts for 1260 orfs and
  thermodynamic information}.
\newblock \emph{\bibinfo{journal}{Molecular Systems Biology}}
  \textbf{\bibinfo{volume}{3}}, \bibinfo{pages}{1} (\bibinfo{year}{2007}).

\bibitem{onnela_structure_2007}
\bibinfo{author}{Onnela, J.-P.} \emph{et~al.}
\newblock \bibinfo{title}{Structure and tie strengths in mobile communication
  networks}.
\newblock \emph{\bibinfo{journal}{PNAS}} \textbf{\bibinfo{volume}{104}},
  \bibinfo{pages}{7332} (\bibinfo{year}{2007}).

\bibitem{gonzales_uncovering_2008}
\bibinfo{author}{Gonzalez, M.~C.}, \bibinfo{author}{Hidalgo, C.~A.} \&
  \bibinfo{author}{Barab\'asi, A.-L.}
\newblock \bibinfo{title}{Understanding individual human mobility patterns}.
\newblock \emph{\bibinfo{journal}{Nature}} \textbf{\bibinfo{volume}{453}},
  \bibinfo{pages}{479} (\bibinfo{year}{2008}).

\bibitem{radicchi-definition-2004}
\bibinfo{author}{Radicchi, F.}, \bibinfo{author}{Castellano, C.},
  \bibinfo{author}{Cecconi, F.}, \bibinfo{author}{Loreto, V.} \&
  \bibinfo{author}{Parisi, D.}
\newblock \bibinfo{title}{Defining and identifying communities in networks}.
\newblock \emph{\bibinfo{journal}{Proceedings of the National Academy of
  Sciences of the United States of America}} \textbf{\bibinfo{volume}{101}},
  \bibinfo{pages}{2658--2663} (\bibinfo{year}{2004}).

\bibitem{newman_finding_2004}
\bibinfo{author}{Newman, M. E.~J.} \& \bibinfo{author}{Girvan, M.}
\newblock \bibinfo{title}{Finding and evaluating community structure in
  networks}.
\newblock \emph{\bibinfo{journal}{Physical Review E}}
  \textbf{\bibinfo{volume}{69}}, \bibinfo{pages}{026113}
  (\bibinfo{year}{2004}).

\bibitem{rosvall_infomap_2008}
\bibinfo{author}{Rosvall, M.} \& \bibinfo{author}{Bergstrom, C.~T.}
\newblock \bibinfo{title}{Maps of random walks on complex networks reveal
  community structure}.
\newblock \emph{\bibinfo{journal}{Proceedings of the National Academy of
  Sciences}} \textbf{\bibinfo{volume}{105}}, \bibinfo{pages}{1118--1123}
  (\bibinfo{year}{2008}).

\bibitem{PhysRevLett.93.218701}
\bibinfo{author}{Reichardt, J.} \& \bibinfo{author}{Bornholdt, S.}
\newblock \bibinfo{title}{Detecting fuzzy community structures in complex
  networks with a potts model}.
\newblock \emph{\bibinfo{journal}{Phys. Rev. Lett.}}
  \textbf{\bibinfo{volume}{93}}, \bibinfo{pages}{218701}
  (\bibinfo{year}{2004}).

\bibitem{Li:arXiv0807.0521}
\bibinfo{author}{Li, D.} \emph{et~al.}
\newblock \bibinfo{title}{Synchronization interfaces and overlapping
  communities in complex networks}.
\newblock \emph{\bibinfo{journal}{Phys. Rev. Lett.}}
  \textbf{\bibinfo{volume}{101}}, \bibinfo{pages}{168701}
  (\bibinfo{year}{2008}).

\bibitem{lancichinetti_detecting_2009}
\bibinfo{author}{Lancichinetti, A.}, \bibinfo{author}{Fortunato, S.} \&
  \bibinfo{author}{Kertesz, J.}
\newblock \bibinfo{title}{Detecting the overlapping and hierarchical community
  structure in complex networks}.
\newblock \emph{\bibinfo{journal}{New Journal of Physics}}
  \textbf{\bibinfo{volume}{11}}, \bibinfo{pages}{033015}
  (\bibinfo{year}{2009}).

\bibitem{FortunatoBarthelemy07_ModularityResolution}
\bibinfo{author}{Fortunato, S.} \& \bibinfo{author}{Barth\'elemy, M.}
\newblock \bibinfo{title}{Resolution limit in community detection}.
\newblock \emph{\bibinfo{journal}{Proceedings of the National Academy of
  Sciences}} \textbf{\bibinfo{volume}{104}}, \bibinfo{pages}{36--41}
  (\bibinfo{year}{2007}).

\bibitem{clauset_2004_finding}
\bibinfo{author}{Clauset, A.}, \bibinfo{author}{Newman, M. E.~J.} \&
  \bibinfo{author}{Moore, C.}
\newblock \bibinfo{title}{Finding community structure in very large networks}.
\newblock \emph{\bibinfo{journal}{Phys. Rev. E}} \textbf{\bibinfo{volume}{70}},
  \bibinfo{pages}{066111} (\bibinfo{year}{2004}).

\bibitem{lancichinetti-comparison-2009}
\bibinfo{author}{Lancichinetti, A.} \& \bibinfo{author}{Fortunato, S.}
\newblock \bibinfo{title}{Community detection algorithms: a comparative
  analysis}.
\newblock \emph{\bibinfo{journal}{Phys. Rev. E}} \textbf{\bibinfo{volume}{80}},
  \bibinfo{pages}{056117} (\bibinfo{year}{2009}).

\bibitem{go}
\bibinfo{author}{{Gene Ontology Consortium}}.
\newblock \emph{\bibinfo{journal}{Nucleic Acids Res.}}
  \textbf{\bibinfo{volume}{36}}, \bibinfo{pages}{D440} (\bibinfo{year}{2008}).

\bibitem{evans_line_2009}
\bibinfo{author}{Evans, T.~S.} \& \bibinfo{author}{Lambiotte, R.}
\newblock \bibinfo{title}{Line graphs, link partitions and overlapping
  communities}.
\newblock \emph{\bibinfo{journal}{Phys. Rev. E}} \textbf{\bibinfo{volume}{80}},
  \bibinfo{pages}{016105} (\bibinfo{year}{2009}).

\bibitem{evans_line2_2009}
\bibinfo{author}{Evans, T.~S.} \& \bibinfo{author}{Lambiotte, R.}
\newblock \bibinfo{title}{{Edge Partitions and Overlapping Communities in
  Complex Networks}}  (\bibinfo{year}{2009}).
\newblock \eprint{arXiv:0912.4389}.

\end{thebibliography}

\begin{thebibliography}{10}
\expandafter\ifx\csname url\endcsname\relax
  \def\url#1{\texttt{#1}}\fi
\expandafter\ifx\csname urlprefix\endcsname\relax\def\urlprefix{URL }\fi
\providecommand{\bibinfo}[2]{#2}
\providecommand{\eprint}[2][]{\url{#2}}

\bibitem{jaccard_1901}
\bibinfo{author}{Jaccard, P.}
\newblock \bibinfo{title}{\'etude comparative de la distribution florale dans
  une portion des alpes et des jura}.
\newblock \emph{\bibinfo{journal}{Bulletin del la Soci\'et\'e Vaudoise des
  Sciences Naturelles}} \textbf{\bibinfo{volume}{37}},
  \bibinfo{pages}{547--579} (\bibinfo{year}{1901}).

\bibitem{newman_finding_2004}
\bibinfo{author}{Newman, M. E.~J.} \& \bibinfo{author}{Girvan, M.}
\newblock \bibinfo{title}{Finding and evaluating community structure in
  networks}.
\newblock \emph{\bibinfo{journal}{Physical Review E}}
  \textbf{\bibinfo{volume}{69}}, \bibinfo{pages}{026113}
  (\bibinfo{year}{2004}).

\bibitem{danon_comparing_2005}
\bibinfo{author}{Danon, L.}, \bibinfo{author}{Diaz-Guilera, A.},
  \bibinfo{author}{Duch, J.} \& \bibinfo{author}{Arenas, A.}
\newblock \bibinfo{title}{Comparing community structure identification}.
\newblock \emph{\bibinfo{journal}{Journal of Statistical Mechanics: Theory and
  Experiment}} \textbf{\bibinfo{volume}{2005}}, \bibinfo{pages}{P09008}
  (\bibinfo{year}{2005}).

\bibitem{shenDetect2009}
\bibinfo{author}{Shen, H.}, \bibinfo{author}{Cheng, X.}, \bibinfo{author}{Cai,
  K.} \& \bibinfo{author}{Hu, M.-B.}
\newblock \bibinfo{title}{Detect overlapping and hierarchical community
  structure in networks}.
\newblock \emph{\bibinfo{journal}{Physica A}} \textbf{\bibinfo{volume}{388}},
  \bibinfo{pages}{1706--1712} (\bibinfo{year}{2009}).

\bibitem{nicosiaExtending2009}
\bibinfo{author}{Nicosia, V.}, \bibinfo{author}{Mangioni, G.},
  \bibinfo{author}{Carchiolo, V.} \& \bibinfo{author}{Malgeri, M.}
\newblock \bibinfo{title}{Extending the definition of modularity to directed
  graphs with overlapping communities}.
\newblock \emph{\bibinfo{journal}{J Stat Mech-Theory E}}
  \bibinfo{pages}{P03024} (\bibinfo{year}{2009}).

\bibitem{PhysRevLett.93.218701}
\bibinfo{author}{Reichardt, J.} \& \bibinfo{author}{Bornholdt, S.}
\newblock \bibinfo{title}{Detecting fuzzy community structures in complex
  networks with a potts model}.
\newblock \emph{\bibinfo{journal}{Phys. Rev. Lett.}}
  \textbf{\bibinfo{volume}{93}}, \bibinfo{pages}{218701}
  (\bibinfo{year}{2004}).

\bibitem{Li:arXiv0807.0521}
\bibinfo{author}{Li, D.} \emph{et~al.}
\newblock \bibinfo{title}{Synchronization interfaces and overlapping
  communities in complex networks}.
\newblock \emph{\bibinfo{journal}{Phys. Rev. Lett.}}
  \textbf{\bibinfo{volume}{101}}, \bibinfo{pages}{168701}
  (\bibinfo{year}{2008}).

\bibitem{lancichinetti_detecting_2009}
\bibinfo{author}{Lancichinetti, A.}, \bibinfo{author}{Fortunato, S.} \&
  \bibinfo{author}{Kertesz, J.}
\newblock \bibinfo{title}{Detecting the overlapping and hierarchical community
  structure in complex networks}.
\newblock \emph{\bibinfo{journal}{New Journal of Physics}}
  \textbf{\bibinfo{volume}{11}}, \bibinfo{pages}{033015}
  (\bibinfo{year}{2009}).

\bibitem{knuth_stanford_1993}
\bibinfo{author}{Knuth, D.~E.}
\newblock \emph{\bibinfo{title}{The Stanford GraphBase: A Platform for
  Combinatorial Computing}} (\bibinfo{publisher}{Addison-Wesley},
  \bibinfo{address}{Reading, MA}, \bibinfo{year}{1993}).

\bibitem{ravasz_science_2002}
\bibinfo{author}{Ravasz, E.}, \bibinfo{author}{Somera, A.~L.},
  \bibinfo{author}{Mongru, D.~A.}, \bibinfo{author}{Oltvai, Z.~N.} \&
  \bibinfo{author}{Barab\'asi, A.-L.}
\newblock \bibinfo{title}{Hierarchical organization of modularity in metabolic
  networks}.
\newblock \emph{\bibinfo{journal}{Science}} \textbf{\bibinfo{volume}{297}},
  \bibinfo{pages}{1551--1555} (\bibinfo{year}{2002}).

\bibitem{palla_cpm_2005}
\bibinfo{author}{Palla, G.}, \bibinfo{author}{Der\'eny, I.},
  \bibinfo{author}{Farkas, I.} \& \bibinfo{author}{Vicsek, T.}
\newblock \bibinfo{title}{Uncovering the overlapping community structure of
  complex networks in nature and society}.
\newblock \emph{\bibinfo{journal}{Nature}} \textbf{\bibinfo{volume}{435}},
  \bibinfo{pages}{814} (\bibinfo{year}{2005}).

\bibitem{rosvall_infomap_2008}
\bibinfo{author}{Rosvall, M.} \& \bibinfo{author}{Bergstrom, C.~T.}
\newblock \bibinfo{title}{Maps of random walks on complex networks reveal
  community structure}.
\newblock \emph{\bibinfo{journal}{Proceedings of the National Academy of
  Sciences}} \textbf{\bibinfo{volume}{105}}, \bibinfo{pages}{1118--1123}
  (\bibinfo{year}{2008}).

\bibitem{newmanFastAlgorithm}
\bibinfo{author}{Newman, M. E.~J.}
\newblock \bibinfo{title}{Fast algorithm for detecting community structure in
  networks}.
\newblock \emph{\bibinfo{journal}{Physical Review E}}
  \textbf{\bibinfo{volume}{69}}, \bibinfo{pages}{066133}
  (\bibinfo{year}{2004}).

\bibitem{clauset_2004_finding}
\bibinfo{author}{Clauset, A.}, \bibinfo{author}{Newman, M. E.~J.} \&
  \bibinfo{author}{Moore, C.}
\newblock \bibinfo{title}{Finding community structure in very large networks}.
\newblock \emph{\bibinfo{journal}{Phys. Rev. E}} \textbf{\bibinfo{volume}{70}},
  \bibinfo{pages}{066111} (\bibinfo{year}{2004}).

\bibitem{palla_directed_2007}
\bibinfo{author}{Palla, G.}, \bibinfo{author}{Farkas, I.~J.},
  \bibinfo{author}{Pollner, P.}, \bibinfo{author}{Derenyi, I.} \&
  \bibinfo{author}{Vicsek, T.}
\newblock \bibinfo{title}{Directed network modules}.
\newblock \emph{\bibinfo{journal}{New Journal of Physics}}
  \textbf{\bibinfo{volume}{9}}, \bibinfo{pages}{186} (\bibinfo{year}{2007}).

\bibitem{PallaQuantifyingSocialGroupEvolution2007}
\bibinfo{author}{Palla, G.}, \bibinfo{author}{Barab\'asi, A.-L.} \&
  \bibinfo{author}{Vicsek, T.}
\newblock \bibinfo{title}{Quantifying social group evolution}.
\newblock \emph{\bibinfo{journal}{Nature}} \textbf{\bibinfo{volume}{446}},
  \bibinfo{pages}{664} (\bibinfo{year}{2007}).

\bibitem{kumpula2008}
\bibinfo{author}{Kumpula, J.~M.}, \bibinfo{author}{Kivel\"a, M.},
  \bibinfo{author}{Kaski, K.} \& \bibinfo{author}{Saram\"aki, J.}
\newblock \bibinfo{title}{Sequential algorithm for fast clique percolation}.
\newblock \emph{\bibinfo{journal}{Phys. Rev. E}} \textbf{\bibinfo{volume}{78}},
  \bibinfo{pages}{026109} (\bibinfo{year}{2008}).

\bibitem{newman_detecting_2004}
\bibinfo{author}{Newman, M. E.~J.}
\newblock \bibinfo{title}{Detecting community structure in networks}.
\newblock \emph{\bibinfo{journal}{The European Physical Journal B}}
  \textbf{\bibinfo{volume}{38}}, \bibinfo{pages}{321--330}
  (\bibinfo{year}{2004}).

\bibitem{newmanGirvanCommsPNAS}
\bibinfo{author}{Girvan, M.} \& \bibinfo{author}{Newman, M. E.~J.}
\newblock \bibinfo{title}{Community structure in social and biological
  networks.}
\newblock \emph{\bibinfo{journal}{Proceedings of the National Academy of
  Sciences}} \textbf{\bibinfo{volume}{99}}, \bibinfo{pages}{7821--7826}
  (\bibinfo{year}{2002}).

\bibitem{lancichinetti-comparison-2009}
\bibinfo{author}{Lancichinetti, A.} \& \bibinfo{author}{Fortunato, S.}
\newblock \bibinfo{title}{Community detection algorithms: a comparative
  analysis}.
\newblock \emph{\bibinfo{journal}{Phys. Rev. E}} \textbf{\bibinfo{volume}{80}},
  \bibinfo{pages}{056117} (\bibinfo{year}{2009}).

\bibitem{newman_whysocial_2003}
\bibinfo{author}{Newman, M. E.~J.} \& \bibinfo{author}{Park, J.}
\newblock \bibinfo{title}{Why social networks are different from other types of
  networks}.
\newblock \emph{\bibinfo{journal}{Physical Review E}}
  \textbf{\bibinfo{volume}{68}}, \bibinfo{pages}{036122}
  (\bibinfo{year}{2003}).

\bibitem{clauset_nature_2008}
\bibinfo{author}{Clauset, A.}, \bibinfo{author}{Moore, C.} \&
  \bibinfo{author}{Newman, M. E.~J.}
\newblock \bibinfo{title}{Hierarchical structure and the prediction of missing
  links in networks}.
\newblock \emph{\bibinfo{journal}{Nature}} \textbf{\bibinfo{volume}{453}},
  \bibinfo{pages}{98} (\bibinfo{year}{2008}).

\bibitem{salespardo_extracting_2007}
\bibinfo{author}{Sales-Pardo, M.}, \bibinfo{author}{Guimera, R.},
  \bibinfo{author}{Moreira, A.} \& \bibinfo{author}{Amaral, L.}
\newblock \bibinfo{title}{Extracting the hierarchical organization of complex
  systems}.
\newblock \emph{\bibinfo{journal}{PNAS}} \textbf{\bibinfo{volume}{104}},
  \bibinfo{pages}{15224--15229} (\bibinfo{year}{2007}).

\bibitem{martinez_grassland_1999}
\bibinfo{author}{Martinez, N.~D.}, \bibinfo{author}{Hawkins, B.~A.} \&
  \bibinfo{author}{adn B.~P.~Feifarek, H. A.~D.}
\newblock \emph{\bibinfo{journal}{Ecology}} \textbf{\bibinfo{volume}{80}},
  \bibinfo{pages}{1044--1055} (\bibinfo{year}{1999}).

\bibitem{vicsek_CP_PRL}
\bibinfo{author}{Der\'enyi, I.}, \bibinfo{author}{Palla, G.} \&
  \bibinfo{author}{Vicsek, T.}
\newblock \bibinfo{title}{Clique percolation in random networks}.
\newblock \emph{\bibinfo{journal}{Phys. Rev. Lett.}}
  \textbf{\bibinfo{volume}{94}}, \bibinfo{pages}{160202}
  (\bibinfo{year}{2005}).

\bibitem{doyonHat2004}
\bibinfo{author}{Doyon, Y.}, \bibinfo{author}{Selleck, W.},
  \bibinfo{author}{Lane, W.~S.}, \bibinfo{author}{Tan, S.} \&
  \bibinfo{author}{C\^ot\'e, J.}
\newblock \bibinfo{title}{Structural and functional conservation of the nua4
  histone acetyltransferase complex from yeast to humans}.
\newblock \emph{\bibinfo{journal}{Mol. Cell. Biol.}}
  \textbf{\bibinfo{volume}{24}}, \bibinfo{pages}{1884} (\bibinfo{year}{2004}).

\bibitem{Dotson12192000}
\bibinfo{author}{Dotson, M.~R.} \emph{et~al.}
\newblock \bibinfo{title}{{Structural organization of yeast and mammalian
  mediator complexes}}.
\newblock \emph{\bibinfo{journal}{Proceedings of the National Academy of
  Sciences of the United States of America}} \textbf{\bibinfo{volume}{97}},
  \bibinfo{pages}{14307--14310} (\bibinfo{year}{2000}).

\bibitem{Wu2004199}
\bibinfo{author}{Wu, P.-Y.~J.}, \bibinfo{author}{Ruhlmann, C.},
  \bibinfo{author}{Winston, F.} \& \bibinfo{author}{Schultz, P.}
\newblock \bibinfo{title}{Molecular architecture of the s. cerevisiae saga
  complex}.
\newblock \emph{\bibinfo{journal}{Molecular Cell}}
  \textbf{\bibinfo{volume}{15}}, \bibinfo{pages}{199 -- 208}
  (\bibinfo{year}{2004}).

\bibitem{AymanSaleh10091998}
\bibinfo{author}{Saleh, A.} \emph{et~al.}
\newblock \bibinfo{title}{{Tra1p Is a Component of the Yeast AdaáSpt
  Transcriptional Regulatory Complexes}}.
\newblock \emph{\bibinfo{journal}{J. Biol. Chem.}}
  \textbf{\bibinfo{volume}{273}}, \bibinfo{pages}{26559--26565}
  (\bibinfo{year}{1998}).

\bibitem{Brown06222001}
\bibinfo{author}{Brown, C.~E.} \emph{et~al.}
\newblock \bibinfo{title}{{Recruitment of HAT Complexes by Direct Activator
  Interactions with the ATM-Related Tra1 Subunit}}.
\newblock \emph{\bibinfo{journal}{Science}} \textbf{\bibinfo{volume}{292}},
  \bibinfo{pages}{2333--2337} (\bibinfo{year}{2001}).

\bibitem{Bhaumik02012004}
\bibinfo{author}{Bhaumik, S.~R.}, \bibinfo{author}{Raha, T.},
  \bibinfo{author}{Aiello, D.~P.} \& \bibinfo{author}{Green, M.~R.}
\newblock \bibinfo{title}{{In vivo target of a transcriptional activator
  revealed by fluorescence resonance energy transfer}}.
\newblock \emph{\bibinfo{journal}{Genes \& Development}}
  \textbf{\bibinfo{volume}{18}}, \bibinfo{pages}{333--343}
  (\bibinfo{year}{2004}).

\bibitem{Baumeister1998367}
\bibinfo{author}{Baumeister, W.}, \bibinfo{author}{Walz, J.},
  \bibinfo{author}{Z{\"{u}}hl, F.} \& \bibinfo{author}{Seem{\"{u}}ller, E.}
\newblock \bibinfo{title}{The proteasome: Paradigm of a self-compartmentalizing
  protease}.
\newblock \emph{\bibinfo{journal}{Cell}} \textbf{\bibinfo{volume}{92}},
  \bibinfo{pages}{367 -- 380} (\bibinfo{year}{1998}).

\bibitem{Boyle12122004}
\bibinfo{author}{Boyle, E.~I.} \emph{et~al.}
\newblock \bibinfo{title}{{GO::TermFinder--open source software for accessing
  Gene Ontology information and finding significantly enriched Gene Ontology
  terms associated with a list of genes}}.
\newblock \emph{\bibinfo{journal}{Bioinformatics}}
  \textbf{\bibinfo{volume}{20}}, \bibinfo{pages}{3710--3715}
  (\bibinfo{year}{2004}).

\bibitem{tanimoto_elementary_1958}
\bibinfo{author}{Tanimoto, T.~T.}
\newblock \bibinfo{title}{An elementary mathematical theory of classification
  and prediction}.
\newblock \bibinfo{type}{Tech. Rep.}, \bibinfo{institution}{IBM Internal
  Report} (\bibinfo{year}{1958}).

\bibitem{bagrowbollt:lcd}
\bibinfo{author}{Bagrow, J.~P.} \& \bibinfo{author}{Bollt, E.~M.}
\newblock \bibinfo{title}{A local method for detecting communities}.
\newblock \emph{\bibinfo{journal}{Phys. Rev. E}} \textbf{\bibinfo{volume}{72}},
  \bibinfo{pages}{046108} (\bibinfo{year}{2005}).

\bibitem{bagrow:EvalLocalMethods}
\bibinfo{author}{Bagrow, J.~P.}
\newblock \bibinfo{title}{Evaluating local community methods in networks}.
\newblock \emph{\bibinfo{journal}{J. Stat. Mech.}}
  \textbf{\bibinfo{volume}{2008}}, \bibinfo{pages}{P05001}
  (\bibinfo{year}{2008}).

\bibitem{clauset:localcomm}
\bibinfo{author}{Clauset, A.}
\newblock \bibinfo{title}{Finding local community structure in networks}.
\newblock \emph{\bibinfo{journal}{Physical Review E}}
  \textbf{\bibinfo{volume}{72}}, \bibinfo{pages}{026132}
  (\bibinfo{year}{2005}).

\bibitem{newman_mixture_2007}
\bibinfo{author}{Newman, M. E.~J.} \& \bibinfo{author}{Leicht, E.~A.}
\newblock \bibinfo{title}{Mixture models and exploratory analysis in networks}.
\newblock \emph{\bibinfo{journal}{Proceedings of the National Academy of
  Sciences}} \textbf{\bibinfo{volume}{104}}, \bibinfo{pages}{9564--9569}
  (\bibinfo{year}{2007}).

\bibitem{guimera_functional_2005}
\bibinfo{author}{Guimer\`a, R.} \& \bibinfo{author}{Amaral, L. A.~N.}
\newblock \bibinfo{title}{Functional cartography of complex metabolic
  networks}.
\newblock \emph{\bibinfo{journal}{Nature}} \textbf{\bibinfo{volume}{433}},
  \bibinfo{pages}{895--900} (\bibinfo{year}{2005}).

\bibitem{fortunato_community_2009}
\bibinfo{author}{Fortunato, S.} \& \bibinfo{author}{Castellano, C.}
\newblock \emph{\bibinfo{title}{Community Structure in Graphs}}
  (\bibinfo{publisher}{Springer}, \bibinfo{year}{2009}).

\bibitem{watts_wsmodel_1998}
\bibinfo{author}{Watts, D.~J.} \& \bibinfo{author}{Strogatz, S.~H.}
\newblock \bibinfo{title}{Collective dynamics of `small-world' networks}.
\newblock \emph{\bibinfo{journal}{Nature}} \textbf{\bibinfo{volume}{393}},
  \bibinfo{pages}{440} (\bibinfo{year}{1998}).

\bibitem{yu_ppi_2008}
\bibinfo{author}{Yu, H.} \emph{et~al.}
\newblock \bibinfo{title}{{High-Quality Binary Protein Interaction Map of the
  Yeast Interactome Network}}.
\newblock \emph{\bibinfo{journal}{Science}} \textbf{\bibinfo{volume}{322}},
  \bibinfo{pages}{104--110} (\bibinfo{year}{2008}).

\bibitem{feist_ecoli_2007}
\bibinfo{author}{Feist, A.~M.} \emph{et~al.}
\newblock \bibinfo{title}{A genome-scale metabolic reconstruction for
  \emph{Escherichia coli} k-12 mg1655 that accounts for 1260 orfs and
  thermodynamic information}.
\newblock \emph{\bibinfo{journal}{Molecular Systems Biology}}
  \textbf{\bibinfo{volume}{3}}, \bibinfo{pages}{1} (\bibinfo{year}{2007}).

\bibitem{onnela_structure_2007}
\bibinfo{author}{Onnela, J.-P.} \emph{et~al.}
\newblock \bibinfo{title}{Structure and tie strengths in mobile communication
  networks}.
\newblock \emph{\bibinfo{journal}{PNAS}} \textbf{\bibinfo{volume}{104}},
  \bibinfo{pages}{7332} (\bibinfo{year}{2007}).

\bibitem{onnela_analysis_2007}
\bibinfo{author}{Onnela, J.-P.} \emph{et~al.}
\newblock \bibinfo{title}{Analysis of a large-scale weighted network of
  one-to-one human communication}.
\newblock \emph{\bibinfo{journal}{New Journal of Physics}}
  \textbf{\bibinfo{volume}{9}}, \bibinfo{pages}{179} (\bibinfo{year}{2007}).

\bibitem{pallaQuantifying_2008}
\bibinfo{author}{Palla, G.}, \bibinfo{author}{Barab\'asi, A.} \&
  \bibinfo{author}{Vicsek, T.}
\newblock \bibinfo{title}{Quantifying social group evolution}.
\newblock \emph{\bibinfo{journal}{Nature}} \textbf{\bibinfo{volume}{446}},
  \bibinfo{pages}{664--667} (\bibinfo{year}{2007}).

\bibitem{gonzales_uncovering_2008}
\bibinfo{author}{Gonzalez, M.~C.}, \bibinfo{author}{Hidalgo, C.~A.} \&
  \bibinfo{author}{Barab\'asi, A.-L.}
\newblock \bibinfo{title}{Understanding individual human mobility patterns}.
\newblock \emph{\bibinfo{journal}{Nature}} \textbf{\bibinfo{volume}{453}},
  \bibinfo{pages}{479} (\bibinfo{year}{2008}).

\bibitem{IMDB}
\bibinfo{author}{IMDb}.
\newblock \bibinfo{title}{http://www.imdb.com} (\bibinfo{year}{2009}).

\bibitem{fowlerConnectingCongress2006}
\bibinfo{author}{Fowler, J.}
\newblock \bibinfo{title}{Connecting the congress: A study of cosponsorship
  networks}.
\newblock \emph{\bibinfo{journal}{Political Analysis}}
  \textbf{\bibinfo{volume}{14}}, \bibinfo{pages}{456--487}
  (\bibinfo{year}{2006}).

\bibitem{fowlerCosponsorNets2006}
\bibinfo{author}{Fowler, J.}
\newblock \bibinfo{title}{Legislative cosponsorship networks in the u.s. house
  and senate}.
\newblock \emph{\bibinfo{journal}{Social Networks}}
  \textbf{\bibinfo{volume}{28}}, \bibinfo{pages}{454--465}
  (\bibinfo{year}{2006}).

\bibitem{pooleRecovering1998}
\bibinfo{author}{Poole, K.~T.}
\newblock \bibinfo{title}{Recovering a basic space from a set of issue scales}.
\newblock \emph{\bibinfo{journal}{American Journal of Political Science}}
  \textbf{\bibinfo{volume}{42}}, \bibinfo{pages}{954--993}
  (\bibinfo{year}{1998}).

\bibitem{PooleBook2005}
\bibinfo{author}{Poole, K.~T.}
\newblock \emph{\bibinfo{title}{Spatial Models of Parliamentary Voting}}
  (\bibinfo{publisher}{Cambridge University Press}, \bibinfo{address}{New
  York}, \bibinfo{year}{2005}).

\bibitem{fellbaum:wordnet}
\bibinfo{author}{Fellbaum, C.}
\newblock \emph{\bibinfo{title}{WordNet: An Electronical Lexical Database}}
  (\bibinfo{publisher}{The MIT Press}, \bibinfo{address}{Cambridge, MA},
  \bibinfo{year}{1998}).

\bibitem{go}
\bibinfo{author}{{Gene Ontology Consortium}}.
\newblock \emph{\bibinfo{journal}{Nucleic Acids Res.}}
  \textbf{\bibinfo{volume}{36}}, \bibinfo{pages}{D440} (\bibinfo{year}{2008}).

\bibitem{yu_tam_2007}
\bibinfo{author}{Yu, H.}, \bibinfo{author}{Jansen, R.},
  \bibinfo{author}{Stolovitzky, G.} \& \bibinfo{author}{Gerstein, M.}
\newblock \bibinfo{title}{{Total ancestry measure: quantifying the similarity
  in tree-like classification, with genomic applications}}.
\newblock \emph{\bibinfo{journal}{Bioinformatics}}
  \textbf{\bibinfo{volume}{23}}, \bibinfo{pages}{2163--2173}
  (\bibinfo{year}{2007}).

\bibitem{kegg}
\bibinfo{author}{Kanehisa, M.} \& \bibinfo{author}{Goto, S.}
\newblock \emph{\bibinfo{journal}{Nucleic Acids Res.}}
  \textbf{\bibinfo{volume}{28}}, \bibinfo{pages}{27--30}
  (\bibinfo{year}{2000}).

\bibitem{nelson1998}
\bibinfo{author}{Nelson, D.~L.}, \bibinfo{author}{McEvoy, C.~L.} \&
  \bibinfo{author}{Schreiber, T.~A.}
\newblock \bibinfo{title}{The university of south florida word association,
  rhyme, and word fragment norms (http://www.usf.edu/freeassociation/)}
  (\bibinfo{year}{1998}).

\bibitem{LeskobitchLarge}
\bibinfo{author}{Leskovec, J.}, \bibinfo{author}{Lang, K.~J.},
  \bibinfo{author}{Dasgupta, A.} \& \bibinfo{author}{Mahoney, M.~W.}
\newblock \bibinfo{title}{Community structure in large networks: Natural
  cluster sizes and the absence of large well-defined clusters}.
\newblock \emph{\bibinfo{journal}{CoRR}}
  \textbf{\bibinfo{volume}{abs/0810.1355}} (\bibinfo{year}{2008}).

\end{thebibliography}

\end{document}